\RequirePackage{rotating}
\RequirePackage{fix-cm}
\documentclass[12pt]{article}    
\usepackage{geometry} 
\usepackage{graphicx}
\usepackage{natbib,mathrsfs,amsmath,amssymb,soul,amsthm}
\usepackage[dvipsnames]{xcolor}
\usepackage{authblk}

\newtheorem{theorem}{Theorem}
\newtheorem{corollary}{Corollary}
\newtheorem{lemma}{Lemma}

\definecolor{darkred}{RGB}{150,50,50}
\definecolor{brown}{RGB}{250,100,100}
\definecolor{green}{RGB}{000,150,100}
\definecolor{purple}{RGB}{250,000,180}

\newcommand{\bSigma}{\mbox{\boldmath $\Sigma$}}
\def\bse{\begin{eqnarray*}}
\def\ese{\end{eqnarray*}}
\def\be{\begin{eqnarray}}
\def\ee{\end{eqnarray}}
\def\bsq{\begin{equation*}}
\def\esq{\end{equation*}}
\def\bq{\begin{equation}}
\def\eq{\end{equation}}
\def\th{^{th}}
\def\var{\hbox{var}}

\def\wh{\widehat}
\def\wt{\widetilde}

\def\th{^{\rm th}}

\def\cE{{\cal E}}
\def\n{\nonumber}

\def\argmin{\mbox{argmin}}
\def\argmax{\mbox{argmax}}
\def\min{{\rm min}}
\def\max{{\rm max}}

\def\sumi{\sum_{i=1}^n}

\def\trans{^{\sf \scriptscriptstyle T}}

\def\bb{{\boldsymbol\beta}}
\def\bg{{\boldsymbol\gamma}}

\def\bnu{{\boldsymbol\nu}}

\def\btheta{{\boldsymbol \theta}}

\def\btau{{\boldsymbol\tau}}

\def\a{{\bf a}}
\def\A{{\bf A}}
\def\B{{\bf B}}
\def\C{{\bf C}}

\def\e{{\bf e}}
\def\k{{\bf k}}
\def\I{{\bf I}}
\def\bfJ{{\bf J}}

\def\m{{\bf m}}

\def\r{{\bf r}}
\def\T{{\bf T}}
\def\U{{\bf U}}
\def\V{{\bf V}}
\def\v{{\bf v}}
\def\W{{\bf W}}

\def\Z{{\bf Z}}

\def\bS{{\bf S}}
\def\calE{{\cal E}}

\def\calG{{\cal G}}
\def\Normal{\hbox{Normal}}
\def\MNormal{\hbox{MNormal}}
\def\Uniform{\hbox{Uniform}}
\def\Gamma{\hbox{Gamma}}

\def\Poisson{\hbox{Poisson}}
\def\0{{\bf 0}}
\def\pr{\hbox{pr}}
\def\sup{\hbox{sup}}
\def\wh{\widehat}
\def\wt{\widetilde}

\def\log{\hbox{log}}

\def\exp{\hbox{exp}}

\def\boxit#1{\vbox{\hrule\hbox{\vrule\kern6pt\vbox{\kern6pt#1\kern6pt}\kern6pt\vrule}\hrule}}

\definecolor{purple}{rgb}{0.84, 0.17, 0.89}

\def\supone{^{\scriptscriptstyle [1]}}
\def\supj{^{\scriptscriptstyle [j]}}
\def\supg{^{\scriptscriptstyle [g]}}
\def\supq{^{\scriptscriptstyle [q]}}

\def\Asc{\mathcal{A}}

\def\Hsc{\mathcal{H}}
\def\bHsc{\boldsymbol{\Hsc}}

\def\Nscr{\mathscr{N}}
\def\bNscr{\boldsymbol{\Nscr}}
\def\Nsc{\mathcal{N}}
\def\bNsc{\boldsymbol{\Nsc}}

\def\bWhat{\widehat{\W}}
\def\bZhat{\widehat{\Z}}

\def\Uscr{\mathscr{U}}
\def\Lscr{\mathscr{L}}

\def\bWhat{\widehat{\W}}
\def\bZhat{\widehat{\Z}}
\def\subplus{_{\scriptscriptstyle \sf +}}
\def\subpplus{_{\scriptscriptstyle \sf ++}}
\def\submu{_{\scriptscriptstyle \mu}}
\def\subG{_{\scriptscriptstyle G}}

\def\subMLE{_{\scriptscriptstyle \sf MLE}}
\def\subMLEg{_{\scriptscriptstyle \sf MLE,[g]}}
\def\suphmu{^{\scriptscriptstyle h_{\mu}\supj}}
\def\suphg{^{\scriptscriptstyle h_{g}\supj}}

\def\subglasso{_{\scriptscriptstyle \sf glasso}}

\def\fpr{\mbox{FPR}}

\def\ape{\mbox{APE}}

\def\Cscr{\mathscr{C}}
\def\sumin{\sum_{i=1}^n}
\def\ninv{n^{-1}}
\def\subDelta{_{\scriptscriptstyle \Delta}}

\begin{document}

\author[1]{Liang Liang}
\author[1,2]{Jue Hou}
\author[2,3]{Hajime Uno}
\author[2,4]{Kelly Cho}
\author[5]{Yanyuan Ma}
\author[1,2,6]{Tianxi Cai}
\affil[1]{Department of Biostatistics, Harvard T.H. Chan School of Public Health}
\affil[2]{Massachusetts Veterans Epidemiology Research and Information Center, US Department of Veteran Affairs}
\affil[3]{Department of Medical Oncology, Dana-Farber Cancer Institute}
\affil[4]{Brigham and Women’s Hospital, Harvard Medical School}
\affil[5]{Department of Statistics, Penn State University}
\affil[6]{Department of Biomedical Informatics, Harvard Medical School}

\title{Semi-supervised Approach to Event Time Annotation Using
Longitudinal Electronic Health Records}

\maketitle

\begin{abstract}
Large clinical datasets derived from insurance claims and electronic health record (EHR) systems
are valuable sources for precision medicine research. These datasets can be used to
develop models for personalized prediction of risk or treatment response. Efficiently deriving
prediction models using real world data, however, faces practical and methodological challenges.
Precise information on important clinical outcomes such as time to cancer progression are not readily
available in these databases. The true clinical event times typically cannot be approximated well based on simple extracts of billing or procedure codes. Whereas, annotating event times manually is time and resource prohibitive. In this paper, we propose a two-step semi-supervised
multi-modal automated time annotation (MATA) method leveraging multi-dimensional
longitudinal EHR encounter records. In step I, we employ a functional principal component analysis
approach to estimate the underlying intensity functions based on
observed point processes from the unlabeled patients. In step II, we fit a penalized proportional odds model to
the event time outcomes with features derived in step I in the labeled data where the non-parametric
baseline function is approximated using B-splines. Under regularity conditions, the resulting estimator of the feature effect vector is shown as root-$n$ consistent. We demonstrate the superiority of our
approach relative to existing approaches through simulations and a real data example on annotating
lung cancer recurrence in an EHR cohort of lung cancer patients from Veteran Health Administration.
\end{abstract}

\textbf{Keywords: } censoring,  electronic health records,  functional principle component analysis, point process, proportional odds model, Semi-supervised learning. 

\section{Introduction}\label{sec1}

While clinical trials and traditional cohort studies remain critical sources of data for clinical research, they have limitations including the generalizability of the study findings and the limited ability to test broader hypotheses. In recent years, real world clinical data derived from disease registry, insurance claims and electronic health record (EHR) systems are increasingly used for precision medicine research. These real word data (RWD) open opportunities for developing accurate personalized risk prediction models, which can be easily incorporated into clinical practice and ultimately realize the promise of precision medicine.
Efficiently deriving prediction models for the risk of developing future clinical events using RWD, however, faces practical and methodological challenges. Precise event time information such as time to cancer recurrence is not readily available in RWD such as EHR and claims data. Simple proxies to the event time based on the encounter time of first diagnosis or procedure codes may poorly approximate the true event time \citep{uno2018determining}. On the other hand, annotating event times manually via chart review is time and resource prohibitive.

Growing efforts have been made in recent years to predict the onset time of clinical events using a large-scale medical encounter data set that lacks precise onset time and a small training set with gold standard labels on the true onset time. Several algorithms exist for predicting cancer recurrence time by extracting features from the encounter pattern of relevant codes. For example, \cite{chubak2015electronic} proposed a rule based algorithm that classifies the recurrence status, $R \in \{+, -\}$, based on decision tree, and assign the recurrence time for those with predicted $R = +$ as the earliest encounter time of one or more specific codes.  \cite{hassett2015detecting} proposed  two-step algorithms where a logistic
regression was used to classify $R$ in step I and then the recurrence time for those with $R = +$ is estimated as a weighted average of the times that the counts of several pre-specified codes peaked. Instead of peak time, \cite{uno2018determining} focuses on the time at which an empirically estimated encounter intensity function has the sharpest change, referred to as the change point throughout the paper.
The recurrence time is approximated as a weighted average of the change point times associated with a few selected codes. Despite of their reasonable empirical performance, these ad hoc procedures have several major limitations. First, only a very small number of codes are selected according to domain knowledge. Second, intensity function estimated based on finite difference may yield substantial variability in the resulting peak or change times due to the sparsity of encounter data.

In this paper, we frame the question of annotating event time with longitudinal encounter records as a statistical question of predicting an event time $T$  using baseline covariates $\U$ as well as features derived from a $p$-variat point process, $\bNscr = (\Nscr_1, ..., \Nscr_p)$.
Specifically, with a small labeled set $\Lscr$ containing observations on $\{T, \U, \bNscr\}$ and a large unlabeled set $\Uscr$ containing observations
on $\U$ and $\bNscr$ only, we propose a two-step semi-supervised
multi-modal automated time annotation (MATA)
procedure by first estimating the underlying subject specific intensity
functions associated with $\bNscr$ and deriving summaries of the intensity functions, denoted by $\bWhat$, as features for predicting $T$. In the second
step, we predict $T$ using $\bZhat = (\U\trans,\bWhat\trans)\trans$ by fitting a penalized proportional odds (PO) model which approximates the non-parametric
baseline function via B-splines. Estimating individualized intensity functions is a challenging task in the current setting because the encounter data is often sparse and the
shape of the intensity functions can vary greatly across subjects. As such traditional multiplicative intensity models
\citep{lawless1987regression, dean1997efficiency, nielsen2005regression} fail to provide accurate approximations. To overcome those difficulties,
we employed a non-parametric FPCA method by \cite{wu2013functional} to estimate the subject specific intensity functions using the large unlabeled set
$\Uscr$. We demonstrate that when the size of $\Uscr$ is sufficiently large relative to the size of $\Lscr$,  the approximation error of $\|\bWhat-\W\|$ can
is ignorable compared to the estimation error from fitting the spline model in $\Lscr$.

Even though the idea of employing a spline-based approach is straightforward and intuitive,
our method differs from the classical B-spline works in the sense that
B-splines are used on the outcome model, i.e., the failure time, rather than the preprocessing of the predictors.
Special attention is devised to accommodate this fact.
We established the novel consistency results and asymptotic convergence rates for the proposed estimator, both the parametric and nonparametric part.
There are some existing literature adopting a spline-based approach in a similar context as ours,
including \cite{shen1998propotional,royston2002flexible,zhang2010spline,younes1997link}. However, \cite{royston2002flexible} and \cite{younes1997link}
did not address the asymptotic properties of their estimators at all; \cite{shen1998propotional} and \cite{zhang2010spline} employed a seive maximum likelihood based approach which considers spline as a special case but only provided theoretical justification on the asymptotics of
the parametric part.

One great advantage of the proposed MATA approach is
the easy implementation of classical variable selection algorithms such as LASSO.
In comparison, \cite{chubak2015electronic,hassett2015detecting,uno2018determining}
exhaust all possible combinations of selected encounters and select the optimal
one under certain criteria, which brings great computational complexity.
No variable selection method has been developed for classical estimating equation based estimators,
e.g., \cite{cheng1995analysis,cheng1997predicting}. Besides, compared to the non-parametric maximum likelihood-estimator (NPMLE), e.g., \cite{zeng2005maximum}, which approximates the non-parametric function by a right-continuous step function with jumps only at observed failure time, our approach is computationally more efficient and stable.

The rest of the paper is organized as follows. In Section 2, we introduce
the proposed MATA approach and prediction accuracy evaluation
measures.
The asymptotic properties of the proposed estimator are
discussed in Section 3. In Section 4, we conduct various simulation studies to explore the performance
of our approach under small labeled sample. In Section 5, we apply our approach to a lung cancer data set.
Section 6 contains a short discussion. Technical details and proofs are provided in the Supplementary Material.

\section{Semi-supervised MATA}\label{sec2}

Let $T$ denote the continuous event time of interest which is observable up to $(X, \Delta)$ in $\Lscr$, where $X = \min(T, C)$, $\Delta = I(T \le C)$ and $C$ is the follow up time.
Let $\bNsc = (\Nsc\supone, ..., \Nsc\supq)\trans$ denote the $q$-variat point processes and $\U$ denote
baseline covariates observable in both $\Lscr$ and $\Uscr$, where $\Nsc\supj$ is a point process associated with the $j$th clinical code whose occurrence times
are $\{t_1\supj, t_2\supj, ...\}$ with $\Nsc\supj(\Asc) = \sum_s I(t_s\supj \in \Asc)$ for any set $\Asc$ in the Borel $\sigma$-algebra of
the positive half of the real line and $I(\cdot)$ is the indicator function. If $T$ denotes the true event time of heart failure, examples of $\Nsc\supj$ include longitudinal encounter processes of diagnostic code for heart failure and NLP mentions of heart failure in clinical notes.
The local intensity function for $\Nsc\supj$ is
$\lambda\supj(t) = E\{d\Nsc\supj(t)\} / dt$, $t\geq0$. Here we assume $\lambda\supj(t)$
is integrable, i.e., $\tau\supj = \int_0^{\infty} \lambda\supj(u)du<\infty$, for $j=1,\cdots,q$. Then the corresponding random density trajectory is $f\supj(t) = \lambda\supj(t)/\tau\supj$,
$t\geq 0$. Equivalently, $\lambda\supj(t)=\tau\supj f\supj(t)
=E\{\Nsc\supj[0,\infty)\}f\supj(t)$, i.e.
the intensity function $\lambda\supj(t)$ is the product of the
density trajectory $f\supj(t)$ and  the
expected lifetime encounters.

The encounter times of the point processes are also only observable up to the end of
follow up $C$ and we let $M\supj = \Nsc\supj([0,C])$ denote the total number of occurrences for $\Nsc\supj$ up to $C$ and $\bHsc_{t}$ denote the history of
$\bNsc$ up to $t$ along with the baseline covariate vector $\U$. Thus, the observed data
consist of
$$
\Lscr = \{(X_i, \Delta_i, C_i, \bHsc_{iC_i}): i = 1, \cdots, n\}, \quad \mbox{and} \quad
\Uscr = \{(C_i,\bHsc_{iC_i}): i = n+1, \cdots, n+N\} ,
$$
where $i$ indexes the subject and we assume that $N \gg n$.

\subsection{Models}

Our proposed MATA procedure involves two models, one for the point processes and another for the
survival function of $T$.
We connect two models by including the underlying intensity functions for the point processes
as the part of the covariates for survival time.

\paragraph{Point Process Model}
The intensity function $\lambda\supj(t)$ for $t\geq 0$ is treated as a
realization of a non-negative valued
stochastic intensity process $\Lambda\supj(t)$.
Conditional on $\Lambda\supj=\lambda\supj$,
the number of observed events is assumed to be a
non-homogeneous Poisson process
with local intensity function $\lambda\supj(t)$ that satisfies
$E\{\Nsc\supj(a,b)\mid \Lambda\supj=\lambda\supj\}=\int_a^b
\lambda\supj(u)du$, where $0\leq a\leq b<\infty$. Thus,
$\tau\supj=E\{\Nsc\supj[0,\infty)\mid \Lambda\supj=\lambda\supj\}$.
Define the truncated random density
\bse
f_C\supj(t) = f\supj(t)/\int_0^C f\supj(t)dt=\lambda\supj(t)/\int_0^C\lambda\supj(t)dt, ~t\in[0,C];
\ese
and its scaled version
\bse
f_{C,\rm scaled}\supj(t) = Cf_C\supj(Ct) = \lambda\supj(Ct)/\int_0^1 \lambda\supj(Ct)dt, ~t\in[0,1].
\ese
As we only observe the point process $\Nsc\supj$ up to $C$, our goal is to estimate the truncated density function $f_C\supj(t)$ or equivalently the scaled density function $f_{C,\rm scaled}\supj(t)$ rather than the density function $f\supj(t)$. Note the scaling is done to meet the uniform endpoint requirement of the FPCA approach by \cite{wu2013functional}.

\paragraph{Event Time Model}
We next relate features derived from the intensity functions to the distribution of $T$. Define $\W\supj = \calG\circ f\supj$, where $\calG$ is a known functional.
For example, if the local
intensity function $f\supj(x)$ follows the exponential distribution with rate $\theta^{-1}$, then we
may set $\calG\circ f\supj = \int xf\supj(x)dx = \theta$. Other
potential features include peaks or change points of intensity
functions $\lambda\supj$. Here peak is defined as the time that the intensity
(or density) curve reaches maximum, while change point is defined
as the time of the largest increases in the intensity (or density) curve.
Due to censoring, we instead have $\W_C\supj = \calG\circ f_C\supj$.
For features like peak and change point, $\W\supj$ and $\W_C\supj$ would be identical
if they were reached before censoring time $C$. We assume that $T \mid \Z=(\U\trans,\W\trans)\trans$ follows an PO model \citep{klein2006survival}
\be\label{eq:surv}
F(t\mid\Z) \equiv \pr(T\leq t\mid \Z) &=& \frac{\exp(\bb\trans\Z) \alpha(t)}{1+\exp(\bb\trans\Z)\alpha(t) }\quad
\mbox{with}\quad \alpha(t) =  \int_0^t \exp\{m(s)\}ds,
\ee
where  $\W=({\W\supone}\trans,\cdots,$ ${\W\supq}\trans)\trans$, $\bb$ is the unknown effect vector of the derived features $\Z$, and $m(t)$ is an unknown smooth function of $t$. This formulation ensures that $\alpha(t)=\int_0^t \exp\{m(s)\}ds$ is  positive and increasing for $t \in (0, \infty)$.

\subsection{Estimation}\label{sec-est}
To derive a prediction rule for $T$ based on the proposed MATA procedure, we first estimate truncated density function $f_C\supj(t)$ from the longitudinal encounter data using the FPCA method proposed by \cite{wu2013functional} to obtain estimates for the derived features $\W$ using unlabeled data $\Uscr$, denoted by $\bWhat$. Then we estimate $\alpha(t)$ and $\bb$ using synthetic observations in the labeled set $\{(X_i, \Delta_i, \bZhat_i\trans), i = 1, ..., n\}$ via penalized estimation with regression spline.

\subsubsection{Step I: Estimation of  $f\supj$}\label{sec-est-intensity}

We estimate the mean $f\supj_{\mu, \rm scaled}(t)$ and variance $G\supj_{\rm scaled}(t,s)$ of the
scaled density function $f_{C, \rm scaled}\supj(t)$
according to the FPCA approach by \cite{wu2013functional}.
Using the estimators $\wh f\supj_{\mu,\rm scaled}(t)$
and $\wh G\supj_{\rm scaled}(t,s)$,
we predict the
scaled density function by $\wh f\supj_{iK,\rm scaled}(t)$ with truncation at zero to ensure nonnegativity of the density function.
The index $K$ in the subscript is the number of basis
functions selected according to the proportion of variation
 explained.  We obtain the truncated density function by
$$
\wh f\supj_{iK}(t)= \wh f\supj_{iK,\rm scaled}(t/C_i)/
\int_0^{C_i} \wh f\supj_{iK,\rm scaled}(t/C_i) dt.
$$
For the $i$-th patient and its $j$-th point process $\Nsc_i\supj$, we
only observe one realization of its expected number of encounters on
$[0,C_i]$, i.e., $M_i=\Nsc_i\supj([0,C_i])$. We approximate the expected numbers of encounters with observed encounters, and  estimate $\lambda_i(t)$ as $\wh\lambda_i\supj(t)=M_i\wh f_{iK}\supj(t)$, for $t\in[0,C_i]$.
We further estimate the derived feature $\W_i$ as
$\wh \W_i = \calG\circ \wh f_i\supj$. Detailed form of these estimators are given in \ref{app:FPCA}. We also establish the rate of convergence for the estimated loadings of the functional PCA, which can be subsequently used as potential derived features.

\subsubsection{Step II. PO Model Estimation with B-spline
  Approximation to $m(\cdot)$}\label{sec-est-PO}

To estimate $m(t)$ and $\bb$ in the PO  model  (\ref{eq:surv}), we approximate $m(t)$
via B-splines with order $r$ (degree $r-1$) described as
follows. Divide the support of censoring time $C$, denoted as $[0,\cE]$, into $(R_n+1)$ subintervals  $\{(\xi_p,\xi
_{p+1}),p=r,r+1,\ldots ,R_n+r-1\}$, where $(\xi
_p)_{p=r+1}^{R_n+r}$ is a sequence of interior knots,
$\xi_{1}=\cdots =\xi_{r}=0<\xi_{r+1}<\cdots \xi_{r+R_n}<\calE=\xi
_{R_n+r+1}=\cdots =\xi_{R_n+2r}.$
Let the basis functions be
$\B_{r}(t)=\{B_{r,1}(t),\dots, B_{r,P_n}(t)\}\trans$ where the number of B-spline basis functions $P_n=R_n+r$. Then $m(t)$
can be approximated by
\bse
m(t;\bg) &=& \bg\trans\B_r(t) = \sum_{p=1}^{P_n}B_{r,p}(t)\gamma_p,
\ese
where $\bg=(\gamma_1,\dots ,\gamma_{P_n})\trans$ is the vector of coefficients for the spline basis functions
$\B_{r}(t)$.

With the B-spline formulation and features $\W_i\supj = \calG \circ f_i\supj$ estimated as $\wh \W_i\supj = \calG \circ \wh f_i\supj$, we can estimate $m(\cdot)$ and $\bb$
by maximizing an estimated likelihood. Specifically, let
\be
l_n(\bb,\bg )&=&\sumi\log\{\wt H_i(\bb,\bg)\}\n\\
&=&\sumi\left[
\Delta_i\{\B_r\trans(X_i)\bg+\wh\Z_i\trans\bb\}\right.\n\\
&&\hskip12mm\left.-
(1+
\Delta_i)\log\left\{1+\exp(\wh\Z_i\trans\bb)  \int_0^{X_i}\exp\{\bg\trans\B_r(t)\}dt\right\}\right]
\label{eq:lnbg}
\ee
where $\wh\Z_i = (\U_i\trans,\wh\W_i{\supone}\trans, ..., \wh\W_i{\supq}\trans)\trans$,
\bse
\wt{H}_i(\bb,\bg)&=&
\frac{\exp[\Delta_i\{\B_r\trans(X_i)\bg+\wh\Z_i\trans\bb\}]}
{[1+\exp(\wh\Z_i\trans\bb)\int_0^{X_i}\exp\{\B_r\trans(t)\bg\}dt]^{(1+\Delta_i)}}.
\ese
Then we may estimate $\bb$ by maximizing the approximated profile likelihood
$$\wh\bb\subMLE = \argmax_{\bb} l_n(\left\{\bb,\wh\bg(\bb)\right\} ,$$
where $\wh\bg(\bb) = \argmax_{\bg} l_n(\bb,\bg)$ and MLE stands for maximum likelihood estimator. Subsequently, we may estimate $m(t)$ as
$$
\wh m\subMLE(t) = \wh\bg\subMLE\trans \B_r(t), \quad \mbox{where}\quad \wh\bg\subMLE = \wh\bg(\wh\bb\subMLE).
$$

\subsubsection{Feature Selection}
When the dimension of $\Z$ is not small, the MLE $\wh\btheta = (\wh\bg\trans, \wh\bb\trans)\trans$ may suffer from high variability. On the other hand, it is highly likely that only a small number of codes are truly predictive of the event time. To overcome this challenge, one may employ standard penalized regression approach by imposing a penalty for the model complexity. To efficiently carry out penalized estimation under the B-spline PO model, we borrow the least square approximation (LSA) strategy proposed in \cite{wang2007unified}
and estimate $(\bg,\bb)$ as 
\be\label{eq:LSA}
\wh\btheta\subglasso =\underset{\btheta}{\argmin}
(\btheta - \wh\btheta\subMLE)\trans
\left\{-n^{-1}\ell_n''(\wh\btheta\subMLE)  \right\}(\btheta - \wh\btheta\subMLE)+\lambda\sum_g\left \| \bb\supg \right \|_2/ \| \wh\bb\subMLEg  \|_2,
\ee
where $\btheta = (\bg\trans,\bb\trans)\trans$, $\wh\btheta\subglasso=(\wh\bg\subglasso\trans,\wh\bb\subglasso\trans)\trans$,
$\bb\supg$ is a subvector of $\bb$ that corresponds to the features in group $g$, $[g]$ represents indices associated with group $g$, and $\|\cdot\|_2$ is the $L_2$ norm. All features associated with a specific code can be joined to create a group. The adaptive group lasso (glasso) penalty \citep{wang2008note} employed in (\ref{eq:LSA}) enables the removal of all features related to a code.
The tuning parameter $\lambda$ can be chosen by standard methods including the Akaike information criterion (AIC), Bayesian information criterion (BIC),
or the cross-validation.

With $\wh\btheta\subglasso$, we may obtain a glasso estimator for $m(t)$ as
$$
\wh m\subglasso(t) = \wh\bg\subglasso\trans \B_r(t)
$$
For any patient with derived feature $\wh\Z$, his/her probability of having an event by $t$ can be predicted as
$$
\wh F(t|\wh\Z) = \frac{e^{\wh\bb\trans\wh\Z}\int_0^t e^{\wh m\subglasso(s)ds}}{1+e^{\wh\bb\trans\wh\Z}\int_0^t e^{\wh m\subglasso(s)ds}}
$$

\subsection{Evaluation of Prediction Performance}\label{sec2.3}

Based on $\wh\pi_t = \wh F(t \mid \wh\Z)$, one may derive subject specific prediction rules for the event status and/or time. For example, one may predict the binary event status  $D_t = I(T \le t)$ using  $\wh\pi_t$ and $\Delta$ using $\wh\pi_C$.  One may also predict $X = \int_0^C I(T \ge t)dt$ based on
$\wh X_u = C (1-\wh\Delta_u) + \wh \Delta_u \wh T_u$ for some $u$ chosen to satisfy a desired sensitivity or specificity level of classifying $\Delta$ based on $\wh\Delta_u = I(\wh \pi_C \ge u)$, where $\wh T = \int_0^C \{1-\wh F(t \mid \wh\Z)\}dt$.

To summarize the overall performance of $(\wh X_u,\wh\Delta_u)$  in predicting $(X, \Delta)$, we may consider the Kendall's-$\tau$ type rank correlation summary measures, e.g.,
$$
\Cscr_{u} = P(\wh X_{ui} \le \wh X_{uj} \mid X_i \le X_j), \mbox{and} \
\Cscr_{u}^+ = P( \wh \Delta_{u,i} =1, \wh X_{ui} \le \wh X_{uj} \mid \Delta_i =1, X_i \le X_j) .
$$
To account for calibration, we propose to examine the absolute
prediction error (APE) measure via
\bse
 \ape_u & =&  E \int_0^{C_i} \left|I(\wh X_{ui}  \ge t)  - I(X_{i}  \ge
          t)  \right|  dt
=  E \int_0^{C_i} \left| I(\wh T_{ui}  \le t) \wh\Delta_{ui} - I(T_{i}  \le t)  \Delta_i \right|  dt \\
&=& E|\wh X_{ui} - X_i|
\ese
$\ape_u$ is an important summary measure for the quality of annotating
$(X, \Delta)$ since most survival estimation procedures rely on the at
risk function $I(X_i \ge t)$ and the counting process $I(T_i \le
t)\Delta_i$. The cut-off value $u$ can be selected also to minimize
$\ape_u$.

These accuracy measures can be estimated empirically
\begin{gather*}
 \wh{\ape_u} = \ninv\sumin \left|\wh X_{ui} - X_i\right|,
 \quad \wh\Cscr_u = \frac{\sum_{i < j} I(\wh X_{ui} \le \wh X_{uj}, X_i \le X_j)}{\sum_{i < j} I(X_i \le X_j)} ,\\
 \wh\Cscr_u^+ = \frac{\sum_{i < j} \wh\Delta_{ui}\Delta_i I(\wh X_{ui} \le \wh X_{uj}, X_i \le X_j)}{\sum_{i < j} \Delta_{i}I(X_i \le X_j)}.
\end{gather*}
Since $\wh X_u$ and $\wh \Delta_u$ are estimated using the same training data, such plug in accuracy estimate may suffer overfitting bias especially when
$n$ is not large. For such cases, standard cross-validation procedures can be used for bias correction.

\section{Asymptotic Results}\label{sec3}

The asymptotic distribution of the proposed MATA estimator is given in theorems \ref{th:Theorem1} and \ref{th:Theorem2} below, with proofs
provided in the Appendix. We assume the following regularity conditions.
\begin{enumerate}
\item[(C1)]
The density function $f_C(t)$
of the random variable $C$ is bounded and bounded away from 0 on
$[0,\calE)$ and satisfies the Lipschitz condition of order $1$ on
$[0,\calE)$. Additionally, $S_C(\calE-)=\lim_{\Delta\to 0+}S_C(\calE-\Delta)>0$.

\item[(C2)] $m(\cdot )\in \C^{(q)}([0,\calE] )$ for $q\geq 2$,
 and the spline order satisfies $r\geq q$.

\item[(C3)] There exists $0<c<\infty$, such that the distances between
neighboring knots satisfy
\bse
\max_{r\le p\le R_n+r}| h_{p+1}-h_p| =o(R_n^{-1})\text{
and }\max_{r\le p\le R_n+r}h_p/\min_{r\le p\le R_n+r}h_p\le c.
\ese
Furthermore, the number of knots satisfies $R_n\rightarrow \infty $, as $
n\rightarrow \infty $, $R_n^{-2}n\rightarrow \infty $ and $R_nn^{-1/(2q)}
\rightarrow \infty$.

\item[(C4)]
The function $m(t)$ is bounded  on $[0,\calE]$. The pdf
of the covariate $\Z$ is bounded and has a compact support.
\end{enumerate}

Here condition C1 assumes that
$S_C(t)$, the survival function of $C$, is discontinuous
at $\calE$. In practice, most studies
have a preselected ending time $\calE$, when all patients that have not
experienced failure are censored. This automatically leads to the discontinuity of
$S_C(t)$ at $\calE$. Besides, for some studies that keep tracking patients until the
last patient is censored or experience failure, the performance of the estimated survival
curve near the tail can be highly uncertain. A straightforward solution is manually censoring
all the patients to at least the last failure time $\calE$, which results in the discontinuity at
this point. Conditions C2 and C3 are standard smoothness and knots
conditions in B-spline approximation.
Condition C4 implies that both $S_T(t)$, the survival function of $T$,
and $f_T(t)$, the density function of $T$,
are bounded away from 0 on $[0,\calE]$. Hence, $[0,\calE]$ is
strictly contained in the support of the failure time $T$, i.e.,
$[0,\calE]\subset {\rm support}(T)$.

\begin{theorem}
\label{th:Theorem1}
Under the Conditions C1-C4, when
$\bb$ equals the truth $\bb_0$ or equals a $\sqrt{n}$-consistent
estimator of $\bb_0$, then $|\wh m(u,\bb
)-m(u)|=O_p\{(nh)^{-1/2}+h^q\}=O_p\{(nh)^{-1/2}\}$ uniformly in
$u\in[0,\calE]$ and as $n\to \infty$,
$\wh\sigma^{-1}(u,\bb_0)\{\wh{m}(u,\bb )-m(u)\}\to
\Normal(0,1)$ in distribution.
\end{theorem}

\begin{corollary}\label{cor:cumu}
Under the Conditions C1-C4,
the  function estimation error
satisfies
\bse
\exp\left[\int_0^t\exp\{\B_r\trans(u)\wh\bg\}du\right]-\exp\left[\int_0^t\exp\{m(u)\}du\right]
=O_p(n^{-1/2}+h^q),
\ese
 and the error is asymptotically normally distributed.
\end{corollary}

\begin{theorem}
\label{th:Theorem2}
Under the Conditions C1-C4, $\| \wh
\bb -\bb_0\|_{2}=O_p( n^{-1/2}) $, and
\bse
n^{1/2}(\wh{\bb }-\bb_0) \to
\Normal\{\0,\A^{-1}\bSigma(\A^{-1})\trans\},
\ese
 where
\bse
\A&=&E\left\{\bS_{\bb\bb,i}(\bb_0,m)\right\}
+E\{\bS_{\bb\bg,i}(\bb_0,m)\}\V_n(\bb_0)^{-1}
E\{\bS_{\bb\bg,i}(\bb_0,m)\}\trans;\\
\bSigma&=&E\left(\left[
\bS_{\bb,i}(\bb_0,m)+E\left\{\bS_{\bb\bg,i}(\bb_0,m)\right\}
\V_n(\bb_0)^{-1}
\bS_{\bg,i}(\bb_0,m)\right]^{\otimes2}\right),
\ese
and
\bse
\bS_{\bg,i}(\bb,m)
&=&
\Delta_i\B_r(X_i)-
(1+\Delta_i)
\frac{\exp(\Z_i\trans\bb)\int_0^{X_i}\exp\{m(u)\} \B_r(u) du
}
{1+\exp(\Z_i\trans\bb)\int_0^{X_i}\exp\{m(u)\}du},\\
\bS_{\bb,i}(\bb,m)
&=&
\Delta_i\Z_i-
(1+\Delta_i)
\frac{\Z_i\exp(\Z_i\trans\bb)\int_0^{X_i}\exp\{m(u)\}du}
{
1+\exp(\Z_i\trans\bb)\int_0^{X_i}\exp\{m(u)\}du},\\
\bS_{\bb\bb,i}(\bb,m)
&=&
-(1+\Delta_i)
\frac{\Z_i^{\otimes2}\exp(\Z_i\trans\bb)\int_0^{X_i}\exp\{m(u)\}du}
{[1+\exp(\Z_i\trans\bb)\int_0^{X_i}\exp\{m(u)\}du]^2},\\
\bS_{\bg\bg,i}(\bb,m)
&=&
 -(1+\Delta_i)
\frac{
\exp(\Z_i\trans\bb)\int_0^{X_i}\exp\{m(u)\} \B_r(u)^{\otimes2} du}
{1+\exp(\Z_i\trans\bb)\int_0^{X_i}\exp\{m(u)\}du}\\
&&+(1+\Delta_i)\frac{\exp(2\Z_i\trans\bb)
[\int_0^{X_i}\exp\{m(u)\} \B_r(u) du]^{\otimes2}}
{[1+\exp(\Z_i\trans\bb)\int_0^{X_i}\exp\{m(u)\}du]^2},\\
\bS_{\bb\bg,i}(\bb,m)
&=&
\frac{-(1+\Delta_i)\Z_i\exp(\Z_i\trans\bb)\int_0^{X_i}\exp\{m(u)\}\B_r\trans(u)du}
{[1+\exp(\Z_i\trans\bb)\int_0^{X_i}\exp\{m(u)\}du]^2}.
\ese
Here and throughout the text, $
a^{\otimes 2}\equiv \a\a\trans$ for any matrix or vector $\a$.
\end{theorem}

All the theoretical results here are derived in the general context
without taking into account
the errors associated with the feature formulation process. In
practice, when specific features are formed, different error
patterns may occur which will require additional analysis.

\section{Simulation}\label{sec4}
We have conducted extensive simulations to evaluate the performance of our proposed MATA
procedure and compare to existing methods including (a) the nonparametric MLE (NLPMLE) approach
by \cite{zeng2005maximum} using the same set of derived features; (b) the tree-based method by \cite{chubak2015electronic} which first uses the decision tree to classify patients
as experienced failure event or not, and then among the patients who are determined to have events,
assign the event time by the earliest arrival time of all groups of
medical encounters used in the decision tree; and (c) the two-step procedure by \cite{hassett2015detecting} and \cite{uno2018determining}, which first fit a logistic regression with group lasso to classify the patients and select significant groups of encounters, and then assign the failure time to patients experiencing event as the
weighted average of the peak time of the significant encounters with adjustment to correct the systematic bias.
Throughout, we fix the total sample size to
be $n+N=4000$ and consider $n \in \{200, 400\}$.

For simplicity, we only consider the case where all patients are
enrolled at the same time as we can always shift each patient's
follow-up period to a preselected origin. The censoring time of the
$i$-th patient, i.e., $C_i$, is simulated from the mixed distribution
$0.909\Uniform[0,\cE)+0.091\delta_{\cE}$, where $\delta_\cE$ is the
Dirac function at $\cE$ and $\cE=20$, for
$i=1,\cdots,n+N$. Intuitively, this imitates a long-term clinical
study that tracks patients up to 20 years, where 90.9\% of the
patients quit the study at uniform rate before the study terminates
and 9.1\% patients stay in the study until the end.
We simulate the number of encounters and encounter arrival times
using the expression $\lambda_i\supj(t) = \tau_i\supj f_i\supj(t)$ for
$t\geq0$.
We consider two sets of density functions $\{f_i\supj(t)\}$ for the point processes:
Gaussian and Gamma. In addition, for each density shape,
we considered both the case that density functions are independent across the
$q=10$ counting processes of the medical encounters
and the case that the densities are correlated. Details on the data
generation mechanisms for the point processes are given in Appendix A of
the Supplementary Materials.
We then set $\tau_i\supj=m_i\supj+5$ with
$\m_i=(m_{i}\supone,\cdots,m_{i}\supq)\trans =
F_{j}^{-1}\{\Phi({\boldsymbol\iota}_{i})\}$, where  $\Phi$ is the cumulative distribution function (CDF)
of the standard normal  and  $F_j$ is the CDF of
a Gamma distribution with shape $k_{2j}$ and scale $\theta_{2j}$,
$\Gamma(k_{2j},\theta_{2j})$. We let $\k_{2}=(k_{21},\cdots,k_{2q})\trans=(0.6, 0.48, 0.36, 1.2, 0.6, 0.9,
0.54, 1.26, 0.45,$
$0.468)\trans$,
$\btheta_{2}=(\theta_{21},\cdots,
\theta_{2q})\trans=(10, 6,
20,  4,  8,
9, 6.5, 5, 16, 14)\trans$ and generate
${\boldsymbol\iota}_i=(\iota_{i1},\cdots,\iota_{iq})\trans$ from $\MNormal(\0,\Sigma_{\boldsymbol\iota})$.
We consider two choices of $\Sigma_{\boldsymbol\iota}$: $\Sigma_{\boldsymbol\iota}=\I_q$  and
$\Sigma_{\boldsymbol\iota}=\{0.5^{\vert m-\ell\vert}\}_{1\leq m,\ell\leq q}$.
We further simulate encounter times $t_{i1}\supj,\cdots,t_{i{M_i\supj}^*}\supj\sim f_i\supj$ with ${M_i\supj}^*\sim\Poisson(m_i\supj)+5$ but only keep the ones that fall into the interval $[0,{\cal E}]$.
The final number of arrival times are thus reduced to $M_i\supj=\#\{k:0\leq t_{ik}\supj\leq {\cal E}\}$ and we relabel the retained
arrival times as $t_{i1}\supj,\cdots,t_{iM_{i}\supj}\supj$.
Simple calculation shows
$E(M_i\supj\mid \tau_i\supj) = \tau_i\supj \pr(\omega\leq \cE)$, where $\omega\sim f_i\supj$.

The event time $T_i$ is simulated from the PO model in (\ref{eq:surv}) where the true features are set to be the log of the peak time and the logit of the ratio between change point time and peak time of the intensity functions $\lambda_i\supj(t)$ for $i=1,\cdots,q$. Intuitively, an early peak time may result in early disease onset; and a relatively close peak time and change point time may imply a quick exacerbation of the disease status.
We set the nonparametric function $\log\alpha(t)=3\log(t)+\alpha_c$ and varies $\alpha_c$ to control the censoring rate. We further set
$\bb = (\bb_1\trans,\cdots,\bb_q\trans)\trans$ with $\bb_1=(-4,-3)\trans$ and $\bb_2=\cdots=\bb_q=\0$.
Consequently, only the first group of medical encounters
affects the recurrence time. The estimated features are set to be the logarithm of the first encounter arrival time and the estimated peak time as well as the logit of the estimated ratio between change point time and peak time.

We summarize results with 400 replications for each configuration. With each simulated dataset, we extract the features for both labeled and unlabeled data of total size $n+N$ whereas the PO model \eqref{eq:surv} is only fitted on the labeled training data of size $n$. The interior knots for B-splines in our approach are chosen to be the $10a\th$ percentile of the observed survival time $X$ with $a=1,2,\cdots,9$
for both Gaussian and Gamma cases.  For the tree approach and the two-step logistic, denoted by "Logi", approach, we use the same input feature space as the $\bZhat$ for the PO model \eqref{eq:surv} of MATA. To evaluate the performance of different procedures, we simulate a validation data of size $n_v = 5000$ in each simulation to evaluate the out-of-sample prediction performance through the accuracy measures discussed in Section \ref{sec2.3}.

\subsection{Results}
The estimated probability of having zero and $\leq3$ arrival times under all settings from a simulation with sample size $500,000$ are given in Table \ref{TabC1} in \ref{app:sim}. As a benchmark, we also present results from fitting the PO model with  true feature sets. For the true feature sets. In Table \ref{Tab1}, we reported the
bias and standard error (se) of the non-zero coefficients, i.e., $\bb_1=(\beta_{11},\beta_{12})\trans$, from MATA and NPMLE.
In general, we find that the MATA procedure performs well with small
sample size regardless of the censoring rate, the correlation
structure between groups of encounters, and the family of the
intensity curves. MASA generally leads to both smaller bias and
smaller standard error compared to the NPMLE.
In the extreme case when $n=200$ and the censoring rate reaches
70\%, leading to an effective sample size of 60,
both estimators deteriorate. However,
the resulting 95\% confidence interval of MATA covers the truth as the absolute bias is less than 1.96 times standard error. In contrast, the NPMLE has smaller standard error in the extreme case but its absolute bias is more then twice of the standard error. These results is consistent with Theorem \ref{th:Theorem2}.

\begin{table}
\centering
\begin{tabular}{rrrrrrrrrr}
\hline\hline
& & \multicolumn{4}{c}{Indp} & \multicolumn{4}{c}{Corr} \\
& & \multicolumn{2}{c}{$\beta_{11}$} & \multicolumn{2}{c}{$\beta_{12}$} & \multicolumn{2}{c}{$\beta_{11}$} & \multicolumn{2}{c}{$\beta_{12}$} \\
& & Bias  & se  & Bias  & se  & Bias  & se  & Bias  & se  \\\hline
\multicolumn{6}{l}{Gaussian, 30\% censoring rate} & \multicolumn{4}{l}{} \\\hline
 $n=200$ & MATA & -0.060    & 0.404   & -0.072    & 0.282   & -0.053    & 0.418   & -0.083    & 0.292   \\
 & NPMLE  & -0.355    & 0.692   & -0.216    & 0.550   & -0.359    & 0.716   & -0.232    & 0.495   \\
 $n=400$ & MATA & 0.017     & 0.271   & -0.020    & 0.183   & 0.030     & 0.258   & -0.027    & 0.177   \\
 & NPMLE  & -0.036    & 0.373   & 0.000   & 0.259   & -0.028    & 0.385   & -0.001    & 0.242   \\\hline
 \multicolumn{6}{l}{Gaussian, 70\% censoring rate} & \multicolumn{4}{l}{} \\\hline
$n=200$ & MATA & -0.408    & 0.893   & -0.305    & 0.582   & -0.352    & 0.776   & -0.277    & 0.532   \\
 & NPMLE  & 1.449     & 0.562   & 1.172     & 0.382   & 1.448     & 0.574   & 1.167     & 0.361   \\
$n=400$ & MATA & -0.082    & 0.440   & -0.081    & 0.279   & -0.088    & 0.403   & -0.095    & 0.280   \\
 & NPMLE  & 1.698     & 0.270   & 1.338     & 0.161   & 1.708     & 0.247   & 1.341     & 0.140  \\\hline\hline
\end{tabular}
\caption{\baselineskip=12pt Displayed are the bias and standard error of the estimation on $\bb_1=(\beta_{11},\beta_{12})\trans$ fitted with the true features from 400 simulations each with $N+n=4,000$ and $n=200$ and 400. Two methods, MATA and NPMLE, are contrasted. Panels from the top to bottom are Gaussian intensities with the subject-specific follow-up duration under 30\% and 70\% censoring rate as discussed in Section \ref{sec4.1}. The results under independent groups of encounters are shown on the left whereas the results for correlated one are shown on the right. }
\label{Tab1}
\end{table}

For both true and estimated feature sets, we computed the
out-of-sample accuracy measures discussed in Section \ref{sec2.3} on a
validation data set. Specifically,
the overall accuracy $\Cscr\subDelta$ of MATA and NPMLE are given in
Table \ref{Tab2}.
All other accuracy measures, i.e.,
Kendall's-$\tau$ type rank correlation summary measures $\Cscr_{u},
\Cscr_{u}^+$, and absolute prediction error $\ape_u$ depend
on $u$, which is easy to control for MATA and NPMLE but not for Tree
and Logi. We therefore minimize the cross-validation error for the
Tree approach and minimize the misclassification rate for the Logi
approach at their first step, i.e., classifying the censoring status
$\Delta$.
For MATA and NPMLE, We calculate these accuracy measures at
$u=0.02\ell$ for $\ell=0,1,\cdots,50$ and pick the $u$ with
minimum $\ape_u$. We then compare these measures at the selected $u$ with Tree and Logi methods in Tables \ref{Tab3} and \ref{Tab4}.

\begin{table}
\begin{center}
\begin{tabular}{ll|ll|ll}
\hline\hline
&        & \multicolumn{2}{c|}{Indep}          & \multicolumn{2}{c}{Corr}   \\
&        & True Ft     & Est Ft     & True Ft      & Est Ft      \\\hline\hline
\multicolumn{2}{l}{Gaussian, 30\%} & \multicolumn{4}{l}{} \\\hline\hline
$n=200$ & MATA & 0.974 (0.002) & 0.875 (0.059) & 0.974 (0.002) & 0.875 (0.059) \\
    & NPMLE  & 0.970 (0.003) & 0.878 (0.011) & 0.970 (0.003) & 0.878 (0.011) \\
$n=400$ & MATA & 0.974 (0.002) & 0.896 (0.018) & 0.974 (0.002) & 0.896 (0.021) \\
    & NPMLE  & 0.973 (0.002) & 0.893 (0.006) & 0.973 (0.002) & 0.894 (0.006) \\\hline\hline
\multicolumn{2}{l}{Gaussian, 70\%} & \multicolumn{4}{l}{} \\\hline\hline
$n=200$ & MATA & 0.973 (0.003) & 0.836 (0.070) & 0.957 (0.005) & 0.757 (0.094) \\
    & NPMLE  & 0.966 (0.004) & 0.832 (0.021) & 0.948 (0.005) & 0.771 (0.033) \\
$n=400$ & MATA & 0.975 (0.002) & 0.879 (0.024) & 0.960 (0.003) & 0.828 (0.043) \\
    & NPMLE  & 0.971 (0.002) & 0.861 (0.010) & 0.955 (0.003) & 0.815 (0.014) \\\hline\hline
\end{tabular}
\end{center}
\caption{\baselineskip=12pt The mean of the $\Cscr\subDelta$ and its standard deviation (in parenthesis) estimated from 400 simulations each with total sample size $N+n=4,000$ and labeled sample size $n=200$ and 400. Two methods, MATA and NPMLE, are contrasted. Both approaches are fitted with true feature set (``True Ft") and estimated feature set (``Est Ft"). Panels from the top to bottom are Gaussian intensities with the subject-specific follow-up duration under 30\% and 70\% censoring rate as discussed in Section \ref{sec4.1}. The results with two different correlation structures on $q$ counting processes of medical encounters, i.e., independent and correlated with specific covariance matrices given in Section \ref{sec4}, are shown on the left and the right, respectively. }
\label{Tab2}
\end{table}

\begin{sidewaystable}
\small
\begin{tabular}{r*{4}{r}*{4}{r}}
\hline\hline
       & \multicolumn{4}{c}{$n=200$} & \multicolumn{4}{c}{$n=400$} \\
       & MATA        & NPMLE         & Tree       & Logi          & MATA        & NPMLE         & Tree       & Logi          \\\hline
\multicolumn{9}{l}{Gaussian, 30\%, independent counting processes, true features} \\\hline
$\Cscr$   & 0.901 (0.002) & 0.894 (0.003) & 0.791 (0.022) & 0.719 (0.042) & 0.901 (0.002) & 0.898 (0.002) & 0.791 (0.018) & 0.718 (0.038) \\
$\Cscr^+$  & 0.868 (0.006) & 0.864 (0.006) & 0.742 (0.027) & 0.718 (0.029) & 0.868 (0.005) & 0.867 (0.005) & 0.747 (0.018) & 0.725 (0.022) \\
$\ape$ & 0.971 (0.032) & 1.049 (0.050) & 1.978 (0.368) & 3.373 (0.947) & 0.962 (0.027) & 1.001 (0.031) & 1.884 (0.125) & 3.406 (0.917) \\\hline
\multicolumn{9}{l}{Gaussian, 70\%, independent counting processes, true features} \\\hline
$\Cscr$   & 0.932 (0.004) & 0.921 (0.005) & 0.867 (0.014) & 0.859 (0.023) & 0.933 (0.002) & 0.926 (0.003) & 0.872 (0.010) & 0.859 (0.021) \\
$\Cscr^+$  & 0.782 (0.021) & 0.757 (0.023) & 0.705 (0.064) & 0.682 (0.086) & 0.789 (0.014) & 0.761 (0.015) & 0.716 (0.051) & 0.699 (0.067) \\
$\ape$ & 0.772 (0.049) & 0.899 (0.057) & 1.479 (0.162) & 1.590 (0.268) & 0.751 (0.030) & 0.840 (0.037) & 1.413 (0.121) & 1.576 (0.234) \\\hline
\multicolumn{9}{l}{Gaussian, 30\%, correlated counting processes, true features} \\\hline
$\Cscr$   & 0.900 (0.003) & 0.894 (0.003) & 0.792 (0.023) & 0.720 (0.042) & 0.901 (0.002) & 0.898 (0.002) & 0.791 (0.018) & 0.724 (0.039) \\
$\Cscr^+$  & 0.866 (0.006) & 0.862 (0.007) & 0.741 (0.029) & 0.719 (0.027) & 0.867 (0.005) & 0.865 (0.005) & 0.748 (0.017) & 0.727 (0.022) \\
$\ape$ & 0.972 (0.035) & 1.050 (0.046) & 1.994 (0.434) & 3.383 (0.962) & 0.963 (0.026) & 1.002 (0.031) & 1.871 (0.114) & 3.278 (0.976) \\\hline
\multicolumn{9}{l}{Gaussian, 70\%, correlated counting processes, true features} \\\hline
$\Cscr$   & 0.932 (0.004) & 0.921 (0.005) & 0.868 (0.015) & 0.859 (0.025) & 0.933 (0.002) & 0.926 (0.003) & 0.873 (0.011) & 0.861 (0.022) \\
$\Cscr^+$  & 0.782 (0.020) & 0.758 (0.022) & 0.698 (0.066) & 0.679 (0.084) & 0.789 (0.013) & 0.762 (0.016) & 0.720 (0.050) & 0.701 (0.068) \\
$\ape$ & 0.771 (0.051) & 0.896 (0.057) & 1.473 (0.175) & 1.599 (0.280) & 0.750 (0.029) & 0.838 (0.038) & 1.395 (0.121) & 1.548 (0.247) \\\hline\hline
\end{tabular}
\caption{
Kendall's-$\tau$ type rank correlation summary measures ($\Cscr$ and $\Cscr^+$), and absolute prediction error ($\ape$) are computed from four methods, MATA, NPMLE, Tree, and Logi, under $q=10$ Gaussian intensities over 400 simulations each with $n+N=4,000$ and $n=200$ or 400. The PO model is fitted with the true features. The upper two panels display the result under independent intensities with 30\% and 70\% censoring rate, respectively; the lower two panels display the result under correlated intensities with 30\% and 70\% censoring rate, respectively.}
\label{Tab3}
\end{sidewaystable}

\begin{sidewaystable}
\small
\begin{tabular}{r*{4}{r}*{4}{r}}
\hline\hline
       & \multicolumn{4}{c}{$n=200$} & \multicolumn{4}{c}{$n=400$} \\
       & MATA        & NPMLE         & Tree       & Logi          & MATA        & NPMLE         & Tree       & Logi          \\\hline
\multicolumn{9}{l}{Gaussian, 30\%, independent counting processes, estimated features} \\\hline
$\Cscr$   & 0.781 (0.020) & 0.768 (0.011) & 0.790 (0.019) & 0.740 (0.036) & 0.791 (0.008) & 0.784 (0.006) & 0.788 (0.022) & 0.744 (0.027) \\
$\Cscr^+$  & 0.701 (0.028) & 0.667 (0.017) & 0.690 (0.019) & 0.654 (0.036) & 0.675 (0.027) & 0.680 (0.010) & 0.692 (0.016) & 0.653 (0.028) \\
\ape & 2.148 (0.332) & 2.184 (0.132) & 1.913 (0.139) & 2.409 (0.540) & 1.960 (0.236) & 2.019 (0.074) & 1.904 (0.125) & 2.272 (0.315) \\\hline
\multicolumn{9}{l}{Gaussian, 70\%, independent counting processes, estimated features} \\\hline
$\Cscr$   & 0.839 (0.018) & 0.833 (0.010) & 0.802 (0.020) & 0.827 (0.015) & 0.853 (0.008) & 0.845 (0.006) & 0.806 (0.019) & 0.832 (0.013) \\
$\Cscr^+$  & 0.397 (0.111) & 0.413 (0.040) & 0.536 (0.107) & 0.500 (0.107) & 0.468 (0.042) & 0.447 (0.025) & 0.555 (0.072) & 0.511 (0.090) \\
$\ape$ & 1.868 (0.253) & 1.938 (0.122) & 2.276 (0.248) & 1.992 (0.216) & 1.685 (0.106) & 1.775 (0.071) & 2.202 (0.223) & 1.907 (0.177) \\\hline
\multicolumn{9}{l}{Gaussian, 30\%, correlated counting processes, estimated features} \\\hline
$\Cscr$   & 0.781 (0.019) & 0.768 (0.011) & 0.789 (0.021) & 0.743 (0.032) & 0.791 (0.010) & 0.783 (0.006) & 0.791 (0.016) & 0.747 (0.022) \\
$\Cscr^+$  & 0.701 (0.028) & 0.669 (0.016) & 0.690 (0.019) & 0.656 (0.036) & 0.672 (0.030) & 0.680 (0.011) & 0.693 (0.014) & 0.654 (0.027) \\
\ape & 2.142 (0.323) & 2.180 (0.122) & 1.915 (0.145) & 2.352 (0.418) & 1.958 (0.257) & 2.018 (0.072) & 1.886 (0.099) & 2.243 (0.194) \\\hline
\multicolumn{9}{l}{Gaussian, 70\%, correlated counting processes, estimated features} \\\hline
$\Cscr$   & 0.838 (0.018) & 0.832 (0.009) & 0.802 (0.021) & 0.827 (0.014) & 0.852 (0.007) & 0.846 (0.005) & 0.804 (0.017) & 0.832 (0.012) \\
$\Cscr^+$  & 0.393 (0.109) & 0.430 (0.042) & 0.533 (0.119) & 0.500 (0.110) & 0.468 (0.041) & 0.450 (0.026) & 0.564 (0.070) & 0.519 (0.083) \\
$\ape$ & 1.879 (0.248) & 1.944 (0.120) & 2.289 (0.253) & 1.987 (0.211) & 1.688 (0.102) & 1.772 (0.071) & 2.225 (0.193) & 1.898 (0.168) \\\hline\hline
\end{tabular}
\caption{
Estimated features, Gaussian.
Kendall's-$\tau$ type rank correlation summary measures ($\Cscr$ and $\Cscr^+$), and absolute prediction error ($\ape$) are computed from four methods, MATA, NPMLE, Tree, and Logi, under $q=10$ Gaussian intensities over 400 simulations each with $n+N=4,000$ and $n=200$ or 400. The PO model is fitted with the estimated features derived from FPCA approach in Section \ref{sec-est-intensity}. The upper two panels display the result under independent intensities with 30\% and 70\% censoring rate, respectively; the lower two panels display the result under correlated intensities with 30\% and 70\% censoring rate, respectively.
}
\label{Tab4}
\end{sidewaystable}

The performance of the MATA estimator when fitted with
  the true features
largely dominates that of NPMLE, Tree, and Logi, with
higher $\Cscr,
\Cscr^+$ and lower $\ape$ in all cases under Gaussian intensities.
When fitted with the estimated features, there is no clear winner
among the four methods
when the labeled data size is $n=200$; however, when the
labeled data size
increased to $n=400$, MATA generally outperforms
the other three approaches in terms of $\ape$.

\section{Example}\label{sec5}
We applied our MATA algorithm to
extraction of cancer recurrence time for Veterans Affair Central Cancer Registry (VACCR).
We obtained from VACCR 36705 lung cancer patients diagnosed with stage I to III lung cancer before 2018 and followed through 2019, among whom 3752 diagnosed in 2000-2018 with cancer stage information had annotated dates for cancer recurrence.
Through the research platform under Office of Research \& Development at Department of Veterans Affairs,
the cancer registry data were linked to the EHR at Veterans Affairs healthcare containing
diagnosis, procedure, medication, lab tests and medical notes.

The gold-standard recurrence status was collected through manual abstraction and
tumor registry for the VACCR data. Besides,
baseline covariates, including age at diagnosis, gender, and cancer stage,
are extracted.
Due to the predominance of male patients in VACCR (2.3\% among the 3752), we excluded gender in the subsequent analysis.
We randomly selected 1000 patients as
training data and used the remaining 2752 as validation data.
To assess the change of performance with the size of training data,
we also considered smaller training data $n=200$ and $n=400$ sub-sampled from the $n=1000$ set.
We ran 400 bootstrap samples from the 1000 train samples for each $n$ to quantify the variability
of analysis.
We selected time unit as month and focused on recurrence within 2 year in the analysis.
Patients without recurrence before 24 months from diagnosis date were censored
at 24 months.
Censoring rate was 39\%.
The diagnosis, procedure, medication codes and mentions in medical notes associated with the following nine events are collected:
lung cancer,
chemotherapy,
computerized tomography (CT) scan,
radiotherapy,
secondary malignant
neoplasm, palliative or hospice care,
recurrence,
medications for systematic therapies (including cytotoxic therapies, targeted therapies or immunotherapies),
biopsy or excision. See Table \ref{TabE1} for the detailed summary
of the sparsity in the nine groups of medical encounters.

For each of the nine selected events except the hospice,
we estimate the subject-specific intensity
function on the training and unlabeled data sets by applying the FPCA approach
described in Section \ref{sec-est-intensity}, and then use the resulting basis functions to
project the intensity functions for the validation set. The peak and change point time
of the estimated intensity functions are then extracted as features. In addition,
first code arrival time, first FPCA score, and the total number of diagnosis and procedure code are added as features for each event. All those features except the FPCA score are transformed in log-scale to reduce the skewness.
The radiotherapy, medication for systematic therapies and biopsy/excision has a zero code rate of 77.8\%,
70.7\% and 96.4\%, respectively.
Consequently, the estimated peak and largest
increase time of these features are identical as the associated first occurrence time for most patients. Thus, only the
first occurrence time and the total number of diagnosis and procedure code are considered for these features. Finally, to overcome the potential collinearity of the extracted features from
the same group (i.e., event), we further run the principal component analysis on each group of features and keep the first few principal components with proportion of variation exceeds 90\%.

Similarly as the simulation, we fit the decision tree to minimize the cross-validation error for the Tree approach and fit the logistic regression model. For MATA,  NPMLE and Logi, we take a fine grid on false positive rate ($\fpr$) on $\Delta$ and compute all other accuracy measures in Section \ref{sec2.3} for each value of $\fpr$. Then we pick the result which matches the $\fpr$ from Tree.

The prediction accuracy is summarized in Table \ref{TabE2}.
For the measurements regarding the timing of recurrence, our MATA estimator dominates the other three approaches with larger $\Cscr$, $\Cscr^+$ and yet smaller $\ape$.

Through its variable selection feature, MATA excluded stage II cancer from the $n=1000$ analysis
and additionally, stage III cancer, age at diagnosis, medication for systematic therapies
from the $n=200$ and $n=400$ analyses.
The selection is consistent with the NPMLE result of the $n=1000$ analysis, as the excluded features coincide
with the feature groups with no p-value $<0.05$.
Additional details for feature selection is given in \ref{app:data}.

\begin{table}
\begin{center}
\begin{tabular}{lllll}
\hline\hline
\multicolumn{5}{c}{$n=1000$} \\
       & MATA         & MLE             & Tree       & Logi          \\\hline
  $\Cscr$ & 0.810 (0.010) & 0.809 (0.009) & 0.755 (0.032) & 0.762 (0.015) \\
  $\Cscr^+$ & 0.688 (0.016) & 0.690 (0.017) & 0.646 (0.049) & 0.650 (0.010) \\
  $\ape$ & 3.326 (0.051) & 3.399 (0.074) & 5.693 (0.960) & 4.317 (0.152) \\
  \hline
\multicolumn{5}{c}{$n=400$} \\
       & MATA         & MLE             & Tree       & Logi          \\\hline
  $\Cscr$ & 0.796 (0.013) & 0.795 (0.013) & 0.748 (0.033) & 0.762 (0.059) \\
  $\Cscr^+$ & 0.663 (0.026) & 0.662 (0.027) & 0.644 (0.044) & 0.646 (0.044) \\
  $\ape$ & 3.576 (0.094) & 3.615 (0.129) & 5.880 (1.159) & 4.653 (0.577) \\
  \hline
\multicolumn{5}{c}{$n=200$} \\
       & MATA         & MLE             & Tree       & Logi          \\\hline
  $\Cscr$ & 0.791 (0.025) & 0.784 (0.024) & 0.748 (0.039) & 0.859 (0.127) \\
  $\Cscr^+$ & 0.624 (0.057) & 0.624 (0.043) & 0.640 (0.044) & 0.681 (0.090) \\
  $\ape$ & 3.842 (0.317) & 3.945 (0.286) & 6.094 (1.418) & 5.736 (1.080) \\
  \hline \hline
\end{tabular}
\end{center}
\caption{\baselineskip=12pt
Mean and bootstrap standard deviation of  Kendall's-$\tau$ type rank correlation summary measures ($\Cscr$ and $\Cscr^+$), and absolute prediction error ($\ape$) for the medical encounter data analyzed in Section \ref{sec5} under the four approaches, i.e., MATA, NPMLE, Tree, and Logi, over 400 bootstrap simulations.}
\label{TabE2}
\end{table}

\section{Discussion}\label{sec6}
We proposed an MATA method to auto-extract patients' longitudinal characteristics from the medical encounter data, and further build a risk prediction model on clinical events occurrence time with the extracted features. Such an approach
integrates both labeled and unlabeled data to obtain the longitudinal
features, thus tackled the sparsity of the medical encounter data. In
addition,  the FPCA approach preserves the flexibility of the resulting subject-specific intensity function to a great extent. Specifically, the intensity functions are often of different shapes between female and male, or between young patients and elder patients in practice.
Therefore, multiplicative intensity model
that assumes the heterogeneity among patients only results from the subject-specific
random effects may not be adequate.
The fitted risk prediction model is chosen to be the proportional odds model, whereas the nonparametric function is approximated by B-splines under certain transformation to ensure its monotonicity. The resulting estimator for the parametric part is shown to be root-$n$ consistent, whereas the non-parametric function is consistent,
under the correctly-specified model. Though the proportional odds model is adopted, our proof can be extended to other semiparametric transformation models such as the proportional hazard model easily. In the presence of large feature sets, we propose to use the group lasso with LSA for feature selection. The finite sample performance of our approach are studied under various settings.

Here, the FPCA is employed
on each group of medical encounters separately for feature extraction. However, different
groups are potentially connected, and separate estimation may fail to capture such relationship. A potential future work is to use the multivariate FPCA to directly address potential covariation among
different groups. Though various multivariate FPCA approaches exist, none of them can
handle the case in the medical encounters, where the encounter arrival times rather than the
underlying intensity functions are observed. Much effort is needed on
developing the applicable multivariate FPCA methodology and theories
in this setting.

The spline model works very well with only a few knots.
Small-sample performance of our
estimator is studied in various simulations,
with the prediction accuracy examined by C-statistics \citep{uno2011c} and Brier Score on
simulated validation data sets.

The adoption of the PO
model is for the simplicity of the illustration, while the theory of our estimator
can be easily generated to arbitrary linear transformation models. The focus of this paper lies in annotating rather than predicting the clinical events of interest. The medical encounter data is assumed to be possibly available even after the failure event.

\bibliographystyle{spbasic}      
\bibliography{ref}   

\begin{appendix}
\setcounter{section}{0}
\renewcommand\thesection{Appendix \Alph{section}}

\renewcommand\thefigure{A\arabic{figure}}
\renewcommand\theequation{A.\arabic{equation}}
\renewcommand\thetable{A\arabic{table}}
\renewcommand\thelemma{A\arabic{lemma}}
\renewcommand\thesubsection{\Alph{section}\arabic{subsection}}
\renewcommand\thedefinition{A\arabic{definition}}
\renewcommand\theremark{A\arabic{remark}}

\setcounter{equation}{0}
\setcounter{lemma}{0}
\setcounter{table}{0}

\section*{Appendices}

In \ref{app:sim}, we present additional simulation studies with Gamma intensities,
as well as extra information on the simulation settings.
In \ref{app:data}, we offer additional details on the data example of lung cancer recurrence
with VACCR data.
In \ref{app:FPCA}, we provide the theoretical properties for the derived features.
In \ref{app:PO}, we provide the theoretical properties for the MATA estimator based on the proportional
odds model.

\section{Additional Simulation Details}\label{app:sim}

\subsection{Simulation Settings for the Gaussian Intensities}\label{sec4.1}
We first simulate Gaussian shape density, i.e., $f_i\supj$ is the density function of $\Normal(\mu_{ij},\sigma_{ij}^2)$ truncated at 0.

Set $\mu_{ij}$ to be ${F_j}^{-1}\{\Phi(\nu_{ij})\}$, $F_j$ is the CDF of $\Gamma(k_{1j},\theta_{1j})$, with
$k_{1j}\sim \Uniform(3,6)$ and $\theta_{1j}\sim\Uniform(2,3)$ for
$j=1,\cdots,q$, and
$\bnu_i = (\nu_{i1},\cdots,\nu_{iq})\trans\sim\MNormal(\0,\Sigma_{\bnu})$, i.e., the multivariate normal distribution with mean $\0$ and variance $\Sigma_{\bnu}$.
For simplicity, we set $\Sigma_\bnu=\Sigma_{\boldsymbol\iota}$.
We further set $\mu_{ij}$ to be one if it is less than one.

Simulate $\sigma_{ij}\sim\Uniform(0.5,s_j)$ with $s_j=\min\{0.9\mu_{ij}, {F_j}^{-1}(0.5)\}$, where $F_j$ is the CDF of $\Gamma(k_{1j},\theta_{1j})$.
The way we simulate $\mu_{ij}$ and $\sigma_{ij}$ guarantees that the largest change in the
intensity functions only occurs after patients enter the study, i.e.,
$\mu_{ij}-\sigma_{ij}>0$, as expected in practice. Besides, the
simulated $\sigma_{ij}$ is not only controlled by the value of
$\mu_{ij}$ but also the median of  $\Gamma(k_{1j},\theta_{1j})$. Thus
$\sigma_{ij}$ will not get too extreme even with a large peak time
$\mu_{ij}$. In other words, the corresponding largest change in the
intensity function $\mu_{ij}-\sigma_{ij}$ is more likely to occur near
the peak time $\mu_{ij}$ than much earlier than $\mu_{ij}$ .

Finally, we set $\alpha_c$, the constant in the nonparametric function $\alpha(t)$, to be 7.5 and 1.1 to obtain an approximately $30\%$ and $70\%$ censoring rate.

We also consider gamma shape density, i.e.,
$f_i\supj(t)$ is the density function of
$\Gamma(k_{ij},\theta_{ij})$,
truncated at 0. Due to the limit of space, we place the Gamma intensities settings in \ref{app:sim} of the Supplementary
Material.

\subsection{\underline{Additional Simulations with Gamma Intensity functions}}

\paragraph{Simulation Settings with Gamma Intensities}\label{sec4.1}
For the setting with gamma shape density, we let
$f_i\supj(t)$ be the density function of
$\Gamma(k_{ij},\theta_{ij})$,
truncated at 0.
Set $k_{ij}=F_j^{-1}\{\Phi(\nu_{ij})\}$, where $F_j$ is the CDF of
$\Uniform(k_{\ell,j},k_{u,j})$, with $k_{\ell,j}\sim\Uniform(2,4)$,
and $k_{u,j}\sim\Uniform(4,6)$, and $\bnu_i=(\nu_{i1},\cdots,\nu_{iq})\trans\sim\MNormal(\0,\Sigma_{\bnu})$.
Generate $\theta_{ij}$ from Gamma $(a_{j},b_{j})$ truncated at its third quartile with $a_j\sim \Uniform(3,6)$, and $b_j\sim\Uniform(2,4)$.
We set $\alpha_c=6.8$ and 1.9 to obtain the approximate 30\% and 70\%
censoring rates.

\paragraph{Results}
For the true feature sets, we reported the
bias and standard error (se) of the non-zero coefficients, i.e., $\bb_1=(\beta_{11},\beta_{12})\trans$, from MATA and NPMLE in Table \ref{Tab1-gamma}.
Similar to the Gaussian intensities settings, we find that the MATA procedure performs well with small
sample size regardless of the censoring rate, the correlation
structure between groups of encounters, and the family of the
intensity curves. MASA generally leads to both smaller bias and
smaller standard error compared to the NPMLE. In the extreme case when $n=200$ and the censoring rate reaches
70\%,
both estimators deteriorate. However,
the resulting 95\% confidence interval of MATA covers the truth as the absolute bias is less than 1.96 times standard error. In contrast, the NPMLE tends to be numerically unstable. We observe the estimation bias of NPMLE for $n=400$ setting is larger than its own standard error and the the bias at $n=200$ setting. These results is consistent with Theorem \ref{th:Theorem2}.

\begin{table}
\small
\centering
\begin{tabular}{rrrrrrrrrr}
\hline\hline
& & \multicolumn{4}{c}{Indp} & \multicolumn{4}{c}{Corr} \\
& & \multicolumn{2}{c}{$\beta_{11}$} & \multicolumn{2}{c}{$\beta_{12}$} & \multicolumn{2}{c}{$\beta_{11}$} & \multicolumn{2}{c}{$\beta_{12}$} \\
& & Bias  & se  & Bias  & se  & Bias  & se  & Bias  & se  \\\hline
 \multicolumn{6}{l}{Gamma, 30\% censoring rate} & \multicolumn{4}{l}{} \\\hline
$n=200$   & MATA    & -0.086    & 0.443   & -0.084    & 0.410   & -0.091    & 0.447   & -0.126    & 0.436   \\
 & NPMLE     & -0.480     & 0.581   & -0.301    & 0.510   & -0.482    & 0.549   & -0.349    & 0.541   \\
$n=400$   & MATA    & -0.006    & 0.296   & -0.032    & 0.271   & 0.019     & 0.283   & -0.067    & 0.266   \\
 & NPMLE     & -0.223    & 0.374   & -0.138    & 0.315   & -0.190     & 0.346   & -0.167    & 0.340    \\\hline
 \multicolumn{6}{l}{Gamma, 70\% censoring rate} & \multicolumn{4}{l}{} \\\hline
$n=200$   & MATA    & -0.383    & 0.743   & -0.299    & 0.676   & -0.400    & 0.783   & -0.339    & 0.666   \\
 & NPMLE     & 0.336     & 0.731   & 0.284     & 0.592   & 0.325     & 0.866   & 0.267     & 0.670    \\
$n= 400$   & MATA    & -0.109    & 0.399   & -0.070    & 0.328   & -0.074    & 0.410   & -0.112    & 0.344 \\
& NPMLE & 0.708 & 0.429 & 0.551 & 0.330 & 0.744 & 0.385 & 0.533 & 0.352 \\
\hline\hline
\end{tabular}
\caption{\baselineskip=12pt Displayed are the bias and standard error of the estimation on $\bb_1=(\beta_{11},\beta_{12})\trans$ fitted with the true features from 400 simulations each with $N+n=4,000$ and $n=200$ and 400. Two methods, MATA and NPMLE, are contrasted. Panels from the top to bottom are Gamma intensities with the subject-specific follow-up duration under 30\% and 70\% censoring rate as discussed in Section \ref{sec4.1}. The results under independent groups of encounters are shown on the left whereas the results for correlated one are shown on the right. }
\label{Tab1-gamma}
\end{table}

For both true and estimated feature sets, we computed the
out-of-sample accuracy measures discussed in Section \ref{sec2.3} on a
validation data set. Specifically,
the overall accuracy $\Cscr\subDelta$ of MATA and NPMLE are given in
Table \ref{Tab2-gamma}.
All other accuracy measures, i.e.,
Kendall's-$\tau$ type rank correlation summary measures $\Cscr_{u},
\Cscr_{u}^+$, and absolute prediction error $\ape_u$ depend
on $u$, which is easy to control for MATA and NPMLE but not for Tree
and Logi. We therefore minimize the cross-validation error for the
Tree approach and minimize the misclassification rate for the Logi
approach at their first step, i.e., classifying the censoring status
$\Delta$.
For MATA and NPMLE, We calculate these accuracy measures at
$u=0.02\ell$ for $\ell=0,1,\cdots,50$ and pick the $u$ with
minimum $\ape_u$. We then compare these measures at the selected $u$ with Tree and Logi methods in Tables \ref{Tab5} and \ref{Tab6}.

\begin{table}
\begin{center}
\small
\begin{tabular}{llllll}
\hline\hline
&        & \multicolumn{2}{c}{Indep}          & \multicolumn{2}{c}{Corr}   \\
&        & True Ft     & Est Ft     & True Ft      & Est Ft      \\\hline
\multicolumn{6}{l}{Gamma, 30\% censoring rate} \\\hline
$n=200$ & MATA & 0.952 (0.003) & 0.793 (0.098) & 0.757 (0.094) & 0.957 (0.005) \\
    & NPMLE  & 0.945 (0.004) & 0.818 (0.015) & 0.771 (0.033) & 0.948 (0.005) \\
$n=400$ & MATA & 0.953 (0.003) & 0.840 (0.050) & 0.828 (0.043) & 0.960 (0.003) \\
    & NPMLE  & 0.950 (0.003) & 0.839 (0.009) & 0.815 (0.014) & 0.955 (0.003) \\\hline
\multicolumn{6}{l}{Gamma, 70\% censoring rate} \\\hline
$n=200$ & MATA & 0.957 (0.005) & 0.755 (0.097) & 0.957 (0.005) & 0.757 (0.094) \\
    & NPMLE  & 0.947 (0.006) & 0.772 (0.032) & 0.948 (0.005) & 0.771 (0.033) \\
$n=400$ & MATA & 0.960 (0.003) & 0.827 (0.046) & 0.960 (0.003) & 0.828 (0.043) \\
    & NPMLE  & 0.955 (0.003) & 0.815 (0.014) & 0.955 (0.003) & 0.815 (0.014) \\\hline\hline
\end{tabular}
\end{center}
\caption{\baselineskip=12pt The mean of the $\Cscr\subDelta$ and its standard deviation (in parenthesis) estimated from 400 simulations each with total sample size $N+n=4,000$ and labeled sample size $n=200$ and 400. Two methods, MATA and NPMLE, are contrasted. Both approaches are fitted with true feature set (``True Ft") and estimated feature set (``Est Ft"). Panels from the top to bottom are Gamma intensities with the subject-specific follow-up duration under 30\% and 70\% censoring rate as discussed in Section \ref{sec4.1}. The results with two different correlation structures on $q$ counting processes of medical encounters, i.e., independent and correlated with specific covariance matrices given in Section \ref{sec4}, are shown on the left and the right, respectively. }
\label{Tab2-gamma}
\end{table}

\begin{sidewaystable}
\setlength{\tabcolsep}{5pt}
\renewcommand{\arraystretch}{0.75}
\small
\begin{tabular}{r*{4}{r}*{4}{r}}
\hline\hline
       & \multicolumn{4}{c}{$n=200$} & \multicolumn{4}{c}{$n=400$} \\
       & MATA        & NPMLE         & Tree       & Logi          & MATA        & NPMLE         & Tree       & Logi          \\\hline
\multicolumn{9}{l}{Gamma, 30\%, independent counting processes, true features} \\\hline
$\Cscr$   & 0.872 (0.003) & 0.864 (0.004) & 0.720 (0.040) & 0.678 (0.074) & 0.873 (0.003) & 0.869 (0.003) & 0.731 (0.023) & 0.683 (0.071) \\
$\Cscr^+$  & 0.814 (0.008) & 0.812 (0.008) & 0.617 (0.047) & 0.658 (0.048) & 0.814 (0.006) & 0.815 (0.007) & 0.636 (0.026) & 0.667 (0.043) \\
$\ape$ & 1.318 (0.038) & 1.410 (0.048) & 3.826 (1.011) & 4.937 (1.844) & 1.307 (0.031) & 1.350 (0.036) & 3.434 (0.578) & 4.784 (1.835) \\\hline
\multicolumn{9}{l}{Gamma, 70\%, independent counting processes, true features} \\\hline
$\Cscr$   & 0.922 (0.004) & 0.914 (0.004) & 0.876 (0.010) & 0.837 (0.042) & 0.924 (0.002) & 0.919 (0.003) & 0.880 (0.008) & 0.841 (0.040) \\
$\Cscr^+$  & 0.735 (0.021) & 0.701 (0.024) & 0.618 (0.078) & 0.604 (0.106) & 0.743 (0.014) & 0.723 (0.017) & 0.630 (0.053) & 0.627 (0.085) \\
$\ape$ & 0.892 (0.049) & 0.995 (0.054) & 1.436 (0.125) & 1.986 (0.539) & 0.869 (0.030) & 0.930 (0.037) & 1.388 (0.090) & 1.917 (0.518) \\\hline
\multicolumn{9}{l}{Gamma, 30\%, correlated counting processes, true features} \\\hline
$\Cscr$   & 0.872 (0.003) & 0.864 (0.004) & 0.720 (0.040) & 0.685 (0.072) & 0.873 (0.002) & 0.869 (0.003) & 0.731 (0.026) & 0.684 (0.068) \\
$\Cscr^+$  & 0.814 (0.008) & 0.813 (0.009) & 0.617 (0.047) & 0.662 (0.049) & 0.816 (0.006) & 0.813 (0.007) & 0.635 (0.031) & 0.668 (0.041) \\
$\ape$ & 1.320 (0.041) & 1.408 (0.049) & 3.819 (1.022) & 4.826 (1.842) & 1.307 (0.030) & 1.350 (0.034) & 3.500 (0.689) & 4.808 (1.810) \\\hline
\multicolumn{9}{l}{Gamma, 70\%, correlated counting processes, true features} \\\hline
$\Cscr$   & 0.921 (0.005) & 0.913 (0.004) & 0.875 (0.011) & 0.841 (0.044) & 0.924 (0.003) & 0.919 (0.003) & 0.879 (0.007) & 0.848 (0.038) \\
$\Cscr^+$  & 0.733 (0.025) & 0.702 (0.024) & 0.616 (0.072) & 0.612 (0.099) & 0.742 (0.015) & 0.724 (0.016) & 0.628 (0.053) & 0.622 (0.086) \\
$\ape$ & 0.900 (0.056) & 0.996 (0.053) & 1.447 (0.126) & 1.933 (0.572) & 0.872 (0.032) & 0.929 (0.036) & 1.390 (0.086) & 1.833 (0.488) \\\hline\hline
\end{tabular}
\linespread{0.75}\selectfont{}
\caption{True features, Gamma.
Kendall's-$\tau$ type rank correlation summary measures ($\Cscr$ and $\Cscr^+$), and absolute prediction error ($\ape$) are computed from four methods, MATA, NPMLE, Tree, and Logi, under $q=10$ Gamma intensities over 400 simulations each with $n+N=4,000$ and $n=200$ or 400. The PO model is fitted with the true features. The upper two panels display the result under independent intensities with 30\% and 70\% censoring rate, respectively; the lower two panels display the result under correlated intensities with 30\% and 70\% censoring rate, respectively.}
\label{Tab5}
\end{sidewaystable}

\begin{sidewaystable}
\setlength{\tabcolsep}{5pt}
\renewcommand{\arraystretch}{0.75}
\small
\begin{tabular}{r*{4}{r}*{4}{r}}
\hline\hline
       & \multicolumn{4}{c}{$n=200$} & \multicolumn{4}{c}{$n=400$} \\
       & MATA        & NPMLE         & Tree       & Logi          & MATA        & NPMLE         & Tree       & Logi          \\\hline
\multicolumn{9}{l}{Gamma, 30\%, independent counting processes, estimated features} \\\hline
$\Cscr$   & 0.728 (0.027) & 0.720 (0.010) & 0.650 (0.043) & 0.668 (0.059) & 0.749 (0.011) & 0.737 (0.007) & 0.667 (0.045) & 0.659 (0.056) \\
$\Cscr^+$  & 0.555 (0.037) & 0.578 (0.015) & 0.456 (0.074) & 0.570 (0.078) & 0.573 (0.018) & 0.589 (0.010) & 0.480 (0.072) & 0.558 (0.063) \\
$\ape$ & 2.732 (0.544) & 2.698 (0.121) & 4.164 (1.191) & 5.018 (1.659) & 2.443 (0.218) & 2.528 (0.074) & 3.840 (1.095) & 4.858 (1.330) \\\hline
\multicolumn{9}{l}{Gamma, 70\%, independent counting processes, estimated features} \\\hline
$\Cscr$   & 0.833 (0.016) & 0.827 (0.011) & 0.829 (0.017) & 0.822 (0.019) & 0.849 (0.010) & 0.841 (0.006) & 0.831 (0.018) & 0.829 (0.014) \\
$\Cscr^+$  & 0.277 (0.115) & 0.325 (0.043) & 0.485 (0.139) & 0.425 (0.119) & 0.371 (0.058) & 0.374 (0.027) & 0.519 (0.070) & 0.450 (0.089) \\
$\ape$ & 2.002 (0.223) & 2.040 (0.136) & 2.006 (0.202) & 2.086 (0.254) & 1.766 (0.141) & 1.866 (0.077) & 1.964 (0.195) & 1.973 (0.166) \\\hline
\multicolumn{9}{l}{Gamma, 30\%, correlated counting processes, estimated features} \\\hline
$\Cscr$   & 0.731 (0.024) & 0.720 (0.011) & 0.656 (0.046) & 0.670 (0.053) & 0.749 (0.010) & 0.737 (0.007) & 0.672 (0.045) & 0.665 (0.057) \\
$\Cscr^+$  & 0.568 (0.038) & 0.577 (0.016) & 0.453 (0.076) & 0.560 (0.073) & 0.579 (0.020) & 0.588 (0.011) & 0.485 (0.072) & 0.561 (0.063) \\
$\ape$ & 2.681 (0.494) & 2.691 (0.127) & 4.133 (1.176) & 4.770 (1.522) & 2.451 (0.244) & 2.522 (0.073) & 3.778 (1.076) & 4.874 (1.354) \\\hline
\multicolumn{9}{l}{Gamma, 70\%, correlated counting processes, estimated features} \\\hline
$\Cscr$   & 0.833 (0.016) & 0.826 (0.011) & 0.828 (0.018) & 0.822 (0.017) & 0.849 (0.009) & 0.840 (0.006) & 0.831 (0.017) & 0.829 (0.012) \\
$\Cscr^+$  & 0.283 (0.107) & 0.322 (0.044) & 0.484 (0.136) & 0.421 (0.124) & 0.366 (0.060) & 0.388 (0.028) & 0.515 (0.076) & 0.442 (0.094) \\
$\ape$ & 1.996 (0.214) & 2.053 (0.135) & 2.017 (0.213) & 2.088 (0.236) & 1.766 (0.128) & 1.868 (0.075) & 1.963 (0.180) & 1.975 (0.156) \\\hline\hline
\end{tabular}
\linespread{0.75}\selectfont{}
\caption{Estimated features, Gamma.
 Kendall's-$\tau$ type rank correlation summary measures ($\Cscr$ and $\Cscr^+$), and absolute prediction error ($\ape$) are computed from four methods, MATA, NPMLE, Tree, and Logi, under $q=10$ Gamma intensities over 400 simulations each with $n+N=4,000$ and $n=200$ or 400. The PO model is fitted with the estimated features derived from FPCA approach in Section \ref{sec-est-intensity}. The upper two panels display the result under independent intensities with 30\% and 70\% censoring rate, respectively; the lower two panels display the result under correlated intensities with 30\% and 70\% censoring rate, respectively.}
\label{Tab6}
\end{sidewaystable}

Similar to the Gaussian intensities setting, the performance of the MATA estimator when fitted with
  the true features
largely dominates that of NPMLE, Tree, and Logi, with
higher $\Cscr,
\Cscr^+$ and lower $\ape$ in all cases except when the encounters are
simulated from independent Gamma counting processes with 30\%
censoring rate.  In this exceptional case, our MATA
  estimator has very minor advantage in $\Cscr^+$ compared to
  NPMLE, and is still better in terms of $\Cscr$ and
  $\ape$.
When fitted with the estimated features, there is no clear winner
among the four methods
when the labeled data size is $n=200$; however, when the
labeled data size
increased to $n=400$, MATA generally outperforms
the other three approaches in terms of $\ape$.

\subsection*{\underline{Supplementary Results on Simulations}}

We show the sparsity in the simulated data in Tables \ref{TabC1}.
We show the Average Model Size and MSE of Estimation in Table \ref{TabA2}.

\begin{table}
\small
\centering
\begin{tabular}{lllllllllll}
\hline\hline
 & $\Nsc\supone$ & $\Nsc^{[2]}$ & $\Nsc^{[3]}$ & $\Nsc^{[4]}$ & $\Nsc^{[5]}$ & $\Nsc^{[6]}$ & $\Nsc^{[7]}$ & $\Nsc^{[8]}$ & $\Nsc^{[9]}$ & $\Nsc^{[10]}$ \\\hline
 \multicolumn{11}{c}{Probability of zero encounters} \\\hline
Indp Gaussian & 0.508 & 0.802 & 0.793 & 0.767 & 0.878 & 0.758 & 0.474 & 0.594 & 0.818 & 0.755 \\
Corr Gaussian & 0.508 & 0.802 & 0.792 & 0.766 & 0.879 & 0.758 & 0.474 & 0.595 & 0.817 & 0.755 \\
Indp Gamma    & 0.761 & 0.805 & 0.716 & 0.786 & 0.750 & 0.944 & 0.712 & 0.810 & 0.939 & 0.750 \\
Corr Gamma    & 0.761 & 0.806 & 0.715 & 0.786 & 0.749 & 0.943 & 0.713 & 0.812 & 0.938 & 0.750 \\\hline
 \multicolumn{11}{c}{Probability of $\leq 3$ encounters} \\\hline
 Indp Gaussian & 0.720 & 0.945 & 0.926 & 0.930 & 0.971 & 0.918 & 0.680 & 0.778 & 0.938 & 0.902 \\
Corr Gaussian & 0.721 & 0.946 & 0.926 & 0.930 & 0.972 & 0.918 & 0.681 & 0.778 & 0.938 & 0.902 \\
Indp Gamma    & 0.938 & 0.962 & 0.913 & 0.939 & 0.933 & 0.995 & 0.935 & 0.970 & 0.995 & 0.931 \\
Corr Gamma    & 0.938 & 0.962 & 0.913 & 0.939 & 0.933 & 0.995 & 0.936 & 0.970 & 0.995 & 0.931 \\
\hline\hline
\end{tabular}
\caption{
Estimated probability of having zero or $\leq 3$ encounter arrival times
under each counting process $\Nsc\supj$ for $j=1,\cdots,10$ from a simulation with sample size $500,000$.}
\label{TabC1}
\end{table}

\begin{table}
\begin{center}
\small
\begin{tabular}{ll|ll|ll|ll|ll}
\hline\hline
 & & \multicolumn{2}{c|}{30\% Indp} & \multicolumn{2}{c}{70\% Indp}
  & \multicolumn{2}{c|}{30\% Corr} & \multicolumn{2}{c}{70\% Corr} \\
  & & Tr Ft & Est Ft & Tr Ft & Est Ft & Tr Ft & Est Ft & Tr Ft & Est Ft \\\hline\hline
\multicolumn{10}{l}{Gaussian}               \\\hline
$n=200$ & AIC & 13.24  & 15.57  & 13.74  & 17.09   & 13.30  & 15.91  & 13.75  & 17.41 \\
 & BIC & 13.07  & 15.45  & 13.45   & 16.87 & 13.09  & 15.83    & 13.39   & 17.27  \\
$n=400$ & AIC & 13.15  & 14.40  & 13.30   & 14.57 & 13.18   & 14.39    & 13.31  & 14.77   \\
 & BIC & 13.00    & 14.05  & 13.01  & 14.17  & 13.00    & 14.12    & 13.00  & 14.22 \\\hline
\multicolumn{10}{l}{Gamma}               \\\hline
$n=200$ & AIC    & 13.38  & 18.37  & 13.88   & 19.35  & 13.41  & 18.04  & 13.94  & 19.57 \\
 & BIC    & 13.23   & 18.22  & 13.65   & 19.01  & 13.29   & 17.88  & 13.72   & 19.21   \\
$n=400$ & AIC    & 13.24  & 15.02    & 13.20    & 15.09  & 13.21   & 15.04   & 13.34   & 14.97 \\
 & BIC    & 13.00    & 14.80   & 13.01  & 14.78  & 13.01  & 14.72    & 13.02  & 14.72   \\\hline\hline
\end{tabular}
\end{center}
\caption{  Average model sizes selected by MATA.}
\label{TabA2}
\end{table}

\section{Additional Details on Data Example}\label{app:data}

We show the sparsity of features in Table \ref{TabE1}.
The radiotherapy, medication for systematic therapies and biopsy/excision has a zero code rate of 77.8\%,
70.7\% and 96.4\%, respectively.
Consequently, the estimated peak and largest
increase time of these features are identical as the associated first occurrence time for most patients. Thus, only the
first occurrence time and the total number of diagnosis and procedure code are considered for these features.

\begin{table}
\begin{center}
\small
\begin{tabular}{l l l}
\hline
\hline
Feature & Zero & $\le 3$ times\\
\hline
 Lung Cancer & 0.014 & 0.087 \\
  Chemotherapy & 0.567 & 0.736 \\
  CT Scan & 0.127 & 0.363 \\
  Radiotherapy & 0.778 & 0.912 \\
  Secondary Maligant Neoplasm & 0.554 & 0.856 \\
  Palliative or Hospice Care & 0.576 & 0.888 \\
  Recurrence & 0.279 & 0.723 \\
  Medication & 0.707 & 0.824 \\
  Biopsy or Excision & 0.964 & 1.000 \\

\hline
\hline
\end{tabular}
\end{center}
\caption{Sparsity of the nine groups of medical encounter data analyzed in Section \ref{sec5}.}
\label{TabE1}
\end{table}

We show the MATA and NPMLE coefficients for $n=1000, 400, 200$
in Tables \ref{tab:coef1000}-\ref{tab:coef200}.
Similar
as in Section \ref{sec4}, our MATA estimator has smaller
bootstrap standard error compared to the NPMLE.
For the analysis with $n=1000$, both
MATA and NPMLE showed a significant impact of
first arrival time and peak time of lung cancer code,
first arrival time and first FPCA score of chemotherapy code,
first arrival time of radiotherapy code,
total number of secondary malignant neoplasm code,
peak and change point times of palliative or hospice care in medical notes,
first FPCA score and total number of recurrence in medical notes
and first arrival time of biopsy or excision.
MATA additionally finds the change point time of lung cancer code to have strong association with high risk of lung cancer recurrence. Furthermore, MATA excludes the stage II cancer, which coincides with the large p-values on those four group of encounters under NPMLE.
For the analyses with $n=200$ and $n=400$, MATA excludes cancer stage, age at diagnosis and medication for systematic therapies, which coincides with the groups without any significant feature from the $n=1000$ NPMLE analysis.

\begin{table}
\begin{center}
\renewcommand{\arraystretch}{1.35}
\linespread{0.75}\selectfont{}
\footnotesize
\caption{Analysis with $n=1000$. Estimated coefficient (``est"), bootstrap standard error (``boot.se"), and p-value (``pval") over 400 bootstraps for the extracted feature sets, including first code time (1stCode), peak time (Pk), change point time (ChP), first FPC score (1stScore), and log of total number of codes (logN), from the nine groups of medical encounter data in 5. For each group, its group p-value (``group pval") is calculated via chi-square test. All features regarding time are transformed in log-scale. The result for the proposed MATA estimator is given in the left panel and that of NPMLE is shown in the right panel.
}\label{tab:coef1000}
\begin{tabular}{ll|rrr|rrr}
\hline\hline
      &          & \multicolumn{3}{c|}{MATA} & \multicolumn{3}{c}{NPMLE} \\
Group  & Feature       & mean    & boot.se  & pval  & mean    & boot.se & pval \\ \hline
  & Stage II & -- & -- & -- &  0.075 & 0.144 & 0.604 \\
   & Stage III &  0.168 & 0.168 & 0.319 &  0.160 & 0.181 & 0.379 \\
   & Age &  0.013 & 0.008 & 0.111 &  0.013 & 0.007 & 0.069 \\
   \hline
Lung Cancer & 1stCode & -0.277 & 0.116 & $\mathbf{0.017}$ & -0.294 & 0.116 & $\mathbf{0.011}$ \\
   & Pk &  0.213 & 0.084 & $\mathbf{0.012}$ &  0.213 & 0.089 & $\mathbf{0.016}$ \\
   & ChP &  0.135 & 0.065 & $\mathbf{0.040}$ &  0.131 & 0.068 & 0.054 \\
   & 1stScore & -0.091 & 0.183 & 0.619 & -0.028 & 0.204 & 0.891 \\
   & logN &  0.072 & 0.108 & 0.502 &  0.070 & 0.121 & 0.561 \\
   \hline
Chemotherapy & 1stCode & -0.140 & 0.065 & $\mathbf{0.032}$ & -0.146 & 0.067 & $\mathbf{0.029}$ \\
   & Pk & -0.162 & 0.106 & 0.127 & -0.169 & 0.111 & 0.127 \\
   & ChP &  0.019 & 0.067 & 0.773 &  0.019 & 0.073 & 0.799 \\
   & 1stScore &  0.652 & 0.180 & $\mathbf{<0.001}$ &  0.678 & 0.188 & $\mathbf{<0.001}$ \\
   & logN &  0.073 & 0.092 & 0.424 &  0.076 & 0.103 & 0.463 \\
   \hline
CT scan & 1stCode &  0.020 & 0.076 & 0.789 &  0.017 & 0.093 & 0.858 \\
   & Pk &  0.104 & 0.093 & 0.262 &  0.115 & 0.103 & 0.266 \\
   & ChP &  0.046 & 0.043 & 0.286 &  0.047 & 0.048 & 0.329 \\
   & 1stScore & -0.244 & 0.132 & 0.065 & -0.266 & 0.131 & $\mathbf{0.042}$ \\
   & logN & -0.019 & 0.096 & 0.847 & -0.034 & 0.112 & 0.763 \\
   \hline
Radiotherapy & 1stCode & -0.327 & 0.157 & $\mathbf{0.037}$ & -0.382 & 0.163 & $\mathbf{0.019}$ \\
   & logN & -0.057 & 0.056 & 0.311 & -0.068 & 0.062 & 0.275 \\
   \hline
Secondary & 1stCode &  0.013 & 0.127 & 0.921 & -0.008 & 0.141 & 0.954 \\
  Malignant & Pk & -0.135 & 0.113 & 0.230 & -0.130 & 0.126 & 0.299 \\
  Neoplasm & ChP & -0.067 & 0.049 & 0.168 & -0.067 & 0.054 & 0.217 \\
   & 1stScore & -0.197 & 0.122 & 0.105 & -0.205 & 0.128 & 0.109 \\
   & logN &  0.333 & 0.077 & $\mathbf{<0.001}$ &  0.335 & 0.079 & $\mathbf{<0.001}$ \\
   \hline
Palliative & 1stCode & -0.055 & 0.085 & 0.517 & -0.066 & 0.089 & 0.457 \\
  or Hospice & Pk & -0.942 & 0.187 & $\mathbf{<0.001}$ & -1.009 & 0.205 & $\mathbf{<0.001}$ \\
  Care & ChP & -0.704 & 0.140 & $\mathbf{<0.001}$ & -0.753 & 0.153 & $\mathbf{<0.001}$ \\
   & 1stScore &  0.068 & 0.095 & 0.470 &  0.070 & 0.098 & 0.471 \\
   & logN &  0.017 & 0.061 & 0.785 &  0.002 & 0.064 & 0.979 \\
   \hline
Recurrence & 1stCode &  0.121 & 0.081 & 0.138 &  0.122 & 0.084 & 0.147 \\
   & Pk & -0.105 & 0.093 & 0.259 & -0.099 & 0.097 & 0.310 \\
   & ChP & -0.046 & 0.058 & 0.426 & -0.042 & 0.060 & 0.479 \\
   & 1stScore & -0.281 & 0.119 & $\mathbf{0.018}$ & -0.288 & 0.122 & $\mathbf{0.018}$ \\
   & logN &  0.234 & 0.076 & $\mathbf{0.002}$ &  0.255 & 0.075 & $\mathbf{<0.001}$ \\
   \hline
Medication & 1stCode &  0.173 & 0.118 & 0.143 &  0.185 & 0.113 & 0.104 \\
   & logN &  0.062 & 0.071 & 0.384 &  0.071 & 0.081 & 0.380 \\
   \hline
Biopsy & 1stCode & -0.865 & 0.411 & $\mathbf{0.035}$ & -0.968 & 0.399 & $\mathbf{0.015}$ \\
   & logN & -0.423 & 0.502 & 0.399 & -0.478 & 0.523 & 0.360 \\
   \hline

\hline
\hline
\end{tabular}
\end{center}
\end{table}

\begin{table}
\begin{center}
\renewcommand{\arraystretch}{1.35}
\linespread{0.75}\selectfont{}
\footnotesize
\caption{Analysis with $n=400$. Estimated coefficient (``est"), bootstrap standard error (``boot.se"), and p-value (``pval") over 400 bootstraps for the extracted feature sets, including first code time (1stCode), peak time (Pk), change point time (ChP), first FPC score (1stScore), and log of total number of codes (logN), from the nine groups of medical encounter data in 5. For each group, its group p-value (``group pval") is calculated via chi-square test. All features regarding time are transformed in log-scale. The result for the proposed MATA estimator is given in the left panel and that of NPMLE is shown in the right panel.
}\label{tab:coef400}
\begin{tabular}{ll|rrr|rrr}
\hline\hline
      &          & \multicolumn{3}{c|}{MATA} & \multicolumn{3}{c}{NPMLE} \\
Group  & Feature       & mean    & boot.se  & pval  & mean    & boot.se & pval \\ \hline
  & Stage II & -- & -- & -- &  0.067 & 0.254 & 0.790 \\
   & Stage III & -- & -- & -- &  0.189 & 0.349 & 0.587 \\
   & Age & -- & -- & -- &  0.014 & 0.012 & 0.264 \\
   \hline
Lung Cancer & 1stCode & -0.232 & 0.178 & 0.192 & -0.311 & 0.189 & 0.101 \\
   & Pk &  0.191 & 0.133 & 0.150 &  0.232 & 0.144 & 0.108 \\
   & ChP &  0.115 & 0.106 & 0.279 &  0.133 & 0.117 & 0.258 \\
   & 1stScore & -0.098 & 0.266 & 0.712 & -0.075 & 0.332 & 0.821 \\
   & logN &  0.078 & 0.163 & 0.633 &  0.074 & 0.203 & 0.715 \\
   \hline
Chemotherapy & 1stCode & -0.120 & 0.109 & 0.270 & -0.150 & 0.126 & 0.232 \\
   & Pk & -0.140 & 0.176 & 0.428 & -0.181 & 0.209 & 0.387 \\
   & ChP &  0.001 & 0.096 & 0.991 &  0.004 & 0.122 & 0.975 \\
   & 1stScore &  0.607 & 0.288 & $\mathbf{0.035}$ &  0.719 & 0.311 & $\mathbf{0.021}$ \\
   & logN &  0.064 & 0.139 & 0.643 &  0.064 & 0.174 & 0.714 \\
   \hline
CT scan & 1stCode &  0.017 & 0.121 & 0.886 &  0.014 & 0.160 & 0.933 \\
   & Pk &  0.068 & 0.143 & 0.634 &  0.110 & 0.179 & 0.538 \\
   & ChP &  0.038 & 0.071 & 0.589 &  0.050 & 0.089 & 0.571 \\
   & 1stScore & -0.207 & 0.204 & 0.310 & -0.291 & 0.222 & 0.190 \\
   & logN & -0.019 & 0.151 & 0.897 & -0.037 & 0.188 & 0.844 \\
   \hline
Radiotherapy & 1stCode & -0.229 & 0.221 & 0.301 & -0.345 & 0.248 & 0.165 \\
   & logN & -0.027 & 0.086 & 0.749 & -0.058 & 0.109 & 0.595 \\
   \hline
Secondary & 1stCode & -0.019 & 0.172 & 0.913 & -0.035 & 0.234 & 0.881 \\
  Malignant & Pk & -0.125 & 0.163 & 0.444 & -0.119 & 0.211 & 0.575 \\
  Neoplasm & ChP & -0.065 & 0.072 & 0.366 & -0.063 & 0.092 & 0.490 \\
   & 1stScore & -0.207 & 0.182 & 0.257 & -0.224 & 0.219 & 0.307 \\
   & logN &  0.302 & 0.128 & $\mathbf{0.018}$ &  0.343 & 0.134 & $\mathbf{0.011}$ \\
   \hline
Palliative & 1stCode & -0.076 & 0.137 & 0.580 & -0.091 & 0.160 & 0.567 \\
  or Hospice & Pk & -0.845 & 0.248 & $\mathbf{<0.001}$ & -0.936 & 0.276 & $\mathbf{<0.001}$ \\
  Care & ChP & -0.631 & 0.185 & $\mathbf{<0.001}$ & -0.699 & 0.206 & $\mathbf{<0.001}$ \\
   & 1stScore &  0.054 & 0.126 & 0.670 &  0.067 & 0.143 & 0.641 \\
   & logN &  0.040 & 0.092 & 0.663 &  0.015 & 0.105 & 0.889 \\
   \hline
Recurrence & 1stCode &  0.089 & 0.116 & 0.443 &  0.125 & 0.134 & 0.351 \\
   & Pk & -0.114 & 0.139 & 0.412 & -0.103 & 0.161 & 0.521 \\
   & ChP & -0.055 & 0.085 & 0.519 & -0.046 & 0.099 & 0.642 \\
   & 1stScore & -0.229 & 0.176 & 0.193 & -0.280 & 0.197 & 0.154 \\
   & logN &  0.199 & 0.122 & 0.104 &  0.266 & 0.124 & $\mathbf{0.033}$ \\
   \hline
Medication & 1stCode & -- & -- & -- &  0.201 & 0.188 & 0.284 \\
   & logN & -- & -- & -- &  0.061 & 0.155 & 0.693 \\
   \hline
Biopsy & 1stCode & -0.814 & 0.689 & 0.238 & -1.127 & 0.734 & 0.125 \\
   & logN & -0.363 & 0.811 & 0.654 & -0.559 & 0.989 & 0.572 \\
   \hline

\hline
\hline
\end{tabular}
\end{center}
\end{table}

\begin{table}
\begin{center}
\renewcommand{\arraystretch}{1.35}
\linespread{0.75}\selectfont{}
\footnotesize
\caption{Analysis with $n=200$. Estimated coefficient (``est"), bootstrap standard error (``boot.se"), and p-value (``pval") over 400 bootstraps for the extracted feature sets, including first code time (1stCode), peak time (Pk), change point time (ChP), first FPC score (1stScore), and log of total number of codes (logN), from the nine groups of medical encounter data in 5. For each group, its group p-value (``group pval") is calculated via chi-square test. All features regarding time are transformed in log-scale. The result for the proposed MATA estimator is given in the left panel and that of NPMLE is shown in the right panel.
}\label{tab:coef200}
\begin{tabular}{ll|rrr|rrr}
\hline\hline
      &          & \multicolumn{3}{c|}{MATA} & \multicolumn{3}{c}{NPMLE} \\
Group  & Feature       & mean    & boot.se  & pval  & mean    & boot.se & pval \\ \hline
  & Stage II & -- & -- & -- &  0.102 & 0.393 & 0.795 \\
   & Stage III & -- & -- & -- &  0.161 & 0.549 & 0.769 \\
   & Age & -- & -- & -- &  0.014 & 0.019 & 0.465 \\
   \hline
Lung Cancer & 1stCode & -0.223 & 0.266 & 0.401 & -0.369 & 0.318 & 0.246 \\
   & Pk &  0.188 & 0.190 & 0.323 &  0.270 & 0.220 & 0.220 \\
   & ChP &  0.112 & 0.148 & 0.451 &  0.160 & 0.180 & 0.375 \\
   & 1stScore & -0.102 & 0.390 & 0.793 & -0.080 & 0.553 & 0.885 \\
   & logN &  0.072 & 0.244 & 0.767 &  0.065 & 0.325 & 0.843 \\
   \hline
Chemotherapy & 1stCode & -0.103 & 0.143 & 0.471 & -0.170 & 0.212 & 0.423 \\
   & Pk & -0.119 & 0.218 & 0.585 & -0.206 & 0.331 & 0.534 \\
   & ChP &  0.006 & 0.160 & 0.972 &  0.027 & 0.243 & 0.913 \\
   & 1stScore &  0.530 & 0.409 & 0.195 &  0.764 & 0.516 & 0.139 \\
   & logN &  0.056 & 0.184 & 0.759 &  0.064 & 0.282 & 0.822 \\
   \hline
CT scan & 1stCode &  0.007 & 0.165 & 0.965 &  0.008 & 0.252 & 0.976 \\
   & Pk &  0.068 & 0.196 & 0.730 &  0.116 & 0.276 & 0.674 \\
   & ChP &  0.037 & 0.109 & 0.730 &  0.055 & 0.143 & 0.700 \\
   & 1stScore & -0.188 & 0.292 & 0.520 & -0.321 & 0.366 & 0.380 \\
   & logN & -0.016 & 0.229 & 0.944 & -0.047 & 0.314 & 0.881 \\
   \hline
Radiotherapy & 1stCode & -0.207 & 0.314 & 0.509 & -0.359 & 0.405 & 0.376 \\
   & logN & -0.029 & 0.114 & 0.798 & -0.059 & 0.163 & 0.718 \\
   \hline
Secondary & 1stCode & -0.036 & 0.273 & 0.896 & -0.056 & 0.415 & 0.893 \\
  Malignant & Pk & -0.095 & 0.232 & 0.683 & -0.118 & 0.349 & 0.735 \\
  Neoplasm & ChP & -0.051 & 0.101 & 0.609 & -0.065 & 0.148 & 0.660 \\
   & 1stScore & -0.161 & 0.248 & 0.516 & -0.197 & 0.348 & 0.571 \\
   & logN &  0.258 & 0.185 & 0.162 &  0.338 & 0.207 & 0.102 \\
   \hline
Palliative & 1stCode & -0.090 & 0.197 & 0.647 & -0.102 & 0.267 & 0.703 \\
  or Hospice & Pk & -0.726 & 0.384 & 0.059 & -0.928 & 0.446 & $\mathbf{0.037}$ \\
  Care & ChP & -0.542 & 0.287 & 0.059 & -0.692 & 0.334 & $\mathbf{0.038}$ \\
   & 1stScore &  0.020 & 0.179 & 0.912 &  0.034 & 0.268 & 0.899 \\
   & logN &  0.041 & 0.124 & 0.740 &  0.024 & 0.173 & 0.890 \\
   \hline
Recurrence & 1stCode &  0.070 & 0.181 & 0.697 &  0.131 & 0.247 & 0.598 \\
   & Pk & -0.094 & 0.183 & 0.608 & -0.105 & 0.240 & 0.661 \\
   & ChP & -0.042 & 0.112 & 0.705 & -0.044 & 0.148 & 0.767 \\
   & 1stScore & -0.237 & 0.264 & 0.369 & -0.332 & 0.338 & 0.326 \\
   & logN &  0.180 & 0.177 & 0.309 &  0.284 & 0.193 & 0.141 \\
   \hline
Medication & 1stCode & -- & -- & -- &  0.174 & 0.321 & 0.589 \\
   & logN & -- & -- & -- &  0.059 & 0.230 & 0.798 \\
   \hline
Biopsy & 1stCode & -0.741 & 0.890 & 0.405 & -1.223 & 1.155 & 0.289 \\
   & logN & -0.467 & 1.551 & 0.763 & -0.876 & 2.311 & 0.705 \\
   \hline

\hline
\hline
\end{tabular}
\end{center}
\end{table}

\clearpage

\section{Convergence Rate of Derived Features}\label{app:FPCA}

Instead of deriving asymptotic properties for truncated density $f_{C_i}$, i.e., random density $f_i$ truncated on $[0,C_i]$, we focus on the scaled densities $f_{C_i,\rm scaled}$, which is $f_{C_i}$ scaled to $[0,1]$. As we assume censoring time $C_i$ has finite support $[0,\cE]$ with $\cE<\infty$, $f_{C_i,\rm scaled}$ and $f_{C_i}$ shared the same asymptotic properties.

Let $f\supj_{\mu,\rm scaled}(t)=E\{f_{C,\rm scaled}\supj(t)\}$ and $G_{\rm scaled}\supj(t,s) = {\rm cov}\{f_{C,\rm scaled}\supj(t),f_{C,\rm scaled}\supj(s)\}$. The Karhunen-Lo\`eve theorem \citep{stark1986probability} states
\bse
f_{C,\rm scaled}\supj(t)= f\supj_{\mu,\rm scaled}(t)+\sum_{k=1}^{\infty}\zeta\supj_{k,\rm scaled}\phi\supj_{k,\rm scaled}(t), {\rm~ for~} t\in[0,1],
\ese
where $\{\phi\supj_{k,\rm scaled}(t)\}$ are the orthonormal eigenfunctions of $G\supj_{\rm scaled}(t,s)$,
$\{\zeta\supj_{k,\rm scaled}\}$ are pairwise uncorrelated random variables with mean 0 and variance $\lambda_{k,\rm scaled}\supj$, and $\{\lambda_{k,\rm scaled}\supj\}$ are eigenvalues of $G\supj_{\rm scaled}(t,s)$.

For the $i$-th patient, conditional on $f\supj_{C_i}(t)$, and $M_i\supj = \Nsc\supj([0,C_i])$, the observed event times $t_{i1}\supj,\cdots, t_{iM_i\supj}\supj$ are assumed to be generated as an i.i.d. sample $t_{ij}\supj\overset{\rm iid}{\sim} f_{C_i}\supj(t)$.
Equivalently, the scaled observed event times $t_{i1}\supj/C_i,\cdots,
t_{iM_i\supj}\supj/C_i\overset{\rm iid}{\sim} f_{C_i,\rm
  scaled}\supj(t)$. Following \cite{wu2013functional}, we estimate
$f\supj_{\mu}(t)$ and $G\supj(t,s)$, which are the mean and covariance
functions of scaled density $f_{C, \rm scaled}\supj(t)$ respectively, as
\bse
\wh f\supj_{\mu,\rm scaled}(t)&=&(M\supj\subplus)^{-1}\sum_{i=1}^{n+N}\sum_{\ell=1}^{M\supj_i}\kappa\submu\suphmu(t-t_{i\ell}\supj/C_i);\\
\wh G\supj_{\rm scaled}(t,s) &=& \wh g_{\rm scaled}\supj(t,s)-\wh f\supj_{\mu,\rm scaled}(t)\wh f\supj_{\mu,\rm scaled}(s),
\ese
for $t,s\in[0,1]$, where
\bse
\wh g_{\rm scaled}\supj(t,s) &=&  (M\supj\subpplus)^{-1} \sum_{i=1}^{n+N}I(M\supj_i\geq2) \sum_{1 \le \ell \ne k \le M_{i}\supj }  \kappa\subG\suphg\left(t-t\supj_{i\ell}/C_i,s-t\supj_{ik}/C_i\right).
\ese
Here $M\supj\subplus=\sum_{i=1}^{n+N} M\supj_i$ is the total number of encounters. $M\supj\subpplus =\sum_{i=1,M\supj_i\geq 2}^{n+N} M\supj_i(M\supj_i-1)$ is the total number of pairs.
$\kappa\submu$ and $\kappa\subG$ are symmetric univariate and bivariate probability density functions, respectively, with $\kappa\submu^h(x) = \kappa\submu(x/h)/h$, $\kappa\subG^h(x_1,x_2) = \kappa\subG(x_1/h, x_2/h)/h^{2}$.
$h_{\mu}\supj$ and $h_g\supj$ are bandwidth parameters.

The estimates of eigenfunctions and eigenvalues, denoted by $\wh\phi_{k,\rm scaled}\supj(x)$ and $\wh\lambda_{k,\rm scaled}\supj$ respectively, are solutions to
\bse
\int_0^1 \wh G_{\rm scaled}\supj(s,t)\wh\phi_{k,\rm scaled}\supj(s)ds = \wh\lambda_{k,\rm scaled}\supj\wh\phi_{k,\rm scaled}\supj(t),
\ese
with constraints $\int_0^1 \wh\phi_{k,\rm scaled}\supj(s)^2ds=1$ and
$\int_0^1 \wh\phi_{k,\rm scaled}\supj(s)\wh\phi_{\ell,\rm scaled}\supj(s)ds=0$.
One can obtain estimated eigenfunctions $\wh\phi_{k,\rm scaled}\supj(x)$ and eigenvalues $\wh\lambda_{k,\rm scaled}\supj$ by numerical spectral decomposition on a properly discretized version of the smooth covariance function $\wh G_{\rm scaled}\supj(t,s)$  \citep{rice1991estimating, capra1997accelerated}. Subsequently, we estimate
\bse
\zeta\supj_{ik,\rm scaled}&=&\int \{f\supj_{C_i,\rm scaled}(t)-f\supj_{\mu,\rm scaled}(t)\}\phi\supj_{k,\rm scaled}(t)dt,
\ese
by
\bse
\wh\zeta\supj_{ik,\rm scaled}=\frac{1}{M\supj_i}\sum_{\ell=1}^{M\supj_i}\wh\phi\supj_{k,\rm scaled}(t_{i\ell}\supj/C_i)-\int\wh f\supj_{\mu,\rm scaled}(t)\wh\phi\supj_{k,\rm scaled}(t)dt.
\ese
Let $\wt\zeta\supj_{ik,\rm scaled} =( M\supj_i)^{-1}\sum_{\ell=1}^{M\supj_i}\phi\supj_{k,\rm scaled}(t_{i\ell}\supj/C_i)-\int f\supj_{\mu,\rm scaled}(t)\phi\supj_{k,\rm scaled}(t)dt$ be the population counterpart of
$\wh\zeta\supj_{ik,\rm scaled}$ constructed with true eigenfunctions.
We show in Lemma \ref{lem:xi} that $\max_i |\wh\zeta_{ik,\rm
  scaled}-\wt\zeta_{ik,\rm scaled}|$
 goes to zero at any $k$ as long as $Nh_\mu^2\to\infty$ and
$Nh_g^4\to\infty$.

We then estimate the scaled density $f\supj_{C_i,\rm scaled}(t)$ as
\bse \textstyle
\wh f\supj_{iK,\rm scaled}(t)= \max\left\{0, \wh f\supj_{\mu,\rm scaled}(t)+\sum_{k=1}^{K\supj}\wh\zeta\supj_{ik,\rm scaled}\wh\phi\supj_{k,\rm scaled}(t) \right\},
\ese
and the truncated density $f\supj_{C_i}(t)$ as
\bse
\wh f\supj_{iK}(t)= \wh f\supj_{iK,\rm scaled}(t/C_i)/
\int_0^{C_i}\wh f\supj_{iK,\rm scaled}(t/C_i) dt.
\ese
For the $i$-th patient and its $j$-th point process $\Nsc_i\supj$, we
only observe one realization of its expected number of encounters on
$[0,C_i]$, i.e., $M_i=\Nsc_i\supj([0,C_i])$. Following
\cite{wu2013functional}, we approximate the expected numbers of encounters with observed encounters, and  estimate $\lambda_i(t)$ as $\wh\lambda_i\supj(t)=M_i\wh f_{iK}\supj(t)$, for $t\in[0,C_i]$.
We further estimate the derived feature $\W_i$ as
$\wh \W_i = \calG\circ \wh f_i\supj$.

For notation simplicity in the proof, we drop the superscript $\supj$, the index for the $j$-th counting process, for $j=1,\cdots,q$ throughout the appendix.

\subsection*{\underline{Derivative of the Mean and Covariance Functions}}
Nonparametric estimation of the mean and covariance function on the scaled
densities are
\bse
\wh f_{\mu,\rm scaled}(t)&=&(M\subplus)^{-1}\sum_{i=1}^{n+N}\sum_{\ell=1}^{M_i}\kappa\submu^{h_\mu}(t-t_{i\ell}/C_i);\\
\wh G_{\rm scaled}(t,s) &=& \wh g_{\rm scaled}(t,s)-\wh f_{\mu,\rm scaled}(t)\wh f_{\mu,\rm scaled}(s),
\ese
for $t,s\in[0,1]$, where
\bse
\wh g_{\rm scaled}(t,s) &=&  (M\subpplus)^{-1} (h_{g})^2 \sum_{i=1}^{n+N}I(M_i\geq2) \sum_{1 \le \ell \ne k \le M_{i} }  \kappa\subG^{h_g}\left(t-t_{i\ell}/C_i,s-t_{ik}/C_i\right).
\ese
Here $M\subplus=\sum_{i=1}^{n+N} M_i$ is the total number of encounters. $M\subpplus =\sum_{i=1,M_i\geq 2}^{n+N} M_i(M_i-1)$ is the total number of pairs.
$\kappa\submu$ and $\kappa\subG$ are symmetric univariate and bivariate probability density functions, respectively, with $\kappa\submu^h(x) = \kappa\submu(x/h)/h$, $\kappa\subG^h(x_1,x_2) = \kappa\subG(x_1/h, x_2/h)/h^{2}$.
$h_{\mu}$ and $h_g$ are bandwidth parameters.

Their derivatives are
\bse
{{\wh f}'_{\mu,\rm scaled}}(t)&=&\frac{1}{M\subplus (h_\mu)^2}\sum_{i=1}^{n+N}\sum_{\ell=1}^{M_i}\kappa_1'\left(\frac{t-t_{i\ell}/C_i}{h_\mu}\right),\\
{\wh G_{\rm scaled}}{}^{(0,1)}(t,s) &=& {\wh g_{\rm scaled}}{}^{(0,1)}(t,s)-\wh f_{\mu,\rm scaled}(t){{\wh f}'_{\mu,\rm scaled}}(s),\\
{\wh G_{\rm scaled}}{}^{(1,0)}(t,s) &=& {\wh g_{\rm scaled}}{}^{(1,0)}(t,s)-{{\wh f}'_{\mu,\rm scaled}}(t)\wh f_{\mu,\rm scaled}(s),\\
\ese
with
\bse
{\wh g_{\rm scaled}}{}^{(\nu,u)}(t,s) &=& \frac{1}{M\subpplus {h_g}^3}\sum_{i=1,M_i\geq2}^{n+N}\sum_{\ell=1}^{M_i}\sum_{k=1,k\neq j}^{M_i} \kappa_2^{(\nu,u)}\left(\frac{t-t_{i\ell}/C_i}{h_g},\frac{s-t_{ik}/C_i}{h_g}\right),
\ese
for $\nu=0,u=1$ and $\nu=1,u=0$, where for an arbitrary bivariate function $h$, $h^{(\nu,u)}(x,y)=\partial^{\nu+u} G(x,y)/\partial^{\nu} x\partial^u y.$

Assume the following regularity conditions holds.
\begin{enumerate}
    \item[(A1)] Scaled random densities $f_{C_i,\rm scaled}$, its mean density $f_{\mu,\rm scaled}$, covariance function $g_{\rm scaled}$ and eigenfunctions $\phi_{k,\rm scaled}(x)$ are thrice continuously differentiable.
    \item[(A2)] $f_{C_i,\rm scaled}$, $f_{\mu,\rm scaled}$ and their first three derivatives are bounded, where the bounds hold uniformly across the set of random densities.
    \item[(A3)] $\kappa_1(\cdot)$ and $\kappa_2(\cdot,\cdot)$ are symmetric univariate and bivariate density function satisfying $$\int u\kappa_1(u)du=\int u\kappa_2(u,v)dudv=\int v\kappa_2(u,v)dudv=0,$$
    $$\int u^2\kappa_1(u)du<\infty,\int u^2\kappa_2(u,v)dudv<\infty, \int v^2\kappa_2(u,v)dudv<\infty$$.
    \item[(A4)] Denote the Fourier transformations $\chi_1(t) = \int \exp(-iut)\kappa_1(u)du$ and
    $\chi_2(s,t)= \int \exp(-ius-ivt)\kappa_2(u,v)dudv$. $\int|\chi_1(u)|du<\infty$ and $\int|u\chi_1(u)|du<\infty$. $\int|\chi_2(u,v)|dudv<\infty$, $\int|u\chi_2(u,v)|dudv<\infty$ and $\int|v\chi_2(u,v)|dudv<\infty$.
    \item[(A5)] The numbers of observations $M_i$ for the $j$-th trajectory of $i$-th object, are i.i.d. r.v.'s that are independent of the densities $f_i$ and satisfy
    \bse
    E(N/M\subplus)<\infty, ~ E\{N/M\subpplus\}<\infty.
    \ese
    \item[(A6)] $h_\mu\to 0,  h_g\to 0,  N{h_\mu}^4\to\infty,N{h_g}^6\to\infty$ as $N\to\infty$.
    \item[(A7)] $M_i,i=1,\cdots,n+N$ are i.i.d positive r.v. generated from a truncated-Poisson distribution with rate $\tau_N$, such that $\pr(M_i=0)=0$, and
    $\pr(M_i=k)={\tau_N}^k\exp(-\tau_N)/[k!\{1-\exp(-\tau_N)\}]$ for $k\geq1$.
    \item[(A8)] $\omega_i = E(M_i\mid C_i) = E(N_i[0,C_i]\mid C_i)$ and $f_{C_i,\rm scaled},i=1,\cdots,n+N$ are independent. $E{\omega_i}^{-1/2}=O(\alpha_N)$, where
    $\alpha_N\to0$ as $N\to\infty$ for $j=1,\cdots,q$.
    \item[(A9)] The number of eigenfunctions and functional principal
      components $K_i$ is a r.v. with $K_i\overset{d}{=}K$, and for
      any $\epsilon>0$, there exists $K_\epsilon^*<\infty$ such that
      $\pr(K>K_\epsilon^*)<\epsilon$ for $j=1,\cdots,q$.
\end{enumerate}

\begin{lemma}
Under the regularity conditions A1 - A6,
\be
\sup_x |\wh f_{\mu,\rm scaled}(x)-f_{\mu,\rm scaled}(x)|&=&O_p\left(h_\mu^2+\frac{1}{\sqrt{N}h_\mu}\right),\\
\sup_x |{\wh f}'_{\mu,\rm scaled}(x)-{f'_{\mu,\rm scaled}}(x)|&=&O_p\left(h_\mu^2+\frac{1}{\sqrt{N}h^2_\mu}\right),\\
\sup_{x,y} |\wh g_{\rm scaled}(x,y)-g_{\rm scaled}(x,y)|&=&O_p\left(h_g^2+\frac{1}{\sqrt{N}h_g^2}\right),\\
\sup_{x,y} |\triangledown \wh g_{\rm scaled}(x,y)-\triangledown {g_{\rm scaled}}(x,y)|&=&O_p\left(h_g^2+\frac{1}{\sqrt{N}h_g^3}\right),\\
\sup_{x,y} |\wh G_{\rm scaled}(x,y)-G_{\rm scaled}(x,y)|&=&O_p\left(h_g^2+\frac{1}{\sqrt{N}h_g^2}+h_\mu^2+\frac{1}{\sqrt{N}h_\mu}\right),\\
\sup_{x,y} |\triangledown \wh G_{\rm scaled}(x,y)-\triangledown{G_{\rm scaled}}(x,y)|&=&O_p\left(h_g^2+\frac{1}{\sqrt{N}h_g^3}+h_\mu^2+\frac{1}{\sqrt{N}h^2_\mu}\right).
\ee
\end{lemma}

\begin{proof}
The proof on the mean density and covariance function can be found in
\cite{wu2013functional}. Here we only obtain the proof for the
derivative of the mean density function.
The proof for the derivative of the covariance function
is similar.

Under conditions A1 and A2, we have
\bse
E\{{\wh f}'_{\mu,\rm scaled} (x)\} &=& E\left[\frac{1}{M\subplus h_\mu^2}\sum_{i=1}^{n+N}M_i E\left\{\kappa_1'\left(\frac{t-t_{i\ell}/C_i}{h_{\mu}}\right)\mid M_i, f_{C_i,\rm scaled}\right\}\right]\\
&=& E\left[\frac{1}{M\subplus }\sum_{i=1}^{n+N}M_i E\{f'_{C_i,\rm scaled}(x)+\frac{1}{2}f'''_{C_i,\rm scaled}(x)\sigma_{\kappa_1}^2h_\mu^2+o(h_\mu^2)\mid M_i\}\right]\\
&=& E\left[\frac{1}{M\subplus}\sum_{i=1}^{n+N}M_i \{f'_{\mu,\rm scaled}(x)+\frac{1}{2}E{f'''_{\mu,\rm scaled}}(x)\sigma_{\kappa_1}^2h_\mu^2+o(h_\mu^2)\}\right]\\
&=& f_{\mu,\rm scaled}'(x)+O(h_\mu^2),
\ese
Hence, $\sup_x|E{\wh f}'_\mu(x)-f'_\mu(x)|=O(h_\mu^2)$.

With inverse Fourier transformation $\kappa_1(t)=(2\pi)^{-1}\int \exp(iut)\chi_1(u)du$, we have
\bse
\kappa_1'(t)=(2\pi)^{-1}i\int u\exp(iut)\chi_1(u)du.
\ese
We further insert this equation into ${\wh f}'_{\mu}$,
\bse
{\wh f}'_{\mu,\rm scaled}(t)&=&\frac{1}{M\subplus h_{\mu}^2}\sum_{k=1}^{n+N}\sum_{\ell=1}^{M_k}\kappa_1'\left(\frac{t-t_{k\ell}/C_k}{h_{\mu}}\right)\\
&=&\frac{1}{M\subplus h_{\mu}^2}\sum_{k=1}^{n+N}\sum_{\ell=1}^{M_k}(2\pi)^{-1}i\int u\exp\{iu(t-t_{k\ell}/C_k)/h_\mu\}\chi_1(u)du\\
&=&\frac{1}{M\subplus}\sum_{k=1}^{n+N}\sum_{\ell=1}^{M_k}(2\pi)^{-1}i\int u\exp\{iu(t-t_{k\ell}/C_k)\}\chi_1(uh_\mu)du\\
&=&(2\pi)^{-1}i\int\varsigma(u)u\exp(iut)\chi_1(uh_\mu)du,
\ese
where
\bse
\varsigma(u) = \frac{1}{M\subplus h_{\mu}^2}\sum_{k=1}^{n+N}\sum_{\ell=1}^{M_k} \exp\{-iut_{k\ell}/C_k\}.
\ese
Therefore,
\bse
&&|{\wh f}'_{\mu,\rm scaled}(t)-E{\wh f}'_{\mu,\rm scaled}(t)|\\
&&\hskip5mm = |(2\pi)^{-1}i\int\{\varsigma(u)-E\varsigma(u)\}u\exp(iut)\chi_1(uh_\mu)du|\\
&&\hskip5mm \leq (2\pi)^{-1}\int|\varsigma(u)-E\varsigma(u)||u\chi_1(uh_\mu)|du.
\ese
Note that the right-hand side of the above inequality is free of $t$. Thus,
\bse
\sup_t|{\wh f}'_{\mu,\rm scaled}(t)-E{\wh f}'_{\mu,\rm scaled}(t)| \leq  (2\pi)^{-1}\int|\varsigma(u)-E\varsigma(u)||u\chi_1(uh_\mu)|du.
\ese
As an intermediate result of the Proof of Theorem 1 in \cite{wu2013functional},
we have
\bse
\var\{\varsigma(u)\}\leq \frac{1}{n+N}\left\{1+2E\left(\frac{n+N}{M\subplus}\right)\right\}.
\ese
This further lead to
\bse
&&E\{\sup_t|{\wh f}'_{\mu,\rm scaled}(t)-E\wh f'_{\mu,\rm scaled}(t)|\}\\
&&\hskip5mm \leq (2\pi)^{-1}E\{\int|\varsigma(u)-E\varsigma(u)||u\chi_1(uh_\mu)|du\}\\
&&\hskip5mm = (2\pi)^{-1}\int E\{|\varsigma(u)-E\varsigma(u)|\}|u\chi_1(uh_\mu)|du\\
&&\hskip5mm \leq (2\pi)^{-1}\int [\var\{\varsigma(u)\}]^{1/2}|u\chi_1(uh_\mu)|du\\
&&\hskip5mm \leq (2\pi)^{-1}\sqrt{\frac{1}{n+N}\left\{1+2E\left(\frac{n+N}{M\subplus}\right)\right\}}\int |u\chi_1(uh_\mu)|du\\
&&\hskip5mm = O\left(\frac{1}{\sqrt{N}h_\mu^2}\right).
\ese
Thus, $\sup_t|{\wh f}'_{\mu,\rm scaled}(t)-E{\wh f}'_{\mu,\rm scaled}(t)|=O_p\left(\frac{1}{\sqrt{N}h_\mu^2}\right)$.
Furthermore,
\bse
\sup_t|\wh f'_{\mu,\rm scaled}(t)-f'_{\mu,\rm scaled}(t)|&\leq& \sup_x|\wh f'_{\mu,\rm scaled}(t)-E\wh f'_{\mu,\rm scaled}(t)|+\sup_t|E\wh f'_{\mu,\rm scaled}(t)-f'_{\mu,\rm scaled}(t)|\\
&=&O_p\left(h_\mu^2+\frac{1}{\sqrt{N}h_\mu^2}\right).
\ese

\end{proof}

\subsection*{\underline{Derivative of the Eigenfunctions}}
\begin{lemma}
Under the regularity conditions A1 - A6,
\be
|\wh\lambda_{k,\rm scaled}-\lambda_{k,\rm scaled}|&=&O_p\left(h_g^2+\frac{1}{\sqrt{N}h_g^2}+h_\mu^2+\frac{1}{\sqrt{N}h_\mu}\right),\label{eq:FPCAlam}\\
\sup_x|\wh\phi_{k,\rm scaled}(x)-\phi_{k,\rm scaled}(x)|&=&O_p\left(h_g^2+\frac{1}{\sqrt{N}h_g^2}+h_\mu^2+\frac{1}{\sqrt{N}h_\mu}\right),\\
\sup_x|\wh\phi'_{k,\rm scaled}(x)-\phi'_{k,\rm scaled}(x)|&=&O_p\left(h_g^2+\frac{1}{\sqrt{N}h_g^3}+h_\mu^2+\frac{1}{\sqrt{N}h^2_\mu}\right).\label{eq:FPCAEigenDeriv}
\ee
\end{lemma}
\begin{proof}
The first two equations are direct result of Theorem 2 in \cite{yao2005functional}. Note that
\bse
\wh\lambda_{k,\rm scaled}\wh\phi'_{k,\rm scaled}(x) &=& \int \wh G_{\rm scaled}^{(1,0)}(x,y)\wh\phi_{k,\rm scaled}(y)dy,\\
{\lambda_{k,\rm scaled}}{\phi_{k,\rm scaled}}'(x) &=& \int G_{\rm scaled}^{(1,0)}(x,y)\phi_{k,\rm scaled}(y)dy,
\ese
where $G_{\rm scaled}^{(1,0)}(x,y)=\partial G_{\rm scaled}(x,y)/\partial x.$
Thus,
\bse
&&|\wh\lambda_{k,\rm scaled}\wh\phi'_{k,\rm scaled}(x)-{\lambda_{k,\rm scaled}}{\phi_{k,\rm scaled}}'(x)| \\
&&\hskip5mm = |\int \wh G_{\rm scaled}^{(1,0)}(x,y)\wh\phi_{k,\rm scaled}(y)dy-\int G_{\rm scaled}^{(1,0)}(x,y)\phi_{k,\rm scaled}(y)dy| \\
&&\hskip5mm \leq \int |\wh G_{\rm scaled}^{(1,0)}(x,y)- G^{(1,0)}_{\rm scaled}(x,y)||\wh\phi_{k,\rm scaled}(y)|dy\\
&&\hskip10mm +\int |G_{\rm scaled}^{(1,0)}(x,y)||\wh\phi_{k,\rm scaled}(y)-\phi_{k,\rm scaled}(y)|dy \\
&&\hskip5mm \leq \{\int |\wh G_{\rm scaled}^{(1,0)}(x,y)- G_{\rm scaled}^{(1,0)}(x,y)|^2dy\}^{1/2}\\
&&\hskip10mm+\{\int |G_{\rm scaled}^{(1,0)}(x,y)|^2dy\}^{1/2}\{\int|\wh\phi_{k,\rm scaled}(y)-\phi_{k,\rm scaled}(y)|^2dy\}^{1/2}.
\ese
Without loss of generality assuming $\lambda_{k,\rm scaled}>0$, then
\bse
\sup_x|(\wh\lambda_{k,\rm scaled}/\lambda_{k,\rm scaled})\wh\phi'_{k,\rm scaled}(x)-\phi'_{k,\rm scaled}(x)| = O_p\left(h_g^2+\frac{1}{\sqrt{N}h_g^3}+h_\mu^2+\frac{1}{\sqrt{N}h^2_\mu}\right).
\ese
Then (\ref{eq:FPCAEigenDeriv}) follows by applying (\ref{eq:FPCAlam}).
\end{proof}

\subsection*{\underline{Derivative of the Estimated Density Functions}}
\begin{lemma}\label{lem:xi}
Under regularity conditions
A1 - A9, for any $\epsilon>0$, there exists an event $A_\epsilon$ with $\pr(A_\epsilon)\geq 1-\epsilon$ such that on $A_\epsilon$ it holds that
\be
|\wh \zeta_{ik,\rm scaled}-\zeta_{ik,\rm scaled}| &=& O_p\left(\alpha_N+\frac{1}{\sqrt{N}h_g^2}+\frac{1}{\sqrt{N}h_\mu}\right),\label{eq:FPCAScore}\\
\sup_x|\wh f_{C_i,\rm scaled}(x)-f_{C_i,\rm scaled}(x)| &=& O_p\left(\alpha_N+\frac{1}{\sqrt{N}h_g^2}+\frac{1}{\sqrt{N}h_\mu}\right),\label{eq:FPCAFuns}\\
\sup_x|\wh f'_{C_i,\rm scaled}(x)-f'_{C_i,\rm scaled}(x)| &=& O_p\left(\alpha_N+h_g^2+\frac{1}{\sqrt{N}h_g^3}+h_\mu^2+\frac{1}{\sqrt{N}h_\mu^2}\right)\label{eq:FPCADeriv}.
\ee
\end{lemma}

\begin{proof}

The existence of $A_\epsilon$ for (\ref{eq:FPCAScore}) - (\ref{eq:FPCAFuns}) are guaranteed by the Theorem 3 in \cite{wu2013functional}.
We followed their definition of $A_\epsilon$, i.e.,
$A_\epsilon^c = \{K>K_\epsilon^*\}\cup \{M_i=1,i=1,\cdots,n+N\}$,
and prove for (\ref{eq:FPCADeriv}).

Note that
\bse
|\wh f'_{C_i,\rm scaled}(x)-f'_{C_i,\rm scaled}(x)| &\leq& |\wh f'_{C_i,\rm scaled}(x)-f^{'K}_{C_i,\rm scaled}(x)|+|f^{'K}_{C_i,\rm scaled}(x)-f'_{C_i,\rm scaled}(x)|.
\ese

We have
\bse
\sup_x E|f^{'K}_{C_i,\rm scaled}(x)-f'_{C_i,\rm scaled}(x)|^2 &=& \sup_x E|\sum_{k=K+1}^\infty \zeta_{ik,\rm scaled}{\phi_{k,\rm scaled}}'(x)|^2\\
&=&\sup_x\sum_{k=K+1}^\infty \lambda_{k,\rm scaled}|{\phi_{k,\rm scaled}}'(x)|^2\to0,
\ese
as $K\to\infty$.
Hence,  $|f^{'K}_{C_i,\rm scaled}(x)-f'_{C_i,\rm scaled}(x)|=o_p(1)$.

Furthermore, on $A_\epsilon$
\bse
&&\sup_x|\wh f'_{C_i,\rm scaled}(x)-f^{'K}_{C_i,\rm scaled}(x)|\\
&&\hskip5mm\leq\sup_x|\wh f'_{\mu,\rm scaled}(x)-f'_{\mu,\rm scaled}(x)|+\sum_{k=1}^K\sup_x|\wh\zeta_{ik,\rm scaled} \wh\phi'_{k,\rm scaled}(x)-\zeta_{ik,\rm scaled}\phi'_{k,\rm scaled}(x)|\\
&&\hskip5mm\leq \sup_x|\wh f'_{\mu,\rm scaled}(x)-f'_{\mu,\rm scaled}(x)|+\sum_{k=1}^K\sup_x|\wh\zeta_{ik,\rm scaled}-\zeta_{ik,\rm scaled}| |\wh\phi'_{k,\rm scaled}(x)|\\
&&\hskip10mm+ \sum_{k=1}^K\sup_x|\zeta_{ik,\rm scaled}||\wh\phi'_{k,\rm scaled}(x)-{\phi_{k}}'(x)|\\
&&\hskip5mm=O_p\left(h_\mu^2+\frac{1}{\sqrt{N}h_\mu^2}\right)+
O_p\left(\alpha_N+\frac{1}{\sqrt{N}h_g^2}+\frac{1}{\sqrt{N}h_\mu}\right)\\
&&\hskip10mm+
O_p\left(h_g^2+\frac{1}{\sqrt{N}h_g^3}+h_\mu^2+\frac{1}{\sqrt{N}h_\mu^2}\right)\\
&&\hskip5mm=O_p\left(\alpha_N+h_g^2+\frac{1}{\sqrt{N}h_g^3}+h_\mu^2+\frac{1}{\sqrt{N}h_\mu^2}\right).
\ese
Therefore (\ref{eq:FPCADeriv}) holds.

\end{proof}

\subsection*{\underline{Peaks and Change Points}}
Assume $f_{C_i,\rm scaled}$ is locally unimodal, i.e., $f'_{C_i,\rm scaled}(x)=0$ has a unique
solution, denoted by $x_{i0}$, in a neighbourhood of $x_{i0}$, denoted by ${\cal B}(x_{i0}) = (x_{i0}-\Delta x_{i0}, x_{i0} + \Delta x_{i0})$. Further assume $|f''_{C_i,\rm scaled}|$ is
bounded away from 0 in $\bigcup_{x_{i0}: f'_{C_i,\rm scaled}(x_{i0})=0}{\cal B}(x_{i0})$, and the bound holds uniformly across
$i=1,\cdots,n+N$. Let $\wh x_{i0}$ be the solution of
$\wh f'_{C_i,\rm scaled}(x)=0$ which is closet to $x_{i0}$. Then
\bse
0&=&\wh f'_{C_i,\rm scaled}(\wh x_{i0})\\
&=&f'_{C_i,\rm scaled}(\wh x_{i0}) + O_p\left(\alpha_N+h_g^2+\frac{1}{\sqrt{N}h_g^3}+h_\mu^2+\frac{1}{\sqrt{N}h_\mu^2}\right)\\
&=&f''_{C_i,\rm scaled}({x_{i0}}^*)(\wh x_{i0}-x_{i0})+ O_p\left(\alpha_N+h_g^2+\frac{1}{\sqrt{N}h_g^3}+h_\mu^2+\frac{1}{\sqrt{N}h_\mu^2}\right),
\ese
where ${x_{i0}}^*$ is an intermediate value between $x_{i0}$ and $\wh x_{i0}$.

Thus, $|\wh x_{i0}-x_{i0}|=O_p\left(\alpha_N+h_g^2+\frac{1}{\sqrt{N}h_g^3}+h_\mu^2+\frac{1}{\sqrt{N}h_\mu^2}\right)$. This further implies $\wh x_{i0}$ is the only solution of $\wh f'_{C_i,\rm scaled}$ in ${\cal B}(x_{i0})$. In other words, there is one-to-one correspondence between estimated peak and the true peak and the estimated peak converges to the true peak uniformly.

The derivation of the change point is similar, and here we only list the order of the absolute difference between estimated change point $\wh y_{i0}$ and the true change point $y_{i0}$.
\bse
|\wh y_{i0}-y_{i0}|=O_p\left(\alpha_N+h_g^2+\frac{1}{\sqrt{N}h_g^4}+h_\mu^2+\frac{1}{\sqrt{N}h_\mu^3}\right).
\ese

\section{B-spline Approximation and Profile-likelihood Estimation}\label{app:PO}

\subsection*{\underline{Some Definitions on Vector and Matrix Norms}}
For any vector $\a =(a_{1},\ldots ,a_{s})\trans
\in R^s $, denote the norm $\|\a\|_r=(|a_1|^r+\dots
+|a_s|^r)^{1/r}$, $1\le r\le \infty $. For positive numbers $a_n$
and $b_n$, $n>1$, let
$a_n\asymp b_n$ denote that $\lim_{n\rightarrow \infty }a_n/b_n=c$,
where $c$ is some nonzero constant. Denote the space of the $q^{th}$ order
smooth functions as $\C^{(q)}([0,\calE] )=\left\{ \phi :
\phi ^{(q)}\in \C[0,\calE] \right\} $. For any $s\times s$
symmetric matrix $\A$, denote its $L_q$ norm as $\|\A\|_q
=\max_{\v \in R^s,\v \ne 0}\|\A\v\|_q\|\v\|_q^{-1}$. Let $\| \A\|_\infty=\max_{1\le i\le s}\sum_{j=1}^s|a_{ij}|$. For a vector $\a$, let $\| \a\|_{\infty
}=\max_{1\le i\le s}|a_i|$.

\subsection*{\underline{Some Definition on Scores and Hessian Matrices}}
Define
\bse
\bS_{\bg,i}(\bb,\bg)
&=&\frac{\partial\log \wt H_i(\bb,\bg)}{\partial\bg}
=
\Delta_i\B_r(X_i)-
(1+\Delta_i)
\frac{\exp(\Z_i\trans\bb)\int_0^{X_i}\exp\{\B_r\trans(u)\bg\} \B_r(u) du
}
{1+\exp(\Z_i\trans\bb)\int_0^{X_i}\exp\{\B_r\trans(u)\bg\}du},\\
\bS_{\bg,i}(\bb,m)
&=&
\Delta_i\B_r(X_i)-
(1+\Delta_i)
\frac{\exp(\Z_i\trans\bb)\int_0^{X_i}\exp\{m(u)\} \B_r(u) du
}
{1+\exp(\Z_i\trans\bb)\int_0^{X_i}\exp\{m(u)\}du},\\
\bS_{\bb,i}(\bb,\bg)
&=&\frac{\partial\log \wt H_i(\bb,\bg)}{\partial\bb}
=
\Delta_i\Z_i-
(1+\Delta_i)
\frac{\Z_i\exp(\Z_i\trans\bb)\int_0^{X_i}\exp\{\B_r\trans(u)\bg\}du}
{
1+\exp(\Z_i\trans\bb)\int_0^{X_i}\exp\{\B_r\trans(u)\bg\}du},\\
\bS_{\bb,i}(\bb,m)
&=&
\Delta_i\Z_i-
(1+\Delta_i)
\frac{\Z_i\exp(\Z_i\trans\bb)\int_0^{X_i}\exp\{m(u)\}du}
{
1+\exp(\Z_i\trans\bb)\int_0^{X_i}\exp\{m(u)\}du}.
\ese
Further, define
\bse
\bS_{\bb\bb,i}(\bb,\bg)
&\equiv&
\frac{\partial\bS_{\bb,i}(\bb,\bg)}{\partial\bb\trans}
=-(1+\Delta_i)
\frac{\Z_i^{\otimes2}\exp(\Z_i\trans\bb)\int_0^{X_i}\exp\{\B_r\trans(u)\bg\}du}
{[1+\exp(\Z_i\trans\bb)\int_0^{X_i}\exp\{\B_r\trans(u)\bg\}du]^2},\\
\bS_{\bb\bb,i}(\bb,m)
&\equiv&
\frac{\partial\bS_{\bb,i}(\bb,\bg)}{\partial\bb\trans}
=-(1+\Delta_i)
\frac{\Z_i^{\otimes2}\exp(\Z_i\trans\bb)\int_0^{X_i}\exp\{m(u)\}du}
{[1+\exp(\Z_i\trans\bb)\int_0^{X_i}\exp\{m(u)\}du]^2},\\
\bS_{\bg\bg,i}(\bb,\bg)
&\equiv&
\frac{\partial\bS_{\bg,i}(\bb,\bg)}{\partial\bg\trans}\\
&=&-(1+\Delta_i)
\frac{
\exp(\Z_i\trans\bb)\int_0^{X_i}\exp\{\B_r\trans(u)\bg\} \B_r(u)^{\otimes2} du}
{1+\exp(\Z_i\trans\bb)\int_0^{X_i}\exp\{\B_r\trans(u)\bg\}du}\\
&&+(1+\Delta_i)\frac{\exp(2\Z_i\trans\bb)
[\int_0^{X_i}\exp\{\B_r\trans(u)\bg\} \B_r(u) du]^{\otimes2}}
{[1+\exp(\Z_i\trans\bb)\int_0^{X_i}\exp\{\B_r\trans(u)\bg\}du]^2},\\
\bS_{\bg\bg,i}(\bb,m)
&\equiv&
 -(1+\Delta_i)
\frac{
\exp(\Z_i\trans\bb)\int_0^{X_i}\exp\{m(u)\} \B_r(u)^{\otimes2} du}
{1+\exp(\Z_i\trans\bb)\int_0^{X_i}\exp\{m(u)\}du}\\
&&+(1+\Delta_i)\frac{\exp(2\Z_i\trans\bb)
[\int_0^{X_i}\exp\{m(u)\} \B_r(u) du]^{\otimes2}}
{[1+\exp(\Z_i\trans\bb)\int_0^{X_i}\exp\{m(u)\}du]^2},\\
\bS_{\bb\bg,i}(\bb,\bg)
&\equiv&
\frac{\partial\bS_{\bb,i}(\bb,\bg)}{\partial\bg\trans}
=
\frac{-(1+\Delta_i) \Z_i\exp(\Z_i\trans\bb)\int_0^{X_i}\exp\{\B_r\trans(u)\bg\}\B_r\trans(u)du}
{[1+\exp(\Z_i\trans\bb)\int_0^{X_i}\exp\{\B_r\trans(u)\bg\}du]^2},\\
\bS_{\bb\bg,i}(\bb,m)
&\equiv&
\frac{-(1+\Delta_i)\Z_i\exp(\Z_i\trans\bb)\int_0^{X_i}\exp\{m(u)\}\B_r\trans(u)du}
{[1+\exp(\Z_i\trans\bb)\int_0^{X_i}\exp\{m(u)\}du]^2}.
\ese

Note that
\bse
\frac{l_n(\bb,\bg)}{\partial\bg}=\sumi \bS_{\bg,i}(\bb,\bg), \ \ \
\frac{l_n(\bb,\bg)}{\partial\bb}=\sumi \bS_{\bb,i}(\bb,\bg).
\ese

For $u\in [0,\calE] $, define
\be
\wh\sigma^2(u,\bb)
=\B_r\trans(u) \{\V_n(\bb_0)\}^{-1}\{
n^{-2}\sumi\bS_{\bg,i}(\bb_0,m)^{\otimes2}\}
\{\V_n(\bb_0)\}^{-1}
\B_r(u), \label{eq:sighat2}
\ee
where
\bse
\V_n(\bb)=-E\{\bS_{\bg\bg,i}(\bb,m)\}.
\ese

In the following  theorems, we establish the consistency, asymptotic
normality  of our procedure.

\subsection*{\underline{Proof of Theorem \ref{th:Theorem1}}}
\begin{proof} For $m\in C^q[0,\calE] $, there exists $\bg_0\in R^{P_n}$, such that
\be
\sup_{u\in [0,\calE] }| m(u)-\wt{m}
(u)| =O(h^q),  \label{eq:m-mtilda}
\ee
where $\wt{m}(u)=\B_r\trans(u)\bg_0$ \citep{deBoor2001}.
 In the following, we prove the results for the nonparametric estimator $
\wh{m}(u,\bb )$ in Theorem \ref{th:Theorem1} when $\bb =\bb_0$. Then
the results also hold when $\bb $ is a $\sqrt{n}$-consistent estimator
of $\bb_0$, since the nonparametric convergence rate in Theorem
\ref{th:Theorem1}
is slower than $n^{-1/2}$.  Define the distance between neighboring knots as $h_p=\xi
_{p+1}-\xi_p,r\leq p\leq R_n+r$, and $h=\max_{r\leq p\leq
  R_n+r}h_p$. Let $\rho_n=n^{-1/2}h^{-1}+h^{q-1/2}$.
We will show that for any given $\epsilon >0$, for $n$ sufficiently large,
there exists a large constant $C>0$ such that
\be
\hbox{pr}\{\sup_{\| \btau \|_{2}=C}l_n(\bb_0,
\bg_0+\rho_n\btau )<l_n(\bb_0,\bg_0)\}\geq 1-6\epsilon. \label{eqA7new}
\ee
This implies that for $n$ sufficiently large, with probability at least $
1-6\epsilon $, there exists a local maximum for (\ref{eq:lnbg}) in the
ball $\{ \bg_0+\rho_n\btau:\| \btau
\|_2\le C\} $. Hence, there exists a local maximizer such
that $\|\wh\bg(\bb_0)-\bg_0\|_2=O_p(\rho_n)$.
Note that
\bse
\frac{\partial^2 l_n(\bb_0,\bg)}{\partial\bg\partial\bg\trans}
=\sumi\bS_{\bg\bg,i}(\bb_0,\bg)
\ese
and
\bse
\bS_{\bg\bg,i}(\bb,\bg)
&=&-(1+\Delta_i)
\frac{
\exp(\Z_i\trans\bb)\int_0^{X_i}\exp\{\B_r\trans(u)\bg\} \B_r(u)^{\otimes2} du}
{[1+\exp(\Z_i\trans\bb)\int_0^{X_i}\exp\{\B_r\trans(u)\bg\}du]^2}\\
&&-(1+\Delta_i)
\frac{
\exp(2\Z_i\trans\bb)\int_0^{X_i}\exp\{\B_r\trans(u)\bg\} \B_r(u)^{\otimes2} du
\int_0^{X_i}\exp\{\B_r\trans(u)\bg\}du}
{[1+\exp(\Z_i\trans\bb)\int_0^{X_i}\exp\{\B_r\trans(u)\bg\}du]^2}\\
&&+(1+\Delta_i)\frac{\exp(2\Z_i\trans\bb)
[\int_0^{X_i}\exp\{\B_r\trans(u)\bg\} \B_r(u) du]^{\otimes2}}
{[1+\exp(\Z_i\trans\bb)\int_0^{X_i}\exp\{\B_r\trans(u)\bg\}du]^2}.
\ese
The first term above is negative-definite, and last two terms are also
negative-definite because of Cauch-Schwartz inequality, hence
$\bS_{\bg\bg,i}(\bb_0,\bg)$
 is negative-definite. Thus,
$l_n(\bb_0,\bg)$ is a concave function of $\bg$, so the
local maximizer is  the global maximizer of (\ref{eq:lnbg}), which
will show the convergence of $\wh\bg(\bb_0)$ to $\bg_0$.

By Taylor expansion, we have
\be
l_n(\bb_0,\bg_0+\rho_n\btau)-l_n(\bb_0,\bg_0)
=\frac{\partial l_n(\bb_0,\bg_0)}{\partial \bg\trans}
\rho_n\btau-
\left\{-\frac{1}{2} \rho_n\btau\trans
\frac{\partial^2l_n(\bb_0,\bg^\ast)}{\partial\bg \partial
\bg\trans}\rho_n\btau\right\},  \label{eq:Ln}
\ee
where $\bg^{\ast }=\rho \bg +(1-\rho )\bg_0$ for some $
\rho \in (0,1)$. Moreover,
\bse
\left|\frac{\partial l_n(\bb_0,\bg_0)}{\partial \bg\trans}
\rho_n\btau \right|\le
\rho_n\left\|\frac{\partial l_n(\bb_0, \bg_0)}{\partial
\bg}\right\|_2\left\|\btau\right\|_2
=C\rho_n\left\|\frac{\partial l_n(\bb_0,\bg_0)}{\partial \bg} \right\|_2
=C\rho_n\|\T_{n1}+\T_{n2}\|_2,
\ese
where
\bse
\T_{n1} &=&\sumi\bS_{\bg,i}(\bb_0,m)\\
\T_{n2} &=&\sumi\bS_{\bg,i}(\bb_0,\bg_0)-\sumi\bS_{\bg,i}(\bb_0,m).
\ese
Recall that $S_C(\cdot)$ and $f_C(\cdot)$ are the censoring process survival
and density functions respectively, we have
\bse
&&E\{\bS_{\bg,i}(\bb_0,m)\mid\Z_i\}\\
&=&E\left[
\Delta_i\B_r(X_i)-
(1+\Delta_i)
\frac{\exp(\Z_i\trans\bb_0)\int_0^{X_i}\exp\{m(u)\} \B_r(u) du
}
{1+\exp(\Z_i\trans\bb_0)\int_0^{X_i}\exp\{m(u)\}du}\right]\\
&=&\exp(\Z_i\trans\bb_0)\int_0^\calE
\left[\B_r(X_i)-\frac{2\exp(\Z_i\trans\bb_0)\int_0^{X_i}\exp\{m(u)\} \B_r(u) du
}
{1+\exp(\Z_i\trans\bb_0)\int_0^{X_i}\exp\{m(u)\}du}\right]\\
&&\times\frac{\exp\{m(X_i)\}}
{[1+\exp(\Z_i\trans\bb_0)\int_0^{X_i}\exp\{m(u)\}du]^2}S_C(X_i\mid\Z_i)dX_i\\
&&-\int_0^{\calE-}
\frac{\exp(\Z_i\trans\bb_0)\int_0^{X_i}\exp\{m(u)\} \B_r(u) du
}
{[1+\exp(\Z_i\trans\bb_0)\int_0^{X_i}\exp\{m(u)\}du]^2}f_C(X_i\mid\Z_i)dX_i\\
&&-
\frac{\int_0^{\calE}\exp\{m(u)\} \B_r(u) du
}
{[1+\exp(\Z_i\trans\bb_0)\int_0^{\calE}\exp\{m(u)\}du]^2}S_C(\calE-\mid\Z_i)\\
&=&
\exp(\Z_i\trans\bb_0)\left[\int_0^{\calE-}
\frac{\exp\{m(X_i)\}\B_r(X_i)}
{[1+\exp(\Z_i\trans\bb_0)\int_0^{X_i}\exp\{m(u)\}du]^2}S_C(X_i\mid\Z_i)dX_i\right.\\
&&-\int_0^{\calE-}\frac{2 \exp\{m(X_i)\}
\exp(\Z_i\trans\bb_0)\int_0^{X_i}\exp\{m(u)\} \B_r(u) du
}
{[1+\exp(\Z_i\trans\bb_0)\int_0^{X_i}\exp\{m(u)\}du]^3}S_C(X_i\mid\Z_i)dX_i
\\
&&\left.-\int_0^{\calE-}
\frac{\int_0^{X_i}\exp\{m(u)\} \B_r(u) du}
{[1+\exp(\Z_i\trans\bb_0)\int_0^{X_i}\exp\{m(u)\}du]^2}f_C(X_i\mid\Z_i)
dX_i\right]\\
&&-
\frac{\int_0^{\calE}\exp\{m(u)\} \B_r(u) du
}
{[1+\exp(\Z_i\trans\bb_0)\int_0^{\calE}\exp\{m(u)\}du]^2}S_C(\calE-\mid\Z_i)\\
&=&\exp(\Z_i\trans\bb_0)\left[\int_0^{\calE-}S_C(X_i\mid\Z_i)
\frac{\partial}{\partial X_i}\frac{\int_0^{X_i}\exp\{m(u)\} \B_r(u) du}
{[1+\exp(\Z_i\trans\bb_0)\int_0^{X_i}\exp\{m(u)\}du]^2}
dX_i\right.\\
&&\left.-\int_0^{\calE-}
\frac{\int_0^{X_i}\exp\{m(u)\} \B_r(u) du}
{[1+\exp(\Z_i\trans\bb_0)\int_0^{X_i}\exp\{m(u)\}du]^2}f_C(X_i\mid\Z_i)
dX_i\right]\\
&&-
\frac{\int_0^{\calE}\exp\{m(u)\} \B_r(u) du
}
{[1+\exp(\Z_i\trans\bb_0)\int_0^{\calE}\exp\{m(u)\}du]^2}S_C(\calE-\mid\Z_i)\\
&=&\exp(\Z_i\trans\bb_0)\left[\int_0^{\calE-}
\frac{\partial}{\partial X_i}\frac{S_C(X_i\mid\Z_i)\int_0^{X_i}\exp\{m(u)\} \B_r(u) du}
{[1+\exp(\Z_i\trans\bb_0)\int_0^{X_i}\exp\{m(u)\}du]^2}dX_i
\right]\\
&&-
\frac{\int_0^{\calE}\exp\{m(u)\} \B_r(u) du
}
{[1+\exp(\Z_i\trans\bb_0)\int_0^{\calE}\exp\{m(u)\}du]^2}S_C(\calE-\mid\Z_i)\\
&=&\0.
\ese
In the following, all the integrals are calculated on $[0,\calE]$, unless otherwise specified.

Thus, $E(\T_{n1})=\0$.
Further
\bse
&&E[\{\e_p\trans\bS_{\bg,i}(\bb_0,m)\}^2\vert\Z_i]\\
&=&E\left(\left[
\Delta_iB_{r,p}(X_i)-
(1+\Delta_i)
\frac{\exp(\Z_i\trans\bb_0)\int_0^{X_i}\exp\{m(u)\} B_{r,p}(u) du}
{1+\exp(\Z_i\trans\bb_0)\int_0^{X_i}\exp\{m(u)\}du}\right]^2\Bigg\vert\Z_i\right)\\
&=& \int\left[
B_{r,p}(X_i)-2
\frac{\exp(\Z_i\trans\bb_0)\int_0^{X_i}\exp\{m(u)\} B_{r,p}(u) du}
{1+\exp(\Z_i\trans\bb_0)\int_0^{X_i}\exp\{m(u)\}du}\right]^2 f_T(X_i\mid\Z_i) S_C(X_i\mid\Z_i) dX_i\\
&&+\int\left[
\frac{\exp(\Z_i\trans\bb_0)\int_0^{X_i}\exp\{m(u)\} B_{r,p}(u) du}
{1+\exp(\Z_i\trans\bb_0)\int_0^{X_i}\exp\{m(u)\}du}\right]^2 f_C(X_i\mid\Z_i) S_T(X_i\mid\Z_i) dX_i\\
&=&\int\left[
B_{r,p}(X_i)-2
\frac{\exp(\Z_i\trans\bb_0)\int_0^{X_i}\exp\{m(u)\} B_{r,p}(u) du}
{1+\exp(\Z_i\trans\bb_0)\int_0^{X_i}\exp\{m(u)\}du}\right]^2 f_T(X_i\mid\Z_i) S_C(X_i\mid\Z_i) dX_i\\
&&+\int_0^{\calE-}\left[
\frac{\exp(\Z_i\trans\bb_0)\int_0^{X_i}\exp\{m(u)\} B_{r,p}(u) du}
{1+\exp(\Z_i\trans\bb_0)\int_0^{X_i}\exp\{m(u)\}du}\right]^2 f_C(X_i\mid\Z_i) S_T(X_i\mid\Z_i) dX_i\\
&&+\left[
\frac{\exp(\Z_i\trans\bb_0)\int_0^{\calE}\exp\{m(u)\} B_{r,p}(u) du}
{1+\exp(\Z_i\trans\bb_0)\int_0^{\calE}\exp\{m(u)\}du}\right]^2 S_C(\calE-\mid\Z_i) S_T(\calE\mid\Z_i)\\
&\le& C_1''\left(\int\left[
B_{r,p}(X_i)-2
\frac{\exp(\Z_i\trans\bb_0)\int_0^{X_i}\exp\{m(u)\} B_{r,p}(u) du}
{1+\exp(\Z_i\trans\bb_0)\int_0^{X_i}\exp\{m(u)\}du}\right]^2 dX_i\right.\\
&&\left.+\int_0^{\calE-}\left[
\frac{\exp(\Z_i\trans\bb_0)\int_0^{X_i}\exp\{m(u)\} B_{r,p}(u) du}
{1+\exp(\Z_i\trans\bb_0)\int_0^{X_i}\exp\{m(u)\}du}\right]^2 dX_i\right.\\
&&\left.+\left[
\frac{\exp(\Z_i\trans\bb_0)\int_0^{\calE}\exp\{m(u)\} B_{r,p}(u) du}
{1+\exp(\Z_i\trans\bb_0)\int_0^{\calE}\exp\{m(u)\}du}\right]^2\right)\\
&\le&C_1''\left(2\int
B_{r,p}(X_i)^2 dX_i+
9\exp(2\Z_i\trans\bb_0)\int\left[
\int\exp\{m(u)\} B_{r,p}(u) du\right]^2 dX_i\right.\\
&&\left. +\exp(2\Z_i\trans\bb_0)
\left[\int_0^{\calE}\exp\{m(u)\} B_{r,p}(u) du\right]^2\right)\\
&\le& C_1''\left(2\int
B_{r,p}(X_i)^2 dX_i+
9\calE\exp(2\Z_i\trans\bb_0)\left[
\int\exp\{m(u)\} B_{r,p}(u) du\right]^2\right.\\
&&\left. +\exp(2\Z_i\trans\bb_0)
\left[\int_0^{\calE}\exp\{m(u)\} B_{r,p}(u) du\right]^2\right)\\
&\le& C_1''\left(2\int
B_{r,p}(X_i)^2 dX_i+
(9\calE+1)\exp(2\Z_i\trans\bb_0)\left[
\int\exp\{2m(u)\}du\int B_{r,p}^2(u) du\right]\right)\\
&\le& C_1'h,
\ese
 for some constant $0<C_1'<\infty$ by Condition (C4).
 Thus, $E(\|n^{-1}\T_{n1}\|_{2}^{2})\le P_nn^{-1}C_1'h$. By Condition
 (C3), we have $h\asymp P_n^{-1}$. Then $E(\|n^{-1}\T_{n1}\|_{2}^{2})\le
C_1n^{-1}$ for some constant $0<C_1<\infty$. Then for any $\epsilon >0$, by
Chebyshev's inequality, we have $\hbox{pr}(\|n^{-1}\T_{n1}\|_{2}\geq
\sqrt{n^{-1}C_1\epsilon ^{-1}})\le \epsilon $,
or equivalently
\be\label{eq:cheb1}
\hbox{pr}(\|\T_{n1}\|_{2}\geq
\sqrt{nC_1\epsilon ^{-1}})\le \epsilon.
\ee
Moreover, by (\ref{eq:m-mtilda}), we have $\sup_u|\B_r\trans(u)\bg_0
-m(u)|=O(h^q)$. Denote
\bse
&&T_{ip}\\
&=&\e_p\trans\{\bS_{\bg,i}(\bb_0,\bg_0)-
\bS_{\bg,i}(\bb_0, m)\}\\
&=&
(1+\Delta_i)\left[
\frac{\exp(\Z_i\trans\bb)\int_0^{X_i}\exp\{m(u)\}B_{r,p}(u) du
}
{1+\exp(\Z_i\trans\bb)\int_0^{X_i}\exp\{m(u)\}du}
-
\frac{\exp(\Z_i\trans\bb)\int_0^{X_i}\exp\{\B_r\trans(u)\bg\} B_{r,p}(u) du
}
{1+\exp(\Z_i\trans\bb)\int_0^{X_i}\exp\{\B_r\trans(u)\bg\}du}\right]\\
&=&
\frac{(1+\Delta_i) \exp(\Z_i\trans\bb)\int_0^{X_i}
[\exp\{m(u)\}
-\exp\{\B_r\trans(u)\bg\}] B_{r,p}(u) du
}
{[1+\exp(\Z_i\trans\bb)\int_0^{X_i}\exp\{m(u)\}du]
[1+\exp(\Z_i\trans\bb)\int_0^{X_i}\exp\{\B_r\trans(u)\bg\}du]
}\\
&&+(1+\Delta_i) \exp(2\Z_i\trans\bb)\left[
\frac{
  \int_0^{X_i}\exp\{m(u)\}B_{r,p}(u) du \int_0^{X_i}[\exp\{\B_r\trans(u)\bg\}
-\exp\{m(u)\}]du
}
{[1+\exp(\Z_i\trans\bb)\int_0^{X_i}\exp\{m(u)\}du]
[1+\exp(\Z_i\trans\bb)\int_0^{X_i}\exp\{\B_r\trans(u)\bg\}du]
}\right]\\
&&+(1+\Delta_i) \exp(2\Z_i\trans\bb)\left[
\frac{
\int_0^{X_i}[\exp\{m(u)\}-\exp\{\B_r\trans(u)\bg\} ]B_{r,p}(u) du \int_0^{X_i}\exp\{m(u)\}du
}
{[1+\exp(\Z_i\trans\bb)\int_0^{X_i}\exp\{m(u)\}du]
[1+\exp(\Z_i\trans\bb)\int_0^{X_i}\exp\{\B_r\trans(u)\bg\}du]
}\right],
\ese
then
\bse
|T_{ip}|&\le&2
 \exp(\Z_i\trans\bb)\int_0^{X_i}
|\exp\{m(u)\}
-\exp\{\B_r\trans(u)\bg\}| B_{r,p}(u) du\\
&&+2 \exp(2\Z_i\trans\bb)
  \int_0^{X_i}\exp\{m(u)\}B_{r,p}(u) du \int_0^{X_i}|\exp\{\B_r\trans(u)\bg\}
-\exp\{m(u)\}|du\\
&&+2\exp(2\Z_i\trans\bb)
\int_0^{X_i}|\exp\{m(u)\}-\exp\{\B_r\trans(u)\bg\}|B_{r,p}(u) du
\int_0^{X_i}\exp\{m(u)\}du\\
&\le&C_2'h^{q+1}
\ese
for a constant $0<C_2'<\infty$ under Condition (C4).
Therefore,
$E(\|\T_{n2}\|_2) \le \{P_n(C_2'h^{q+1}n)^2\}^{1/2}
=P_n^{1/2}C_2'nh^{q+1}\le C_2nh^{q+1/2}
$
for a constant $0<C_2<\infty$,
and
$E(\|\T_{n2}\|_{2}^2) \le P_n(C_2'h^{q+1}n)^2
\le (C_2nh^{q+1/2})^2$.
Again by Chebyshev's inequality,
for $1/4>\epsilon >0$, we have
\be\label{eq:cheb2}
&&\pr(\|\T_{n2}\|_{2}\ge \epsilon
^{-1/2}C_2nh^{q+1/2})\n\\
&\le&\pr\{
|\|\T_{n2}\|_2-E(\|\T_{n2}\|_{2})|\ge \epsilon^{-1/2}C_2nh^{q+1/2}/2\}\n\\
&&+
\pr\{E(\|\T_{n2}\|_2)\ge
\epsilon^{-1/2}C_2nh^{q+1/2}/2\}\n\\
&\le&\pr(|\|\T_{n2}\|_2-E(\|\T_{n2}\|_2)|
\ge
\epsilon^{-1/2}\{\var(\|\T_{n2}\|_2)\}^{1/2}
/2) \n\\
&&+
\pr(
C_2nh^{q+1/2}
\ge
\epsilon^{-1/2}C_2nh^{q+1/2}/2)\n\\
&=&\pr(|\|\T_{n2}\|_2-E(\|\T_{n2}\|_2)|
\ge
\epsilon^{-1/2}\{\var(\|\T_{n2}\|_2)\}^{1/2}
/2) \n\\
&\le& 4\epsilon.
\ee
Combining (\ref{eq:cheb1}) and (\ref{eq:cheb2}), with probablity at
least $1-5\epsilon$,
\be
|\{\partial l_n(\bb_0, \bg_0)/\partial \bg \}\trans
\rho_n\btau |
&\le &
C\rho_n
(\|\T_{n1}\|_{2}+\|\T_{n2}\|_{2}) \n\\
&\le &C\rho_n(\sqrt{C_1\epsilon^{-1}}n^{1/2}+\epsilon
^{-1/2}C_2nh^{q+1/2}).  \label{eq:Lndev1}
\ee

Moreover, Lemma \ref{Lem1} implies there exists a constant $0<C_3<\infty$ such that
\bse
-\frac12\btau\trans\frac{\partial^2l_n(\bb_0,\bg^*)}{\partial \bg \partial \bg \trans}\btau
\geq nC_3C^2h
\ese
for $n$
sufficiently large, with probability approaching 1.
Thus, for any $\epsilon>0$, there is probability at least $1-\epsilon$,
\be\label{eq:Lndev2}
-2^{-1} (\rho_n\btau )\trans\{\partial^2l_n(\bb_0,\bg^{\ast })/\partial\bg \partial \bg\trans\}(\rho_n\btau)\ge\rho_n^2C_3C^2nh.
\ee

Therefore, by (\ref{eq:Ln}), (\ref{eq:Lndev1}) and (\ref{eq:Lndev2}), for $n$
sufficiently large, with probability at least $1-6\epsilon$,
\bse
&&l_n(\bb_0,\bg_0+\rho_n\btau)-l_n(\bb_0,\bg_0) \\
&\le &C\rho_n(\sqrt{C_1\epsilon ^{-1}}n^{1/2}+\epsilon
^{-1/2}C_2nh^{q+1/2})-\rho_n^{2}C_3C^2nh \\
&=&C\rho_nh(\sqrt{C_{1}\epsilon ^{-1}}n^{1/2}h^{-1}+\epsilon
^{-1/2}C_2nh^{q-1/2}-CC_3n\rho_n)\\
&=&C\rho_nh(\sqrt{C_{1}\epsilon ^{-1}}n^{1/2}h^{-1}+\epsilon
^{-1/2}C_2nh^{q-1/2}-CC_3n^{1/2}h^{-1}-CC_3nh^{q-1/2})\\
&<&0,
\ese
when $C>\max (C_3^{-1}\sqrt{C_1\epsilon
  ^{-1}},\epsilon^{-1/2}C_3^{-1}C_2)$. This shows
(\ref{eqA7new}). Hence, we have $\| \wh\bg(\bb_0)-\bg_0\|_{2}=O_p(\rho
_n)=O_p(n^{-1/2}h^{-1}+h^{q-1/2})=o_p(1)$
under Condition (C3).

It is easily seen that
$E\{\|\bS_{\bg,i}(\bb_0,m)\|_\infty^d\}\le C_4^d h$ for a constant
$1<C_4<\infty$ and any $d\ge1$,
by Bernstein's inequality, under condition (C3), we have
\bse
\|n^{-1}\sumi \bS_{\bg,i}(\bb_0,m)\|_\infty=O_p[h+\{h\log(n)\}^{1/2}n^{-1/2}]=O_p(h).
\ese
Also, it is easy to check that
\bse
\|n^{-1}\sumi \bS_{\bg,i}(\bb_0,m)
-n^{-1}\sumi \bS_{\bg,i}(\bb_0,\bg_0)
\|_\infty=O_p(h^{q+1}).
\ese
Thus, combining with Lemma \ref{Lem3}-\ref{Lem4}, we have
\be
&&\left|\B_r(u)\trans\left[ \left\{ -n^{-1}
\frac{\partial^2l_n(\bb_0,\bg_0)}{\partial \bg \partial \bg \trans}\right\}
^{-1}\left\{n^{-1}\frac{\partial l_n(\bb_0,\bg_0)}{\partial\bg} \right\}-
\V_n(\bb_0)^{-1}n^{-1}\sumi\bS_{\bg,i}(\bb_0,m)\right] \right|  \n\\
&\le &r\left\{\|\B_r(u)\|_{\infty }\left\|
\left\{ -n^{-1}
\frac{\partial^2l_n(\bb_0,\bg_0)}{\partial \bg \partial \bg \trans}\right\}
^{-1}\right\|_\infty\|
n^{-1}\sumi\bS_{\bg,i}(\bb_0,\bg_0)
-n^{-1}\sumi\bS_{\bg,i}(\bb_0,m)\|_\infty  \n \right.\\
&&\left.+\|\B_r(u)\|_\infty\left\|
\left\{ -n^{-1}
\frac{\partial^2l_n(\bb_0,\bg_0)}{\partial \bg \partial \bg \trans}\right\}^{-1}
-
\V_n(\bb_0)^{-1}\right\|_\infty\|
n^{-1}\sumi\bS_{\bg,i}(\bb_0,m)
\|_\infty  \right\} \n\\
&=& O_p(h^{-1})O_p(h^{q+1})+O_p(h^{q-1}+n^{-1/2}h^{-1})O_p(h)\n\\
&=&O_p(h^{q}+n^{-1/2}),  \label{eq:VnDn}
\ee
where the inequality above uses the fact that for arbitrary $u$,
only $r$ elements in $\B_r(u)$ are non-zero.

Let $\wh{\e}=\V_n(\bb_0)^{-1}
n^{-1}\sumi\bS_{\bg,i}(\bb_0,m)$.
Let $\Z=(\Z_1\trans,\dots, \Z_n\trans)\trans$.
 By
Central Limit Theorem,
\bse
\left[\B_r\trans(u)\text{var}\left(
\wh\e|\Z\right) \B_r(u)\right]
^{-1/2}\B_r\trans(u)\wh\e\rightarrow \hbox{Normal}(0,1),
\ese
where $\var( \wh\e|\Z)
=\{\V_n(\bb_0)\}^{-1}\{
n^{-2}\sumi\bS_{\bg,i}(\bb_0,m)^{\otimes2}\}
\{\V_n(\bb_0)\}^{-1}
$ and $\B_r\trans(u)\var( \wh\e
| \Z)\B_r(u)=\wh\sigma^2(u,\bb_0)$.
With Lemma \ref{Lem3} and \ref{Lem5}, we can get that
$c_5(nh)^{-1} \|
\B_r(u)\|_2^2\le
\B_r\trans(u)\var( \wh\e|\Z)\B_r(u)
\le C_5(nh)^{-1}  \|
\B_r(u)\|_2^2,
$
for some constants $0<c_5,c_5<\infty$.
So there exist constants $0<c_\sigma\le C_\sigma<\infty$
such that with probability approaching $1$ and for large enough $n$,
\be
c_\sigma(nh)^{-1/2}\le \inf_{u\in [0,\calE] }\wh\sigma(u,\bb_0)\le \sup_{u\in [0,\calE] }\wh\sigma(u,\bb_0)\le C_\sigma(nh)^{-1/2}.  \label{eq:sig2}
\ee
Thus $\B_r\trans(u)\wh\e=O_p\left\{(nh)^{-1/2}\right\} $
uniformly in $u\in [0,\calE] $, and hence
\bse
\B_r\trans(u)\left\{-\partial ^{2}l_n(\bb_0,\bg_0)
/\partial \bg \partial\bg\trans\right\}
^{-1}\{\partial l_n(\bb_0,\bg_0)/\partial\bg\}&=&O_p\left\{
  (nh)^{-1/2}+h^{q}+n^{-1/2}\right\}\\
&=& O_p(h^{q}+n^{-1/2}h^{-1/2}).
\ese
uniformly in $u\in [0,\calE] $ as well.

By Taylor expansion,
\be
\B_r\trans(u)\{\wh{\bg}(\bb_0)-\bg_0\}&=&\B_r\trans(u)\left\{-\partial
^{2}l_n(\bb_0,\bg_0)/\partial \bg\partial \bg\trans
\right\} ^{-1}\{\partial l_n(\bb_0,\bg_0)/\partial \bg
\}\{1+o_p(1)\}\n\\
&=&\B_r\trans(u)\left\{-\partial
^{2}l_n(\bb_0,\bg_0)/\partial \bg\partial \bg\trans
\right\} ^{-1}\{\partial l_n(\bb_0,\bg_0)/\partial \bg
\}\n\\
&&+o_p(h^q+n^{-1/2}h^{-1/2}).  \label{eq:lambdahat}
\ee
Thus by (\ref{eq:VnDn}), (\ref{eq:sig2}), (\ref{eq:lambdahat}) and Condition (C3),
\bse
&&\sup_{u\in [0,\calE] }|\wh\sigma(u,\bb_0)^{-1}\left[\B_r\trans(u)\left\{ \wh\bg(\bb_0)-\bg_0\right\}
-\B_r\trans(u)\wh\e\right] |\\
&=&O_p\{(nh)^{1/2})\}\{O_p(h^{q}+n^{-1/2})+o_p(h^q+n^{-1/2}h^{-1/2})\}\\
&=&O_p(n^{1/2}h^{q+1/2}+h^{1/2})+o_p(1)\\
&=&o_p(1).
\ese
Therefore by Slutsky's theorem $\wh\sigma^{-1}(u,\bb_0)\left\{
\wh{m}(u,\bb_0)-\wt{m}(u)\right\} \rightarrow \hbox{Normal}
(0,1)$ and $\wh{m}(u,\bb_0)-\wt{m}(u)=O_p\left\{
(nh)^{-1/2}\right\} $ uniformly in $u\in [0,\calE] $. By
$\sup_{u\in [0,\calE] }|m(u)-\wt{m}(u)|=O(h^q)$, we have
$|\wh{m}(u,\bb_0)-m(u)|=O_p\{(nh)^{-1/2}+h^q\}$ uniformly in $u\in
[0,\calE]$. By Slutsky's theorem and Condition (C3), we have
\bse
\wh\sigma^{-1}(u,\bb_0)\left\{ \wh{m}(u,\bb_0)-m(u)\right\} \rightarrow \hbox{Normal}(0,1).
\ese
\end{proof}

\subsection*{\underline{Proof of Corollary \ref{cor:cumu}}}
\begin{proof} Using delta method, it is seen that
\bse
&&\exp\left[\int_0^t\exp\{\B_r\trans(u)\wh\bg\}du\right]-\exp\left[\int_0^t\exp\{m(u)\}du\right]\\
&\asymp& \int_0^t\B_r\trans(u)\wh\bg du-\int_0^tm(u)du\\
&\le& \int_0^t\B_r\trans(u)du(\wh\bg-\bg_0)+O_p(h^q)\\
&\asymp&h I_{P_n}\trans(\wh\bg-\bg_0)+O_p(h^q)\\
&\asymp&n^{-1}h I_{P_n}\trans\V_n(\bb_0)^{-1}\sumi\bS_{\bg,i}(\bb_0,\bg_0)+O_p(h^q).
\ese
We have shown $E\bS_{\bg,i}(\bb_0,m)=\0$ and
$\|E\bS_{\bg,i}(\bb_0,\bg_0)\|_\infty=O(h^{q+1})$.
Thus,
\bse
&&|h I_{P_n}\trans \V_n(\bb_0)^{-1}
E\bS_{\bg,i}(\bb_0,\bg_0)|\\
&&\hskip5mm \leq P_n h\|\V_n(\bb_0)^{-1}\|_\infty
\|E\bS_{\bg,i}(\bb_0,\bg_0)\|_\infty\\
&&\hskip5mm = O_p(P_n hh^{-1} h^{q+1})\\
&&\hskip5mm = O_p(h^q),
\ese
where the order of $\|\V_n(\bb_0)^{-1}\|_\infty$ is
prove in Lemma \ref{Lem3}.

Further,
\bse
&& I_{P_n}\trans\V_n(\bb_0)^{-1}E\{\bS_{\bg,i}(\bb_0,\bg_0)^{\otimes2}\}
\V_n(\bb_0)^{-1} I_{P_n}\\
&\le&\| I_{P_n}\|_2\|\V_n(\bb_0)^{-1}\|_2\|E\{\bS_{\bg,i}(\bb_0,\bg_0)^{\otimes2}\}\|_2\|
\V_n(\bb_0)^{-1}\|_2\| I_{P_n}\|_2\\
&=&O_p(P_n^{1/2}h^{-1}hh^{-1}P_n^{1/2})\\
&=&O_p(h^{-2}),
\ese
where the order of $\|E\{\bS_{\bg,i}(\bb_0,\bg_0)^{\otimes2}\}\|_2$
is a direct corollary of Lemma \ref{Lem5}.
Thus,
$$n^{-1}h I_{P_n}\trans\V_n(\bb_0)^{-1}\sumi\bS_{\bg,i}(\bb_0,\bg_0)=O_p(n^{-1/2}+h^q).$$ 
Thus,
by Central Limit Theorem,
$$\exp[\int_0^t\exp\{\B_r\trans(u)\wh\bg\}du]-\exp[\int_0^t\exp\{m(u)\}du]
=O_p(n^{-1/2}+h^q),$$ and is asymptotically normally distributed.
\end{proof}

\subsection*{\underline{Proof of Theorem \ref{th:Theorem2}}}
\begin{proof} Because $\bS_{\bb\bb,i}(\bb,\bg)$ is negative definite and
$E\{\bS_{\bb,i}(\bb_0,m)\}=\0$, similar but simpler derivation as for
Theorem \ref{th:Theorem1} can be used to show the consistency of the
maximizer $\wh\bb$.

Because at any $\bb$, $\sumi\bS_{\bg,i}\{\bb,\wh\bg(\bb)\}=\0$, hence
\bse
\0&=&\sumi\frac{\partial\bS_{\bg,i}\{\bb,\wh\bg(\bb)\}}{\partial\bb\trans}
+\sumi\bS_{\bg\bg,i}\{\bb,\wh\bg(\bb)\}
\frac{\partial\wh\bg(\bb)}{\partial\bb\trans}\\
&=&\sumi\bS_{\bb\bg,i}\trans\{\bb,\wh\bg(\bb)\}
+\sumi\bS_{\bg\bg,i}\{\bb,\wh\bg(\bb)\}
\frac{\partial\wh\bg(\bb)}{\partial\bb\trans}.
\ese
so
\be\label{eq:gb}
\frac{\partial\wh\bg(\bb_0)}{\partial\bb\trans}&=&-
[n^{-1}\sumi\bS_{\bg\bg,i}\{\bb_0,\wh\bg(\bb_0)\}]^{-1}
n^{-1}\sumi\bS_{\bb\bg,i}\trans\{\bb_0,\wh\bg(\bb_0)\}\n\\
&=&\V_n(\bb_0)^{-1}
E\left\{\bS_{\bb\bg,i}\trans(\bb_0,m)\right\}+\r_1,
\ee
where $\r_1$ is the residual term and is of smaller order of $\V_n(\bb_0)^{-1}
E\left\{\bS_{\bb\bg,i}\trans(\bb_0,m)\right\}$ componentwise.
Note that $\bS_{\bb\bg,i}(\bb,\bg)=O_p(h)$ uniformly elementwise. Hence,
\bse
\|\bS_{\bb\bg,i}\trans(\bb_0,m)\|_2 = \|\bS_{\bb\bg,i}(\bb_0,m)\|_2 = O_p(h^{1/2}),\\
\|\bS_{\bb\bg,i}\trans(\bb_0,m)\|_\infty = O_p(h),\\
\|\bS_{\bb\bg,i}(\bb_0,m)\|_\infty = O_p(1).
\ese
Subsequently, we have
\bse
\|\V_n(\bb_0)^{-1}
E\left\{\bS_{\bb\bg,i}\trans(\bb_0,m)\right\}\|_2&\leq&
\|\V_n(\bb_0)^{-1}\|_2
\|E\left\{\bS_{\bb\bg,i}\trans(\bb_0,m)\right\}\|_2\\
&=&O_p(h^{-1})O_p(h^{1/2}) = O_p(h^{-1/2}),
\ese
and
\bse
\|\V_n(\bb_0)^{-1}
E\left\{\bS_{\bb\bg,i}\trans(\bb_0,m)\right\}\|_\infty&\leq&
\|\V_n(\bb_0)^{-1}\|_\infty
\|E\left\{\bS_{\bb\bg,i}\trans(\bb_0,m)\right\}\|_\infty\\
&=&O_p(h^{-1})O_p(h) = O_p(1).
\ese
Here we use the fact that $\|\V_n(\bb_0)^{-1}\|_2 = O_p(h^{-1})$
and $\|\V_n(\bb_0)^{-1}\|_\infty=O_p(h^{-1})$, where the former one is a direct
corollary of Lemma \ref{Lem1} and the latter one is shown in Lemma \ref{Lem4}.
Therefore, $\|\r_1\|_2 = o_p(h^{-1/2})$ and $\|\r_1\|_\infty=o_p(1)$.

By Taylor expansion, for
$\bb^*=\rho\bb_0+(1-\rho)\wh\bb$, $0<\rho<1$,
\be
\0&=&n^{-1/2}\sumi\bS_{\bb,i}\{\wh\bb,\wh\bg(\wh\bb)\}\n\\
&=&n^{-1/2}\sumi\bS_{\bb,i}\{\bb_0,\wh\bg(\bb_0)\}+n^{-1}\sumi\bS_{\bb\bb,i}\{\bb^*,\wh\bg(\bb^*)\}n^{1/2}(\wh\bb-\bb_0)\n\\
&&+n^{-1}\sumi\left[\bS_{\bb\bg,i}\{\bb^*,\wh\bg(\bb^*)\}
\frac{\partial\wh\bg(\bb^*)}{\partial\bb\trans}
\right]n^{1/2}(\wh\bb-\bb_0)\n\\
&=&n^{-1/2}\sumi\bS_{\bb,i}\{\bb_0,\wh\bg(\bb_0)\}+\left[E\left\{\bS_{\bb\bb,i}(\bb_0,m)\right\}+o_p(1)\right]n^{1/2}(\wh\bb-\bb_0)\n\\
&&+\left[E\{\bS_{\bb\bg,i}(\bb_0,m)\}
\frac{\partial\wh\bg(\bb_0)}{\partial\bb\trans}+\r_2
\right]n^{1/2}(\wh\bb-\bb_0),\label{eq:expand}
\ee
where $\r_2$ is the residual term and is of smaller order
  of
$E\{\bS_{\bb\bg,i}(\bb_0,m)\}{\partial\wh\bg(\bb_0)}/{\partial\bb\trans}$ componentwise.
We claim that the residual term $\r_2$ satisfies
$\|\r_2\|_2=o_p(1)$ and $\|\r_2\|_\infty=o_p(1)$.
This is because
\bse
\left\|E\{\bS_{\bb\bg,i}(\bb_0,m)\}
\frac{\partial\wh\bg(\bb_0)}{\partial\bb\trans}\right\|_2&\leq&
\|E\{\bS_{\bb\bg,i}(\bb_0,m)\}\|_2
\left\|\frac{\partial\wh\bg(\bb_0)}{\partial\bb\trans}\right\|_2\\
&=&O_p(h^{1/2})O_p(h^{-1/2}) = O_p(1),
\ese
and
\bse
\left\|E\{\bS_{\bb\bg,i}(\bb_0,m)\}
\frac{\partial\wh\bg(\bb_0)}{\partial\bb\trans}\right\|_\infty&\leq&
\|E\{\bS_{\bb\bg,i}(\bb_0,m)\}\|_\infty
\left\|\frac{\partial\wh\bg(\bb_0)}{\partial\bb\trans}\right\|_\infty\\
&=& O_p(1)O_p(1) = O_p(1),
\ese
which leads to the claimed order of the residual $\r_2$ in  (\ref{eq:expand}).

We further use Taylor expansion to write
\bse
&&n^{-1/2}\sumi\bS_{\bb,i}\{\bb_0,\wh\bg(\bb_0)\}\n\\
&=&n^{-1/2}\sumi\bS_{\bb,i}(\bb_0,\bg_0)
+n^{-1/2}\sumi\bS_{\bb\bg,i}(\bb_0,\bg^*)
\{\wh\bg(\bb_0)-\bg_0\}\\
&=&n^{-1/2}\sumi\bS_{\bb,i}(\bb_0,\bg_0)\\
&&+n^{-1/2}\sumi
\frac{-(1+\Delta_i)
  \Z_i\exp(\Z_i\trans\bb_0)\int_0^{X_i}\exp\{\B_r\trans(u)\bg^*\}
\B_r\trans(u)\{\wh\bg(\bb_0)-\bg_0\}du}
{[1+\exp(\Z_i\trans\bb_0)\int_0^{X_i}\exp\{\B_r\trans(u)\bg^*\}du]^2}\\
&=&n^{-1/2}\sumi\bS_{\bb,i}(\bb_0,\bg_0)\\
&&+n^{-1/2}\sumi
\frac{-(1+\Delta_i)
  \Z_i\exp(\Z_i\trans\bb_0)\int_0^{X_i}\exp\{m(u)\}
\B_r\trans(u)\{\wh\bg(\bb_0)-\bg_0\}du}
{[1+\exp(\Z_i\trans\bb_0)\int_0^{X_i}\exp\{m(u)\}du]^2}\\
&&+n^{1/2}O_p(h^q)O_p(h^q+n^{-1/2}h^{-1/2})\\
&=&n^{-1/2}\sumi\bS_{\bb,i}(\bb_0,\bg_0)\\
&&+\left(E\left[
\frac{-(1+\Delta_i)
  \Z_i\exp(\Z_i\trans\bb_0)\int_0^{X_i}\exp\{m(u)\}
\B_r\trans(u)du}
{[1+\exp(\Z_i\trans\bb_0)\int_0^{X_i}\exp\{m(u)\}du]^2}\right]
+\r\right)
n^{1/2}\{\wh\bg(\bb_0)-\bg_0\}\\
&&+o_p(1)\\
&=&n^{-1/2}\sumi\bS_{\bb,i}(\bb_0,\bg_0)
+E\left\{\bS_{\bb\bg,i}(\bb_0,m)\right\}
n^{1/2}\{\wh\bg(\bb_0)-\bg_0\}
+o_p(1),
\ese
where $\bg^*=\rho\bg_0+(1-\rho)\wh\bg$, $0<\rho<1$, and the residual term
$\r$ in the second last equality satisfies
$\|\r\|_\infty=O_p(n^{-1/2})$
and $\|\r\|_2=O_p(n^{-1/2}h^{1/2})$.

Plugging this and
(\ref{eq:gb}) into (\ref{eq:expand}), recall that
\bse
\A=E\left\{\bS_{\bb\bb,i}(\bb_0,m)\right\}
-E\{\bS_{\bb\bg,i}(\bb_0,m)\}
[E\{\bS_{\bg\bg,i}(\bb_0,m)\}]^{-1}
E\left\{\bS_{\bb\bg,i}\trans (\bb_0,m)\right\},
\ese
we get
\bse
&&n^{1/2}\{-\A+o_p(1)\}(\wh\bb-\bb)\\
&=&
n^{-1/2}\sumi\bS_{\bb,i}(\bb_0,\bg_0)
+E\left\{\bS_{\bb\bg,i}(\bb_0,m)\right\}
n^{1/2}\{\wh\bg(\bb_0)-\bg_0\}
+o_p(1)\\
&=&
n^{-1/2}\sumi\bS_{\bb,i}(\bb_0,\bg_0)
+n^{-1/2}\sumi E\left\{\bS_{\bb\bg,i}(\bb_0,m)\right\}
\V_n(\bb_0)^{-1}
\bS_{\bg,i}(\bb_0,\bg_0)
+o_p(1).
\ese
It is straightforward to check that
\bse
&&E\{\bS_{\bb,i}(\bb_0,m)\}\\
&=&E\left[\Delta_i\Z_i-
(1+\Delta_i)
\frac{\Z_i\exp(\Z_i\trans\bb)\int_0^{X_i}\exp\{m(u)\}du}
{
1+\exp(\Z_i\trans\bb)\int_0^{X_i}\exp\{m(u)\}du}\right]\\
&=&
\int
\left[\Z_i-2 \frac{\Z_i\exp(\Z_i\trans\bb)\int_0^{X_i}\exp\{m(u)\}du}
{1+\exp(\Z_i\trans\bb)\int_0^{X_i}\exp\{m(u)\}du}\right]
\frac{\exp[\{m(X_i)+\Z_i\trans\bb\}]S_C(X_i,\Z_i)}
{[1+\exp(\Z_i\trans\bb)\int_0^{X_i}\exp\{m(u)\}du]^2}dX_i\\
&&-\int\left[\frac{\Z_i\exp(\Z_i\trans\bb)\int_0^{X_i}\exp\{m(u)\}du}{1+\exp(\Z_i\trans\bb)\int_0^{X_i}\exp\{m(u)\}du}\right]
\frac{f_C(X_i,\Z_i)}
{1+\exp(\Z_i\trans\bb)\int_0^{X_i}\exp\{m(u)\}du}
dX_i\\
&=&\int\frac{\partial}{\partial X_i}\left[\frac{\Z_i\exp(\Z_i\trans\bb)\int_0^{X_i}\exp\{m(u)\}du
S_C(X_i,\Z_i)}
{[1+\exp(\Z_i\trans\bb)\int_0^{X_i}\exp\{m(u)\}du]^2}\right]
dX_i\\
&=&\0,
\ese
 and we already have
$E\{\bS_{\bg,i}(\bb_0,m)\}=\0$. Thus,
\bse
&&E[\bS_{\bb,i}(\bb_0,\bg_0)
+E\left\{\bS_{\bb\bg,i}(\bb_0,m)\right\}
\V_n(\bb_0)^{-1}
\bS_{\bg,i}(\bb_0,\bg_0)]\\
&&\hskip 5mm = E\left[\bS_{\bb,i}(\bb_0,\bg_0)-\bS_{\bb,i}(\bb_0,m)
+E\left\{\bS_{\bb\bg,i}(\bb_0,m)\right\}
\V_n(\bb_0)^{-1}
\{\bS_{\bg,i}(\bb_0,\bg_0)-\bS_{\bg,i}(\bb_0,m)\}\right]\\
&&\hskip 5mm =O(h^{q+1})+O_p(\|E\left\{\bS_{\bb\bg,i}(\bb_0,m)\right\}\|_\infty
\|\V_n(\bb_0)^{-1}\|_\infty
\|E\{\bS_{\bg,i}(\bb_0,\bg_0)-\bS_{\bg,i}(\bb_0,m)\}\|_\infty)\\
&&\hskip 5mm =O(h^{q+1})+O_p(1)O_p(h^{-1})O_p(h^{q+1})\\
&&\hskip 5mm =O(h^{q}).
\ese
By Central Limit Theorem,
$n^{1/2}(\wh\bb-\bb)\to
\hbox{Normal}\{\0,\A^{-1}\bSigma(\A^{-1})\trans\}$, where
$\bSigma$ is given in Theorem \ref{th:Theorem2}.
\end{proof}

\section{Matrix Norms}
\begin{lemma}\label{Lem1}
There exists constants $0<c<C<\infty$ such that, for $n$ sufficiently large,
with probability approach 1,
\bse
ch<\left\|-n^{-1}\frac{\partial^2l_n(\bb_0,\bg^*)}{\partial \bg \partial \bg \trans}\right\|_2<Ch,\\
ch<\left\|-n^{-1}\frac{\partial^2l_n(\bb_0,\bg^*)}{\partial \bg \partial \bg \trans}\right\|_\infty<Ch,\\
ch<\left\|\V_n(\bb_0)\right\|_2<Ch,\\
ch<\left\|\V_n(\bb_0)\right\|_\infty<Ch,
\ese
where $\bg^*$ is an arbitrary vector in $R^{P_n}$ with $\|\bg^*-\bg_0\|_2 = o_p(1)$.
Furthermore, for arbitrary $\a\in R^{P_n}$,
\bse
ch\|\a\|_2^2<\a\trans\left\{-n^{-1}\frac{\partial^2l_n(\bb_0,\bg^*)}{\partial \bg \partial \bg \trans}\right\}\a<Ch\|\a\|_2^2,\\
ch\|\a\|_2^2<\a\trans\V_n(\bb_0)\a<Ch\|\a\|_2^2.
\ese
\end{lemma}

\begin{proof} We only prove the result for
$\partial^2l_n(\bb_0,\bg^*)/\partial \bg \partial \bg \trans$. The proof for
$\V_n(\bb_0)$ can be obtained similarly. We have
\be\label{eq:Lndevlower}
&&-n^{-1}\a\trans\frac{\partial^2l_n(\bb_0,\bg^*)}{\partial \bg \partial \bg \trans}\a\n\\
&=&n^{-1}\sumi \a\trans\left(
(1+\Delta_i)
\frac{
\exp(\Z_i\trans\bb_0)\int_0^{X_i}\exp\{\B_r\trans(u)\bg^*\} \B_r(u)^{\otimes2} du}
{1+\exp(\Z_i\trans\bb_0)\int_0^{X_i}\exp\{\B_r\trans(u)\bg^*\}du}\right.\n\\
&&\left.-(1+\Delta_i)\frac{\exp(2\Z_i\trans\bb_0)
[\int_0^{X_i}\exp\{\B_r\trans(u)\bg^*\} \B_r(u) du]^{\otimes2}}
{[1+\exp(\Z_i\trans\bb_0)\int_0^{X_i}\exp\{\B_r\trans(u)\bg^*\}du]^2}\right)\a\n\\
&\ge&n^{-1}\sumi\a\trans\left((1+\Delta_i)
\frac{
\exp(\Z_i\trans\bb_0)\int_0^{X_i}\exp\{\B_r\trans(u)\bg^*\} \B_r(u)^{\otimes2} du}
{[1+\exp(\Z_i\trans\bb_0)\int_0^{X_i}\exp\{\B_r\trans(u)\bg^*\}du]^2}\right)\a\n\\
&\ge&c_1' n^{-1}\sumi\a\trans\left\{(1+\Delta_i)
 \B_r(u)^{\otimes2} du\right\}\a\n\\
&\to&c_1'E\a\trans\left\{
(1+\Delta_i)\int_0^{X_i}\B_r(u)^{\otimes2} du
\right\}\a\n\\
&\ge&c_1'\a\trans\left\{\int_0^{\calE}
\int_0^{x}\B_r(u)^{\otimes2} du
f_C(x)S_T(x)dx \right\}\a\n\\
&=& c_1'\a\trans\left\{\int_0^{\calE-}
\int_0^{x}\B_r(u)^{\otimes2} du
f_C(x)S_T(x)dx +\int_0^{\calE}\B_r(u)^{\otimes2} du
S_C(\calE-)S_T(\calE)\right\}\a\n\\
&\ge&S_T(\calE)S_C(\calE-)c_1'\a\trans\left\{\int_0^{\calE}\B_r(u)^{\otimes2} du
\right\}\a\n\\
&\geq& c_1h\|\a\|_2^2,
\ee
for positive constants $0<c_1', c_1<\infty$
 by conditions (C1) and (C4).

Following a similar proof, we can further obtain
\be\label{eq:Lndevupper}
&&\a\trans\left\{-n^{-1}\frac{\partial^2l_n(\bb_0,\bg^*)}{\partial \bg \partial
\bg \trans}\right\}\a\n\\
&&\hskip 5mm \leq n^{-1}\sumi\a\trans\left[
(1+\Delta_i)
\frac{
\exp(\Z_i\trans\bb_0)\int_0^{X_i}\exp\{\B_r\trans(u)\bg^*\} \B_r(u)^{\otimes2} du}
{1+\exp(\Z_i\trans\bb_0)\int_0^{X_i}\exp\{\B_r\trans(u)\bg^*\}du}\right]\a\n\\
&&\hskip 5mm \leq C_1' n^{-1}\sumi\a\trans\int_0^{X_i}\B_r(u)^{\otimes2} du\a\n\\
&&\hskip 5mm \leq C_1' \a\trans\int_0^{\calE}\B_r(u)^{\otimes2} du\a\n\\
&&\hskip 5mm \leq C_1h\|\a\|_2^2,
\ee
for some constant $0<C_1',C_1<\infty$, because $\int_0^{\calE}\B_r(u)^{\otimes2} du$
is an $r$-banded matrix with diagonal and $j\th$ off-diagonal elements of order
$O(h)$ uniformly elementwise, for $j=1,\cdots,r-1$, and 0 elsewhere.

Combining (\ref{eq:Lndevlower}) and (\ref{eq:Lndevupper}), we have
\bse
c_1h<\left\|-n^{-1}\frac{\partial^2l_n(\bb_0,\bg^*)}{\partial \bg \partial
\bg \trans}\right\|_2<C_1h.
\ese

Next, we investigate the order of
$\|-n^{-1}\{\partial^2l_n(\bb_0,\bg^{\ast })/\partial\bg \partial \bg\trans\}\|_\infty$.
We have
\bse
&&\left\|-n^{-1}\frac{\partial^2l_n(\bb_0,\bg^{\ast })}{\partial\bg \partial \bg\trans}\right\|_\infty \\
&&\hskip5mm = \max_{1\leq j\leq P_n}\sum_{k=1}^{P_n} |\left[-n^{-1}\{\partial^2l_n(\bb_0,\bg^{\ast })/\partial\bg \partial \bg\trans\}\right]_{jk}|\\
&&\hskip5mm \geq\sum_{k=1}^{P_n} \left|\left[-n^{-1}\{\partial^2l_n(\bb_0,\bg^{\ast })/\partial\bg \partial \bg\trans\}\right]_{1k}\right|\\
&&\hskip5mm \geq c_2'\sum_{k=1}^{P_n} \left|n^{-1}\sumi \int_0^{X_i} B_{r,1}(u)B_{r,k}(u)du\right|\\
&&\hskip5mm = c_2'n^{-1}\sumi \int_0^{X_i} \sum_{k=1}^{P_n} B_{r,1}(u)B_{r,k}(u)du\\
&&\hskip5mm = c_2'n^{-1}\sumi \int_0^{X_i} \sum_{k=1}^{r} B_{r,1}(u)B_{r,k}(u)du\\
&&\hskip5mm \to c_2'E\int_0^{X_i} \sum_{k=1}^{r} B_{r,1}(u)B_{r,k}(u)du\\
&&\hskip5mm = c_2 h
\ese
with probability 1 as $n\to\infty$, where $0<c_2,c_2'<\infty$ are
constants. Here, for an arbitrary matrix $\A$ we use $\A_{jk}$ to
denote its element in the $j\th$ row and the $k\th$ column. In the
above inequalities, we use the fact that B-spline basis are all
non-negative and are non-zero on no more than $r$ consecutive
intervals formed by its knots.

On the other hand,
\bse
&&\|-n^{-1}\{\partial^2l_n(\bb_0,\bg^{\ast })/\partial\bg \partial \bg\trans\}\|_\infty \\
&&\hskip5mm = \max_{1\leq j\leq P_n}\sum_{k=1}^{P_n} |\left[-n^{-1}\{\partial^2l_n(\bb_0,\bg^{\ast })/\partial\bg \partial \bg\trans\}\right]_{jk}|\\
&&\hskip5mm\leq C_2'\max_{1\leq j\leq P_n}\sum_{k=1}^{P_n} \left[\left|\left\{n^{-1}\sumi\int_0^{X_i}\B_r(u)^{\otimes2}du\right\}_{jk}\right|+\left|n^{-1}\sumi\left\{\int_0^{X_i}\B_{r}(u)du\right\}^{\otimes2}_{jk}\right| \right]\\
&&\hskip5mm\leq C_2'n^{-1}\sumi\max_{1\leq j\leq P_n}\sum_{k=1}^{P_n}\left[ \left|\left\{\int_0^{X_i}\B_r(u)^{\otimes2}du\right\}_{jk}\right|+\left|\left\{\int_0^{X_i}\B_{r}(u)du\right\}^{\otimes2}_{jk}\right|\right]\\
&&\hskip5mm= C_2'n^{-1}\sumi\max_{1\leq j\leq P_n}\sum_{k=1}^{P_n} \left[\left\{\int_0^{X_i}\B_{r,j}(u)\B_{r,k}(u)du\right\}
+
\int_0^{X_i}\B_{r,j}(u)du\int_0^{X_i}\B_{r,k}(u)du
\right]\\
&&\hskip5mm\leq C_2'n^{-1}\sumi\left[\max_{1\leq j\leq
  P_n}\sum_{k=\max(1,j-r+1)}^{\min(j+r-1,P_n)}
\left\{\int_0^{X_i}\B_{r,j}(u)\B_{r,k}(u)du\right\}\right.\\
&&\hskip10mm\left.+ \max_{1\leq j\leq
  P_n} \sum_{k=1}^{P_n}\int_0^{X_i}\B_{r,j}(u)du\int_0^{X_i}\B_{r,k}(u)du\right]\\
&&\hskip5mm\leq C_2'n^{-1}\sumi\left[\max_{1\leq j\leq
  P_n}\sum_{k=\max(1,j-r+1)}^{\min(j+r-1,P_n)}
\left\{\int_0^{\calE}\B_{r,j}(u)\B_{r,k}(u)du\right\}\right.\\
&&\hskip10mm\left.+ \max_{1\leq j\leq
  P_n} \sum_{k=1}^{P_n}\int_0^{\calE}\B_{r,j}(u)du\int_0^{\calE}\B_{r,k}(u)du\right]\\
&&\hskip5mm \leq C_2h.
\ese
Hence, $c_2h\leq\|-n^{-1}\{\partial^2l_n(\bb_0,\bg^{\ast
})/\partial\bg \partial \bg\trans\}\|_\infty\leq C_2h$.

Therefore, Lemma \ref{Lem1} holds for $c = \min(c_1,c_2)$ and $C = \max(C_1,C_2)$.
\end{proof}

\begin{lemma}\label{Lem2}
\bse
\left\|-n^{-1}\frac{\partial^2l_n(\bb_0,\bg^*)}{\partial \bg \partial \bg \trans}-\V_n(\bb_0)\right\|_2 = O_p(h^{q+1}+n^{-1/2}h),\\
\left\|-n^{-1}\frac{\partial^2l_n(\bb_0,\bg^*)}{\partial \bg \partial \bg \trans}-\V_n(\bb_0)\right\|_\infty = O_p(h^{q+1}+n^{-1/2}h).
\ese
\end{lemma}

\begin{proof}Similarly as the previous derivations,
\be\label{eq:similar}
&&\left\|-n^{-1}\frac{\partial^2 l_n(\bb_0,\bg_0)}{\partial\bg\partial\bg\trans}-\V_n(\bb_0)\right\|_\infty\n\\
&=&\|-n^{-1}\sumi \bS_{\bg\bg,i}(\bb_0,\bg_0) -\V_n(\bb_0)\|_\infty\n\\
 &\le&\left\|n^{-1}\sumi (1+\Delta_i)\left(
\frac{
\exp(\Z_i\trans\bb_0)\int_0^{X_i}\exp\{\B_r\trans(u)\bg_0\} \B_r(u)^{\otimes2} du}
{1+\exp(\Z_i\trans\bb_0)\int_0^{X_i}\exp\{\B_r\trans(u)\bg_0\}du}\right.\right.\n\\
&&-
\frac{
\exp(\Z_i\trans\bb_0)\int_0^{X_i}\exp\{m(u)\} \B_r(u)^{\otimes2} du}
{1+\exp(\Z_i\trans\bb_0)\int_0^{X_i}\exp\{m(u)\}du}\n\\
&&\left.\left.-\frac{\exp(2\Z_i\trans\bb_0)
[\int_0^{X_i}\exp\{\B_r\trans(u)\bg_0\} \B_r(u) du]^{\otimes2}}
{[1+\exp(\Z_i\trans\bb_0)\int_0^{X_i}\exp\{\B_r\trans(u)\bg_0\}du]^2}
+\frac{\exp(2\Z_i\trans\bb_0)
[\int_0^{X_i}\exp\{m(u)\} \B_r(u) du]^{\otimes2}}
{[1+\exp(\Z_i\trans\bb_0)\int_0^{X_i}\exp\{m(u)\}du]^2}
\right)\right\|_\infty\n\\
&&+\left\|n^{-1}\sumi (1+\Delta_i)\left(
\frac{
\exp(\Z_i\trans\bb_0)\int_0^{X_i}\exp\{m(u)\} \B_r(u)^{\otimes2} du}
{1+\exp(\Z_i\trans\bb_0)\int_0^{X_i}\exp\{m(u)\}du}\right.\right.\n\\
&&\left.\left.
-\frac{\exp(2\Z_i\trans\bb_0)
[\int_0^{X_i}\exp\{m(u)\} \B_r(u) du]^{\otimes2}}
{[1+\exp(\Z_i\trans\bb_0)\int_0^{X_i}\exp\{m(u)\}du]^2}
\right)\right.\n\\
&&-E\left\{
(1+\Delta_i)\left(
\frac{
\exp(\Z_i\trans\bb_0)\int_0^{X_i}\exp\{m(u)\} \B_r(u)^{\otimes2} du}
{1+\exp(\Z_i\trans\bb_0)\int_0^{X_i}\exp\{m(u)\}du}\right.\right.\n\\
&&\left.\left.\left.
+\frac{\exp(2\Z_i\trans\bb_0)
[\int_0^{X_i}\exp\{m(u)\} \B_r(u) du]^{\otimes2}}
{[1+\exp(\Z_i\trans\bb_0)\int_0^{X_i}\exp\{m(u)\}du]^2}
\right)
\right\}
\right\|_\infty\n\\
&=&O_p(h^{q+1}+n^{-1/2}h),
\ee
where the second term $O_p(n^{-1/2}h)$ in the
last equality is obtained using both the Central Limit Theorem and
the matrices above are banded to the first order.
Specifically, $-n^{-1}\partial^2
l_n(\bb_0,\bg_0)/\partial\bg\partial\bg\trans-\V_n(\bb_0)$ has
diagonal and $j\th$ off-diagonal element with order
$O_p(h^{q+1}+n^{-1/2}h)$ for $j=1,\cdots,r-1$ and all the other
elements of order $O_p(h^{q+2}+n^{-1/2}h^2)$.
Further,
\bse
\left\|-n^{-1}\frac{\partial^2
  l_n(\bb_0,\bg_0)}{\partial\bg\partial\bg\trans}-
\V_n(\bb_0)\right\|_2=O_p(h^{q+1}+n^{-1/2}h),
\ese
again because the matrices are banded to the first order. In fact for arbitrary vector
$\a\in R^{P_n}$,
\bse
&&\left|\a\trans
\left\{-n^{-1}\frac{\partial^2
  l_n(\bb_0,\bg_0)}{\partial\bg\partial\bg\trans}-
\V_n(\bb_0)\right\}\a\right| \\
&\le& \sum_{j,k}|a_j| \left|\left\{-n^{-1}\frac{\partial^2
  l_n(\bb_0,\bg_0)}{\partial\bg\partial\bg\trans}-
\V_n(\bb_0)\right\}_{jk}\right| |a_k| \\
&=& \sum_{|j-k|\le2r-1}|a_j| \left|\left\{-n^{-1}\frac{\partial^2
  l_n(\bb_0,\bg_0)}{\partial\bg\partial\bg\trans}-
\V_n(\bb_0)\right\}_{jk}\right| |a_k| \\
&&+\sum_{|j-k|>2r-1}|a_j| \left|\left\{-n^{-1}\frac{\partial^2
  l_n(\bb_0,\bg_0)}{\partial\bg\partial\bg\trans}-
\V_n(\bb_0)\right\}_{jk}\right| |a_k|\\
&\leq& C'(h^{q+1}+n^{-1/2}h)\sum_{|j-k|\leq {2r-1}} |a_j| |a_k| +C_3''(h^{q+2}+n^{-1/2}h^2)\sum_{|j-k|> {2r-1}} |a_j| |a_k|\\
&\leq& C'(h^{q+1}+n^{-1/2}h)\sum_{|j-k|\leq {2r-1}} (a_j^2+ a_k^2)/2 +C_3''(h^{q+2}+n^{-1/2}h^2)\sum_{|j-k|> {2r-1}} (a_j^2+ a_k^2)/2\\
&\leq& C'(h^{q+1}+n^{-1/2}h)(2r+hP_n) \|\a\|_2^2\\
&\leq& C(h^{q+1}+n^{-1/2}h)\|\a\|_2^2,
\ese
where $0<C,C'<\infty$ are constants.\end{proof}

\begin{lemma}\label{Lem3}
There exists constant $0<c,C<\infty$, such that for $n$ sufficiently large,
with probability approach 1,
\bse
ch^{-1/2}<\left\|\left\{-n^{-1}\frac{\partial^2
  l_n(\bb_0,\bg_0)}{\partial\bg\partial\bg\trans}\right\}^{-1}\right\|_\infty<Ch^{-1},\\
ch^{-1/2}<\|\V_n(\bb_0)^{-1}\|_\infty<Ch^{-1}.
\ese
\end{lemma}

\begin{proof}
We have $\V_n(\bb_0) = h\V_0(\bb_0)-h^2\V_1(\bb_0)$,
where
\bse
\V_0(\bb_0)=h^{-1}E\left[
(1+\Delta_i)
\frac{
\exp(\Z_i\trans\bb_0)\int_0^{X_i}\exp\{m(u)\} \B_r(u)^{\otimes2} du}
{1+\exp(\Z_i\trans\bb_0)\int_0^{X_i}\exp\{m(u)\}du}
\right]
\ese
is a banded matrix with each nonzero element of order $O(1)$ uniformly and
\bse
\V_1(\bb_0) = h^{-2}E\left\{(1+\Delta_i)\frac{\exp(2\Z_i\trans\bb_0)
[\int_0^{X_i}\exp\{m(u)\} \B_r(u) du]^{\otimes2}}
{[1+\exp(\Z_i\trans\bb_0)\int_0^{X_i}\exp\{m(u)\}du]^2}\right\}
\ese
is a matrix with all elements of order $O(1)$ uniformly.
It is easily seen that $\V_0(\bb)$ is positive definite,
and $\V_1(\bb)$ is semi-positive definite.

According to \cite{demko1977inverses} and Theorem 4.3 in Chapter 13
of \cite{devore1993constructive}, we have
\bse
\|\V_0(\bb_0)^{-1}\|_\infty\leq C',
\ese
for some constant $0<C'<\infty$. Furthermore,
there exists constants
$0<C''<\infty$ and $0<\lambda<1$ such that
$|\{\V_0(\bb)^{-1}\}_{jk}|\leq C''\lambda^{|j-k|}$ for $j,k=1,\cdots,P_n$.

We want to show that
\bse
\|\{\I-h\V_0(\bb_0)^{-1}\V_1(\bb_0)\}^{-1}\|_\infty
\ese
is bounded. As a result,
\bse
\|\V_n(\bb_0)^{-1}\|_\infty &=& h^{-1}\|\{\I-h\V_0(\bb_0)^{-1}\V_1(\bb_0)\}^{-1}\V_0(\bb_0)^{-1}\|_\infty\\
&\leq& h^{-1}\|\{\I-h\V_0(\bb_0)^{-1}\V_1(\bb_0)\}^{-1}\|_\infty\|\V_0(\bb_0)^{-1}\|_\infty \\
&=& O_p(h^{-1}).
\ese

Denote $\W = -\V_0(\bb_0)^{-1}\V_1(\bb_0)$. There exists constant $0<\kappa<\infty$ such that $|\{\V_1(\bb_0)\}_{jk}|<\kappa$
for $j,k=1,\cdots,P_n$. Hence,
\bse
|\W_{ij}| &=& |\{\V_0(\bb_0)^{-1}\V_1(\bb_0)\}_{jk}| \\
&=& |\sum_{\ell=1}^{P_n} \{\V_0(\bb_0)^{-1}\}_{j\ell}\{\V_1(\bb_0)\}_{\ell k}|\\
&\leq& \sum_{\ell=1}^{P_n} C'' \lambda^{|j-\ell|}\kappa \\
&\leq& 2C''\kappa(1-\lambda)^{-1}\leq \kappa_1,
\ese
where $\kappa_1 = \max\{1,2C''\kappa(1-\lambda)^{-1}\}\geq 1$.

Let $P_n h \leq \kappa_2$, where $1\leq \kappa_2<\infty$ is a constant.
Similar derivation as before shows there exists some constant $0<\wt c<\wt C<\infty$,
such that for arbitrary $\a\in R^{P_n}$,
$\wt c\|\a\|_2^2<\a\trans\V_0(\bb_0)\a<\wt C\|\a\|_2^2$ and
$\wt c\|\a\|_2^2<\a\trans\{\V_0(\bb_0)-h\V_1(\bb_0)\}\a<\wt C\|\a\|_2^2$.
Hence,
\bse
\|(\I+h\W)^{-1}\|_2 &=& \|\{\V_0(\bb_0)-h\V_1(\bb_0)\}^{-1}\V_0(\bb_0)\|_2\\
&\leq&\|\{\V_0(\bb_0)-h\V_1(\bb_0)\}^{-1}\|_2\|\V_0(\bb_0)\|_2\\
&\leq& \kappa_3\equiv\wt C/\wt c.
\ese
where $1\le\kappa_3<\infty$.

In the following, we will use induction to show that
\bse
a_{P_n} &\equiv& |{\rm det}(\I_{P_n}+h\W_{P_n})|\leq (1+h\kappa_4)^{P_n-1},\\
b_{P_n} &\equiv& |{\rm det}(\bfJ_{P_n}+h\W_{P_n})|\leq (\kappa_1+2\kappa_1^2\kappa_2\kappa_3)h (1+h\kappa_4)^{P_n-2},
\ese
where $\bfJ_{P_n} = (\bfJ_{ij})_{1\leq i,j\leq P_n}$ with $\bfJ_{ij}=1$ if $j-i=1$ and $\bfJ_{ij}=0$
otherwise. Here $\kappa_4=4\kappa_1^2\kappa_2(1+\kappa_1\kappa_2\kappa_3)$.

When $P_n=2$,
\bse
a_2 &=& |{\rm det}(\I_2+h\W_2)| \\
&=& |(1+h\W_{11})(1+h\W_{22})-h^2\W_{12}\W_{21}|\\
&\leq& |(1+h\W_{11})(1+h\W_{22})|+|h^2\W_{12}\W_{21}|\\
&\leq& (1+h\kappa_1)^2+h^2\kappa_1^2\\
&=&1+2h\kappa_1+2h^2\kappa_1^2\\
&\leq& 1+4h\kappa_1^2 \leq 1+h\kappa_4.
\ese
Similary, we have
\bse
b_2 &=& |{\rm det}(\bfJ_2+h\W_2)| \\
&\leq& h^2\kappa_1^2+h(1+h\kappa_1)\kappa_1\\
&\leq& (\kappa_1+2\kappa_1^2)h\\
&\leq& (\kappa_1+2\kappa_1^2\kappa_2\kappa_3)h.
\ese
Assume the result holds for $2,\cdots,P_n-1$, then for $P_n$,
denote $\W_{P_n,-P_n} = (W_{P_n 1},\cdots,W_{P_n (P_n-1)})\trans$ and
$\W_{-P_n,P_n} = (W_{1 P_n},\cdots,W_{(P_n-1) P_n})\trans$,
we have
\bse
a_{P_n} &=& |{\rm det} (\I_{P_n}+h\W_{P_n})|\\
&=&|{\rm det} (\I_{P_n-1}+h\W_{P_n-1}) (1+hW_{P_n P_n}-h^2\W_{P_n,-P_n}\trans(\I_{P_n-1}+h\W_{P_n-1})^{-1} \W_{-P_n,P_n}|\\
&\leq& a_{P_n-1} \{1+h\kappa_1+ h^2(n-1)\kappa_1^2\|(\I_{P_n-1}+h\W_{P_n-1})^{-1}\|_2\}\\
&\leq& a_{P_n-1} \{1+h\kappa_1+ h \kappa_1^2\kappa_2\kappa_3\}\\
&\leq& (1+h\kappa_4)^{P_n-1},
\ese
and
\bse
b_{P_n} &=& |{\rm det} (\bfJ_{P_n}+h\W_{P_n})|\\
&=&|\{hW_{P_n 1}-h^2 \W_{P_n,-1}\trans(\I_{P_n-1}+h\W_{P_n-1})^{-1}\W_{-P_n,1}\}{\rm det} (\I_{P_n-1}+h\W_{P_n-1})|\\
&\leq& \{h\kappa_1 +h^2\kappa_1^2(n-1)\|(\I_{P_n-1}+h\W_{P_n-1})^{-1}\|_2\}a_{P_n-1}\\
&\leq& h(\kappa_1 +\kappa_1^2\kappa_2\kappa_3) a_{P_n-1}\\
&\leq& h(\kappa_1 +2\kappa_1^2\kappa_2\kappa_3) (1+h\kappa_4)^{P_n-2}.
\ese
where $\W_{P_n,-1}=(W_{P_n 2},\cdots,W_{P_n
P_n})\trans$ and $\W_{-P_n,1}=(W_{1 1},\cdots,W_{(P_n-1)
1})\trans$.

  Therefore,
\bse
&&\|(\I_{P_n}+h\W_{P_n})^{-1}\|_\infty\\
&&\hskip 5mm=\max_j\sum_{k}|\{(\I_{P_n}+h\W_{P_n})^{-1}\}_{jk}|\\
&&\hskip 5mm \leq \frac{a_{P_n-1}+b_{P_n-1}(P_n-1)}{|{\rm det} (\I_{P_n}+h\W_{P_n})|}\\
&&\hskip 5mm \leq \frac{(1+h\kappa_4)^{P_n}+(\kappa_1 +2\kappa_1^2\kappa_2\kappa_3)\kappa_2 (1+h\kappa_4)^{P_n-2}}{|{\rm det} (\I_{P_n}+h\W_{P_n})|}\\
&&\hskip 5mm \leq \frac{2\kappa_4(1+h\kappa_4)^{\frac{\kappa_2\kappa_4}{h\kappa_4}}}{|{\rm det} (\I_{P_n}+h\W_{P_n})|},
\ese
where the numerator converges to $2\kappa_4\exp(\kappa_2\kappa_4)$ as $h\to0$, or equivalently, $P_n\to\infty$.
Here in the first equation above we use the fact that the $(j,k)\th$ element of
the matrix $(\I_{P_n}+h\W_{P_n})^{-1}$ is the determinant of the matrix $\I_{P_n}+h\W_{P_n}$ without its
$j\th$ column and $k\th$ row, divided by the determinant
  of $\I_{P_n}+h\W_{P_n}$ itself.
 Specifically,
when $j=k$, the absolute value of that $(j,k)\th$ element is
$|{\rm det}( \I_{P_n-1}+h\W_{P_n-1})|/|{\rm det} (\I_{P_n}+h\W_{P_n})|=a_{P_n-1}/|{\rm det} (\I_{P_n}+h\W_{P_n})$;
when $j\neq k$, with certain column operations, we obtain $|{\rm det} (\bfJ_{P_n-1}+h\W_{P_n-1})|/|{\rm det} (\I_{P_n}+h\W_{P_n})|=b_{P_n-1}/|{\rm det} (\I_{P_n}+h\W_{P_n})|$.

Now it remains to show that there exists $\kappa_5>0$, such that
\bse
a_{P_n}=|{\rm det}(\I_{P_n}+h\W_{P_n})|\geq\kappa_5,
\ese
for $P_n$ sufficiently large.
This can be seen from
\bse
a_{P_n} &=&|{\rm det}(\I_{P_n}+h\W_{P_n})|\\
&=&|{\rm det} (\I_{P_n-1}+h\W_{P_n-1}) (1+hW_{P_n P_n}-h^2\W_{P_n,-P_n}\trans(\I_{P_n-1}+h\W_{P_n-1})^{-1} \W_{-P_n,P_n}|\\
&\geq& \{1-h\kappa_1-h^2(P_n-1)\kappa_1^2\kappa_3\}a_{P_n-1}\\
&\geq& (1-h\kappa_1-h\kappa_1^2\kappa_2\kappa_3)a_{P_n-1}\\
&\geq& (1-h\kappa_4)a_{P_n-1}\\
&\geq& (1-h\kappa_4)^{P_n-3}a_2\\
&\geq& (1-h\kappa_4)^{P_n-2}\\
&\geq& (1-h\kappa_4)^{\frac{\kappa_2\kappa_4}{h\kappa_4}}\to\exp(-\kappa_2\kappa_4).
\ese
Thus the result holds for $\kappa_5=\exp(-\kappa_2\kappa_4)$.

Therefore, we have $\|\V_n(\bb_0)^{-1}\|_\infty\leq Ch^{-1}$ for some constant
$0<C<\infty$.

On the other hand, we have
\bse
\|\V_n(\bb_0)^{-1}\|_\infty\geq P_n^{-1/2} \|\V_n(\bb_0)^{-1}\|_2\geq ch^{-1/2},
\ese
for some constant $0<c<\infty$
\citep{horn1990matrix, golub1996matrix}.

The proof for $\left\{-n^{-1}\partial^2
  l_n(\bb_0,\bg_0)/\partial\bg\partial\bg\trans\right\}^{-1}$
  is similar, and hence is omitted.
\end{proof}

\begin{lemma}\label{Lem4}
\bse
\left\|\left\{-n^{-1}\frac{\partial^2l_n(\bb_0,\bg^*)}{\partial \bg \partial \bg \trans}\right\}^{-1}-\V_n(\bb_0)^{-1}\right\|_\infty=O_p(h^{q-1}+n^{-1/2}h^{-1}).
\ese
\end{lemma}
\begin{proof} According to Lemma \ref{Lem2} - \ref{Lem3}, we have
\bse
&&\left\|\left\{-n^{-1}\frac{\partial^2
  l_n(\bb_0,\bg_0)}{\partial\bg\partial\bg\trans}\right\}^{-1}-
\V_n(\bb_0)^{-1}\right\|_\infty\\
&&\hskip5mm=\left
\|\V_n(\bb_0)^{-1}\left[\V_n(\bb_0)-\left\{-n^{-1}\frac{\partial^2
  l_n(\bb_0,\bg_0)}{\partial\bg\partial\bg\trans}\right\}\right]\left\{-n^{-1}\frac{\partial^2
  l_n(\bb_0,\bg_0)}{\partial\bg\partial\bg\trans}\right\}^{-1}\right\|_{\infty}\\
&&\hskip 5mm \le\left\|\left\{-n^{-1}\frac{\partial^2
  l_n(\bb_0,\bg_0)}{\partial\bg\partial\bg\trans}\right\}^{-1}\right\|_{\infty}
\|\V_n(\bb_0)^{-1}\|_{\infty}
\left\|-n^{-1}\frac{\partial^2
  l_n(\bb_0,\bg_0)}{\partial\bg\partial\bg\trans}-
\V_n(\bb_0)\right\|_{\infty}\\
&&\hskip5mm=O_p(h^{-2})O_p(h^{q+1}+n^{-1/2}h)\\
&&\hskip5mm=O_p(h^{q-1}+n^{-1/2}h^{-1}).
\ese
\end{proof}

\begin{lemma}\label{Lem5}
There exists constants $0<c<C<\infty$ such that for $n$ sufficiently large,
with probability approach 1,
for arbitrary $\a\in R^{P_n}$,
\bse
ch\|\a\|_2^2<\a\trans\left\{n^{-1}\sumi \bS_{\bg,i}(\bb_0,m)^{\otimes2}\right\}\a<Ch\|\a\|_2^2.
\ese
\end{lemma}
\begin{proof}
 We have
\bse
n^{-1}\sumi\bS_{\bg,i}(\bb_0,m)^{\otimes2}
&=&n^{-1}\sumi\left[\Delta_i\B_r(X_i)-
(1+\Delta_i)
\frac{\exp(\Z_i\trans\bb)\int_0^{X_i}\exp\{m(u)\} \B_r(u) du
}{1+\exp(\Z_i\trans\bb)\int_0^{X_i}\exp\{m(u)\}du}\right]^{\otimes2}\\
&\le&n^{-1}\sumi C'\left[
\B_r(X_i)^{\otimes2}
+\left\{\int \B_r(u) du\right\}^{\otimes2}\right],
\ese
for some constants $0<C'<\infty$.
Similar derivation leads to
\bse
ch\|\a\|_2^2\le
\a\trans\{n^{-1}\sumi\bS_{\bg,i}(\bb_0,m)^{\otimes2}\}\a\le
Ch\|\a\|_2^2,
\ese
for constant $0<c<C<\infty$.
\end{proof}

\end{appendix}

\end{document}